\documentclass[acmsmall]{acmart}

\usepackage{enumitem,xspace,nicefrac,tcolorbox,xifthen,tikz,subcaption,physics,suffix,wrapfig,algorithm,algpseudocode,bbm}

\usetikzlibrary{shapes, arrows, positioning}

\usepackage{thmtools,cancel,dsfont}
\usepackage[bb=dsserif]{mathalpha}
\usepackage[capitalize]{cleveref}
\usepackage{xcolor}
    \definecolor{myred}{HTML}{ea4335}
    \definecolor{mygreen}{HTML}{41a756}
    \definecolor{myblue}{HTML}{4285f4}

\IfPackageLoadedTF{hyperref}{}
{
\usepackage[colorlinks=true,allcolors=myblue]{hyperref}
}

\usepackage{comment} 
\usepackage{thm-restate}

\newcommand{\inner}[2]{\left \langle #1, #2 \right \rangle}
\newcommand{\inZ}{\in \mathbb{Z}}
\newcommand{\inN}{\in \mathbb{N}}

\newcommand{\inR}{\in \mathbb{R}}

\newcommand{\Z}{\mathbb{Z}}
\newcommand{\N}{\mathbb{N}}

\newcommand{\R}{\mathbb{R}}

\newcommand{\States}{\mathbb{S}}

\newcommand{\E}{\mathbb{E}}

\newcommand{\pr}{\mathbb{P}}
\newcommand{\eqas}{\stackrel{a.s.}{=} }

\newcommand{\Win}{\texttt{w}}
\newcommand{\TotalWin}{\texttt{W}}
\newcommand{\AnyReq}{\texttt{k}}
\newcommand{\TotalAnyReq}{\texttt{K}}

\newcommand{\Req}{\texttt{r}}
\newcommand{\U}{\texttt{U}}
\newcommand{\val}{\texttt{V}}
\newcommand{\valDist}{\mathcal{D}}
\newcommand{\mChain}{\textbf{X}}
\newcommand{\mChainY}{\textbf{Y}}

\newcommand{\dmmf}{\textsc{DMMF}\xspace}
\newcommand{\algorithmName}{Win-Rate Matching\xspace}
\newcommand{\allocFunction}{\mathcal{A}}

\newcommand{\str}{\texttt{S}}
\newcommand{\Strats}{\mathcal{S}}
\newcommand{\thr}{\texttt{Thr}}

\newcommand{\Hist}{H}
\newcommand{\ftd}{\texttt{WRM}}
\newcommand{\strThresh}{\str^{\thr}}
\newcommand{\ValueFunc}{\mathcal{V}}

\newcommand{\ind}[1]{\mathds{1}\left[#1\right]}

\usepackage[suppress]{color-edits}

\addauthor[Eva]{et}{blue}
\addauthor[Giannis]{gf}{teal}
\addauthor[Sid]{sb}{magenta}
 \addauthor[Chido]{co}{orange}
\begin{document}

% Articles V9pmc001-V9pmc060 use:
\received{October 2024}
\received[revised]{January 2025}
\received[accepted]{January 2025}

\setcopyright{acmlicensed}
\acmJournal{POMACS}
\acmYear{2025} \acmVolume{9} \acmNumber{1} \acmArticle{2}
\acmMonth{3}\acmDOI{10.1145/3711695}
\title{Allocating Public Goods via Dynamic Max-Min Fairness: Long-Run Behavior and Competitive Equilibria}

%\author{Anonymous Authors}
\author{Chido Onyeze}
\email{chidoonyeze@cs.cornell.edu}
\orcid{0009-0001-7229-3338}
\affiliation{%
    \institution{Cornell University}
    \city{Ithaca}
    \state{New York}
    \country{USA}
}

\author{Siddhartha Banerjee}
\email{sbanerjee@cornell.edu}
\orcid{0000-0002-8954-4578}
\affiliation{%
    \institution{Cornell University}
    \city{Ithaca}
    \state{New York}
    \country{USA}
}

\author{Giannis Fikioris}
\email{gfikioris@cs.cornell.edu}
\orcid{0000-0002-4920-478X}
\affiliation{%
    \institution{Cornell University}
    \city{Ithaca}
    \state{New York}
    \country{USA}
}

\author{\'Eva Tardos}
\email{eva.tardos@cornell.edu}
\orcid{0000-0002-2978-1475}
\affiliation{%
    \institution{Cornell University}
    \city{Ithaca}
    \state{New York}
    \country{USA}
}


\begin{abstract}
    Dynamic max-min fair allocation (DMMF) is a simple and popular mechanism for the repeated allocation of a shared resource among competing agents: in each round, each agent can choose to request or not for the resource, which is then allocated to the requesting agent with the least number of allocations received till then.
Recent work has shown that under DMMF, a simple threshold-based request policy enjoys surprisingly strong robustness properties, wherein each agent can realize a significant fraction of her optimal utility irrespective of how other agents' behave.
While this goes some way in mitigating the possibility of a `tragedy of the commons' outcome, the robust policies require that an agent defend against arbitrary (possibly adversarial) behavior by other agents. This however may be far from optimal compared to real world settings, where other agents are selfish optimizers rather than adversaries. Therefore, robust guarantees give no insight on how agents behave in an equilibrium, and whether outcomes are improved under one.

Our work aims to bridge this gap by studying the existence and properties of equilibria under DMMF. To this end, we first show that despite the strong robustness guarantees of the threshold based strategies, \emph{no Nash equilibrium exists} when agents participate in DMMF, each using some fixed threshold-based policy. 
On the positive side, however, we show that for the symmetric case, a simple data-driven request policy guarantees that no agent benefits from deviating to a different fixed threshold policy. In our proposed policy agents aim to match the historical allocation rate with a vanishing drift towards the rate optimizing overall welfare for all users.
Furthermore, the resulting equilibrium outcome can be significantly better compared to what follows from the robustness guarantees.

Our results are built on a complete characterization of the steady-state distribution under DMMF, as well as new techniques for analyzing strategic agent outcomes under dynamic allocation mechanisms; we hope these may prove of independent interest in related problems.
\end{abstract}

\begin{CCSXML}
<ccs2012>
   <concept>
       <concept_id>10003752.10010070.10010099.10010100</concept_id>
       <concept_desc>Theory of computation~Algorithmic game theory</concept_desc>
       <concept_significance>500</concept_significance>
       </concept>
   <concept>
       <concept_id>10003752.10010070.10010099.10010101</concept_id>
       <concept_desc>Theory of computation~Algorithmic mechanism design</concept_desc>
       <concept_significance>500</concept_significance>
       </concept>
   <concept>
       <concept_id>10003752.10010070.10010099.10010103</concept_id>
       <concept_desc>Theory of computation~Exact and approximate computation of equilibria</concept_desc>
       <concept_significance>500</concept_significance>
       </concept>
   <concept>
       <concept_id>10003752.10010070.10010099.10010104</concept_id>
       <concept_desc>Theory of computation~Quality of equilibria</concept_desc>
       <concept_significance>500</concept_significance>
       </concept>
   <concept>
       <concept_id>10003752.10010070.10010099.10010105</concept_id>
       <concept_desc>Theory of computation~Convergence and learning in games</concept_desc>
       <concept_significance>300</concept_significance>
       </concept>
 </ccs2012>
\end{CCSXML}

\ccsdesc[500]{Theory of computation~Algorithmic game theory}
\ccsdesc[500]{Theory of computation~Algorithmic mechanism design}
\ccsdesc[500]{Theory of computation~Exact and approximate computation of equilibria}
\ccsdesc[500]{Theory of computation~Quality of equilibria}
\ccsdesc[300]{Theory of computation~Convergence and learning in games}


\keywords{Game Theory, Mechanism Design, Analysis of Random Processes, Equilibrium Computation}

\maketitle

\section{Introduction}

Designing mechanisms for the repeated allocation of public goods is a long-studied problem in economics, which has gained recent attention due to its increasing use in practice. By public good here, we refer to any centralized resource which is shared (i.e., collectively owned and administered) among a set of agents -- for example, a central computing cluster in a university or company, or a large scientific apparatus shared by many research groups, or use of shared spaces like conference rooms or parking spots, or even perishable items like donations to food banks. The common tension in such settings is that while all agents have nominal rights to equal use of the resource, their need or value for the resources varies over time and is not known in advance, making it natural for actual usage to be determined dynamically via some mechanism. The aim is to allocate in a way which is ideally both fair and efficient. Moreover, since the resource is shared, these mechanisms typically do not involve money.

A wide variety of mechanisms have been proposed in different contexts for the repeated public goods setting.
On one hand, simple static policies such as round-robin sharing and fixed priority rules are fair by design, but may not be efficient if agents have equal priority and utilities are time-varying and stochastic. This is remedied by dynamic mechanisms such as usage-based priority rules and `pseudo-market’ mechanisms which allow agents to better express when and how much they need the resource. The combination of lack of money and repeated allocation, however, makes it challenging to understand how selfish agents behave in such mechanisms. In this work, we study strategic behavior in the simplest of such settings, and for a natural and widely-studied mechanism. We show that no equilibrium exists under the most natural class of strategies, and offer a bit more complex strategy resulting in approximate equilibrium with high social welfare.

In more detail, the main setting we consider is as follows: a set of $n$ agents compete over multiple rounds for the right to use a single atomic resource in each round. Each agent $i$ has a random private value $\val_{i}[t]$ for receiving the item in each round $t$, with the values being sampled independently across agents and across rounds. The allocation in round $t$ is determined via what is referred to as \emph{Dynamic Max-Min Fair} (\dmmf) in the literature: each agent chooses whether or not to request for the item in that round, following which, in case of multiple requests, the item is awarded to the agent who has received the least number of items till then. This is a natural dynamic variant of a round-robin policy which allows agents to request for the item only when they have high values, while trying to ensure each agent wins the item roughly an equal fraction of the time. While our focus is primarily on this setting with homogeneous agents and symmetric \dmmf, several of our results also extend to the more general asymmetric variant of \dmmf, wherein the priority order over agents is determined by the fraction of rounds won by each agent $i$ divided by an exogenous `fair share' that the agent is endowed with.

The above mechanism and its variants (including more complex pseudo-market mechanisms where agents can bid using artificial tokens) are natural fits for public goods settings, and so are often used in practice. While they appear simple, their state-space (comprising of the historical allocations and/or budgets of all agents) is large, and so considering the full complexity of agents' policies does not seem promising as a practical guide for modeling agent behavior. As a result not much is known about how agents compete in such mechanisms, and what is the efficiency of an equilibrium outcome. Recent work~\cite{DBLP:conf/sigecom/GorokhBI21,DBLP:conf/sigecom/BanerjeeFT23} has made progress on both these issues by showing that a simple class of myopic \emph{threshold strategies}, where an agent requests whenever she has sufficiently high value regardless of system state, has strong robustness guarantees. 
Such results were first shown for pseudo-market mechanisms, where~\cite{DBLP:conf/sigecom/GorokhBI21} showed that playing a particular threshold strategy could help an agent win at least $\nicefrac{1}{2}$ the rounds when her value $\val_i[t]$ is in its top $1/n$-quantile (i.e., her so-called `ideal utility'), irrespective of how other agents behave. This was subsequently generalized to more complex allocation settings as well as to the simpler \dmmf mechanism~\cite{DBLP:conf/sigecom/BanerjeeFT23,fikioris2023online}; the latter work also showed the strategy could be modified to give stronger bounds for smoother value distributions (in particular, a $1-O(1/\sqrt{n})$ guarantee for $\val_i[t]\sim U[0,1]$). 
%\etcomment{I was trying to tone down our love for previous work. To me it seems that such love is counter-productive if we want this paper accepted.}

While the above robustness guarantees are promising, the corresponding policies are conservative as they are designed to defend against arbitrary (possibly adversarial) behavior by other agents. This may be far from optimal in real world settings, where other
agents are selfish optimizers rather than adversaries, and so may allow an agent to realize a much higher utility by requesting more aggressively. Indeed, in related \emph{distribution-aware} models of public resource allocation (where the principal knows the value distribution of each agent and can use it in the mechanism) there are more complex mechanisms which admit approximate equilibrium policies with vanishing regret in welfare~\cite{jackson,guo2010,gorokh2017,balseiro2017,blanchard2024near}. 
In contrast, for distribution-agnostic mechanisms for repeated public resource allocation games, \emph{almost nothing is known about equilibrium outcomes}. 
Our work makes a significant first step in filling this void.

\subsection{Competitive Equilibria in \dmmf: Our Results in Brief}

As mentioned above, we study the use of the \dmmf mechanism for allocating a single resource in each round between agents with identical priority and values drawn $i.i.d$ across time and independent across agents. We consider the problem over an infinite-horizon, though our results can extend to give finite-horizon guarantees. 
%\etdelete{Finally, as in past work, we focus on the simple class of threshold strategies. }\etcomment{I propose to delete this last sentence? Dynamic threshold strategies wont feel like part of what is claimed here. }

Given the strong robustness guarantees for threshold strategies in \dmmf~\cite{DBLP:conf/sigecom/GorokhBI21,DBLP:conf/sigecom/BanerjeeFT23,fikioris2023online}, and in light of the existing approximate equilibrium results for {similar myopic strategies in distribution-aware settings~\cite{gorokh2017,balseiro2017,blanchard2024near}}, one would hope that similar equilibria exist in our setting. Our first main result (\cref{thm: No Pure Nash Eq}) puts an end to these hopes,
showing that even in the simplest variants of our setting (with $n=2$ symmetric agents and a distribution with $2$ non-zero values), \emph{there is no pure-strategy Nash equilibrium in fixed threshold strategies}~\footnote{Note that while mixed NE always exist, these are not satisfying as models of agent behavior in our setting -- in particular, this would correspond to an agent randomly choosing a threshold at the start, and then playing the same threshold in all rounds, irrespective of whether it is good or not.}.    
Indeed, our analysis shows that our setting suffers from the archetypal tragedy-of-the-commons outcome: if all other agents agree to voluntarily restrain their actions by playing some fixed threshold strategy (which may even be the robust strategy), then an agent's best response is to overuse the resource by requesting for it at all times! Of course, \dmmf ensures the deviating agent is not always successful -- nevertheless, she can make her resulting utility {significantly} better via such a deviation.

\cref{thm: No Pure Nash Eq} thus shows that despite their robustness properties, threshold strategies do not lead to any equilibrium outcome. One could perhaps circumvent this by allowing for more complex strategies where an agent uses different thresholds depending on her relative past allocation fraction (and hence, her priority under \dmmf) -- this however destroys the appealing simplicity of threshold strategies. Ideally, one would like to engineer good equilibrium outcomes using only mild modifications of threshold strategies, but it is unclear if this is possible.

Our second main result provides a partial answer to this, in the form of a novel data-driven threshold policy which we call \algorithmName (\cref{alg:WinRateMatching}). 
At a high-level, the mechanism requires agents to adjust their request rate in each round so as to match the overall historical allocation rate, along with a vanishing drift towards the common request rate that optimizes overall welfare for all users. Note that the resulting request threshold is still independent of the current state, but rather, depends only on a single statistic of the history.
While the strategy can be implemented in a fully decentralized way, for the sake of intuition, \algorithmName\ is more naturally viewed as a `mechanism with advice', where the principal recommends a request threshold to all agents, who then can choose to follow it, or to use an independent threshold strategy. 

In our second main result (\cref{thm:winratematchequilibrium}), we show that following the recommended threshold under \algorithmName is an approximate Nash equilibrium. 
The intuition for this is that if an agent now tries to deviate, this affects the overall request rate, leading to other agents to follow the deviation, and hence nullify its effect.
Furthermore, we show that the resulting equilibrium outcome can be significantly better compared to what follows from the robustness guarantees (\cref{thm:utility_comparison}); in particular, we show that under our equilibrium, we get within a $1-O\left(\log n/n\right)$ fraction of the optimal welfare (as opposed to $1-O\left(1/\sqrt{n}\right)$ under the robust policy) when $\val_i[t] \sim U[0,1]$, while in the worst-case, the competitive ratio improves from $1/2$ for robust policies to $1-1/e$.

\subsection{Overview of our Technical Developments}

In order to study competitive equilibria of \dmmf, we first need to characterize the long-run utilities of agents under different strategy profiles. While doing so under completely general strategies seems infeasible, even studying fixed threshold strategies is challenging. One advantage of threshold strategies is that it essentially reduces the problem of characterizing agent utilities to characterizing their win rate (i.e., the rate at which they are allocated the item; (see~\cref{prop: expected utility expression}); however, even this involves analyzing a highly coupled $n$-dimensional Markov chain.

An important intermediate technical result in this work is a \emph{complete characterization of the long-run behavior of \dmmf under arbitrary threshold strategy profiles} (i.e., in terms of each agent's fair share $\alpha_i$ and request probabilities $p_i$).
A first intuition may be that since \dmmf gives higher priority to agents with lower win rate relative to their share, it leads to some form of state-space collapse wherein all agents win the item at roughly the same relative rate. This however is not always the case, as is illustrated in~\cref{fig:dmmf_progression}, where we show a sample path of \dmmf with $4$ agents, equal fair shares, and different request thresholds. As can be seen from the figure, the win rate trajectories of the agents all grow linearly, but cluster into a collection of subgroups, with trajectories within the same subgroup appearing to share the same relative rate, so the state-collapse happens for each group separately. 

Our characterization (in~\cref{sec:analysis_thrs,sec:decomposition}) formally establishes this `subgroup state-space collapse' phenomenon. In more detail, in~\cref{thm: condition for stablity}, we first establish necessary and sufficient conditions for a global state-space collapse and what happens when some agent violates this condition; we use this for our equilibrium characterizations in~\cref{sec:equilibrium}.
Our condition (see~\cref{def:subgpstable}) can be viewed as a variant of a complete resource pooling condition, wherein we require the normalized request rate of every subgroup of agents to be (weakly) greater than that of the full set $[n]$; intuitively, this gives \dmmf enough room to boost the win rates of agents in that set (by giving them higher priority) whenever they fall back from the overall weighted average.
Subsequently, we generalize our characterization in~\cref{thm: characterize number of wins} by describing how to partition agents based on their $(\alpha_i,p_i)$ into subgroups, such that agents within each subgroup have their (normalized) win rate trajectories be mean reverting about a common line. We hope our characterization, and the associated novel drift-based arguments, prove useful for other settings.

Returning to our primary motivation of studying strategic behavior under \dmmf, in~\cref{thm: No Pure Nash Eq}, we show that no fixed threshold strategy equilibrium exists under fixed threshold strategies. This result depends on our state-space collapse characterization, in particular, to determine the best response function for choosing a fixed rate in the special case of 2 players. 
In the case of $n$ symmetric agents, we offer the design and analysis of the \algorithmName policy. Analyzing competitive behavior in a dynamic policy is in general very challenging. Our result is based on two novel ideas: First, in~\cref{Thm:ThresholdConvergesWhenAllRequest}, we show that when all agents follow the policy, then we can use techniques from stochastic approximation to establish that the request rates converge to the optimal request rate. On the other hand, in~\cref{Thm:ThresholdConvergesWhenDeviator} we show that when an agent deviates while others follow the policy, we show that this leads to the request rates converging to the deviating agent's rate, which is therefore suboptimal. Thus, adapting request rates in this simple data-driven manner leads to disincentivizes agents from deviating from the optimal request rate. Again, we believe this idea may prove of independent interest in other dynamic allocation games.

\subsection{Related work} \label{ssec:related}

As we mentioned before, there has been significant recent interest on the repeated public goods setting, owing to several successful deployments of such settings in practice (for example, see~\cite{prendergast2022allocation,budish2017course}).
The initial focus was on pseudo-market mechanisms~\cite{cavallo2014incentive, jackson, guo2010} in distribution-aware settings; this line of work culminated in establishing mechanisms which achieve vanishing regret compared to the first-best, but make extensive use of knowledge of the value distributions~\cite{gorokh2017,balseiro2017,blanchard2024near}.
A later line of work extends these ideas to distribution-agnostic settings.
To make up for the lack of distribution knowledge, \citet{DBLP:conf/sigecom/GorokhBI21} define the ideal utility benchmark, which has become the standard benchmark for individual agent utility guarantees; they then show that under a simple first-price auction, every agent can always guarantee half of her ideal utility robustly for any value distribution.
\citet{DBLP:conf/sigecom/BanerjeeFT23} generalize the previous guarantee to reusable resources, i.e., where agents may want the resource for multiple consecutive rounds.
\citet{fikioris2023online} generalize both the results of \cite{DBLP:conf/sigecom/GorokhBI21, DBLP:conf/sigecom/BanerjeeFT23} using the {much simpler} \dmmf mechanism.
They re-establish the previous half ideal utility guarantee and extend these guarantees to distributions that are not worst-case; for example, they prove that an agent with values sampled from a uniform distribution can guarantee a $1 - O(1/\sqrt{n})$ fraction of her ideal utility.
\cite{fikioris2023online} also studies settings when an agent's values can be dependent across rounds.
While these latter guarantees are weaker compared to the earlier distribution-aware mechanisms, this is in large part because they assume adversarial behavior by the other agents.
Indeed, recent empirical studies of equilibria in pseudo-market mechanisms~\cite{elokda2022carma, elokda2023self} suggest their performance is much better than the robust guarantees.
Our work aims to bridge the two by considering simple distribution-agnostic mechanisms (in particular, \dmmf), but returning to the more reasonable assumption that agents behave under an equilibrium; by doing so, we are able to get much better utility guarantees.

%\gfdelete{
%The \dmmf mechanism is inspired by the classical max-min fairness mechanism (i.e., single shot \dmmf), most commonly used for divisible resources, which led to many heavily used in practice mechanisms like DRF \citet{DBLP:conf/nsdi/GhodsiZHKSS10}. \citet{freeman2017fair} focus on distributing a divisible resource with an incentive compatible mechanism and study the trade-off of different objectives, e.g., envy-freeness, sharing incentives, etc. They also prove that for that case, the \dmmf mechanism is not incentive compatible. \citet{fikioris2024incentives} study how far from incentive compatible \dmmf is; they also focus on agents with Leontief preferences over multiple goods. \citet{DBLP:conf/osdi/VuppalapatiF0CK23} implement the above mechanism and focus on its experimental performance in practice. \citet{elokda2022carma,elokda2023self} focus on a pseudo market mechanism and prove it has a static equilibrium where each agent has a bidding strategy given his remaining budget; they then focus on its empirical performance in practical examples.
%}

{
Another notable feature of our work is that we use novel performance characterizations of Markovian models of resource allocation in order to study strategic outcomes in such settings. The paradigmatic examples of this are the literature on strategic queueing (\citet{hassin2003queue} and \citet{hassin2016rational} provide a comprehensive summary),
and the use of the network utility maximization paradigm (building on the seminal work of \citet{kelly1998rate}) for studying strategic behavior in packet networks (see chapter $6$ of~\citet{shakkottai2008network}).
More recently, performance characterizations have been used to characterize equilibria in decentralized models of switch scheduling~\cite{gaitonde2023price}; the drift arguments we use in our work are closely related to techniques used in this work.
However, a critical difference of our setting from the others is that our associated stochastic model has the form of a \emph{loss network}, wherein agents' needs are instantaneous, and the principal can (and must) reject requests. In contrast, in earlier work on strategic resource allocation in queueing settings, all requests must be accepted, but can be appropriately delayed.}

{
Apart from the above, there is also extensive work looking at a wide variety of mechanisms for static allocations settings and/or settings with divisible resources.
Such models have been studied for a wide variery of applications, including data-center center applications~\cite{bonald2015multi, bonald2001impact, bonald2006queueing, joe2013multiresource}, cloud computing~\cite{DBLP:conf/nsdi/GhodsiZHKSS10, DBLP:conf/osdi/ShueFS12, DBLP:conf/sigcomm/GrandlAKRA14, DBLP:conf/osdi/GrandlKRAK16,DBLP:conf/sigecom/ParkesPS12}, and food-bank allocations~\cite{sinclair2022sequential,banerjee2023online,yin2022optimal}.
Most of these works do not explicitly consider incentives of participants, but rather focus on fairness-efficiency tradeoffs.
More recently, a line of work has tried extending these results to the more complicated dynamic setting, where agents have time-varying utilities and may behave (such as report utilities) to optimize their value; \cite{DBLP:conf/sigmetrics/FreemanZCL18, fikioris2024incentives, DBLP:conf/osdi/VuppalapatiF0CK23, elokda2022carma, elokda2023self} analyze the incentive, efficiency, and fairness properties of simple policies for online divisible resource allocation.
Specifically, \cite{fikioris2024incentives, DBLP:conf/osdi/VuppalapatiF0CK23} focus on the \dmmf mechanism.
}

\section{Preliminaries}
\label{sec:prelim}
In this section, we will formalize the repeated public good allocation problem and the \dmmf mechanism we consider to solve it. 
We also give some basics about the strategies that the agents consider.

%\subsection{The Repeated Public Good Allocation Problem}

The allocation problem consists of a principal and $n$ agents, and takes place over rounds. In each round $t\inN$, the principal receives access to an indivisible resource and must determine which agent (if any) to allocate the resource to on that round. At the start of the round, each agent $i \in [n]$ observes a private value $\val_i[t] \in [0, 1]$ drawn from a distribution $\valDist_i$; the values are independent and identically distributed across rounds, and independent across agents. 

After seeing her value for round $t$, an agent can decide whether or not she wants to request the resource in that round.
We define $\Req_i[t]$ to be the indicator for the event that agent $i$ requests in round $t$, $\Win_i[t]$ to be the indicator of the event that agent $i$ ``wins'' (i.e., is allocated the item) in round $t$, and $\TotalWin_i[t] = \sum_{\tau=1}^t\Win_i[\tau]$ denote the total wins agent $i$ has up to time $t$. 
At the end of the round, agent $i$ receives additional utility $\val_i[t]$ if allocated the resource (i.e., if $\Win_i[t]=1$), and $0$ utility otherwise. 
For any $t\inN$ and agent $i$, we can thus define the total utility that the agent receives after $t$ rounds as $\U_i[t] = \sum_{\tau=1}^t\val_i[\tau]\Win_i[\tau]$.

The aim of the mechanism is to allocate the resource in a way which is simultaneously fair and efficient. The notion of fairness has been formalized in past work as essentially requiring that the fraction of rounds each agent is allocated is approximately proportional to some notion of her priority or \emph{share}~\cite{DBLP:conf/sigecom/GorokhBI21}.
In this work we will focus on agents with equal {share}, but some of our results extend to the case when each agent has an exogenously defined fair share $\alpha_i\ge 0$, with $\sum_i \alpha_i=1$, and in this case, allocations proportional to $\alpha_i$ are viewed as fair.

\subsection{Dynamic Allocation Mechanisms and \dmmf}
\label{ssec:mechanism}

Next, we formalize a general class of allocation mechanisms for our problem, as well as define the Dynamic Max-Min Fair (\dmmf) allocation mechanism, that is the focus of our work.

We define the history $\Hist[T] \in (\{0, 1\}^n)^t$ of the mechanism up to round $T$ to be a tuple consisting of $\Req_j[t]$ for all $j \in [n]$ and $t \in [T]$.
A mechanism is now defined as a function that in each round $t\inN$ maps the history up to round $t$ to an agent  who will be allocated in that round, i.e., 
\begin{equation*}
\allocFunction: \cup_{t=1}^\infty(\{0, 1\}^n)^t \rightarrow [n]\cup \{\emptyset\}
\end{equation*}
Here $\allocFunction(\Hist[t])$ is the agent who gets the resource in round $t$ if $\allocFunction(\Hist[t]) \in [n]$; if $\allocFunction(\Hist[t]) = \emptyset$ then the resource is not allocated in that round. 
We impose the requirement that if $\allocFunction(\Hist[t]) = i \in [n]$, then $\Req_i[t] = 1$, i.e., the resource in a round is only allocated to agents who request it.
Connecting back to the notation introduced earlier, we thus have $\Win_i[t] = 1$ if and only if $\allocFunction(\Hist[t]) = i$.
%\sbdelete{Let $\TotalWin_i[t] = \sum_{s=1}^t\Win_i[s]$ denote the total wins agent $i$ has at time $t$. Note that for any fixed allocation function, $\Win_i[t]$ is a function of $\Hist[t]$ for all $s \leq t$. Hence, with knowledge of the history up to round $t$, it can deduced which agent won the resource on each round up to round $t$. }

In this paper we focus on the Dynamic Max-Min Fair (\dmmf) allocation mechanism. This is an instance of the general allocation mechanism where the allocation function is defined as follows:
\begin{equation*}
    \allocFunction(\Hist[t]) = \arg\min_{i \in [n]}\left\{\TotalWin_i[t-1]: \Req_i[t] = 1\right\}
\end{equation*} 
with ties broken arbitrarily (for simplicity of presentation, we will assume we break ties in lexicographic order). In other words, on any round, among requesting agents, the resource goes to the one that has won the fewest rounds so far. \dmmf also naturally generalizes to incorporate exogenous fair shares $\{\alpha_i\}_{i\in[n]}$~\cite{fikioris2023online}; in~\cref{alg:DMMF}, we formally present this generalized \dmmf mechanism.

The idea behind \dmmf is to link each agent's allocation priority in a given round to how often they have been allocated in previous rounds. This should encourage agents to only request the resource on rounds when they have sufficiently high value for it, incorporating the opportunity cost of not getting the resource in a future round. Hence, this mechanism aims to align the agent's goal of utility maximization with the fair share constraint. 
Furthermore, by virtue of how simple the agent's decision space is (one bit per round) and how simple the allocation rule is, we believe that the \dmmf mechanism has great practical use.
%\gfedit{A successful deployment of \dmmf can be found in \cite{DBLP:conf/osdi/VuppalapatiF0CK23} for achieving long-term fairness with a divisible resource.}
%\gfcomment{The Elodka paper is a pseudo-market, do we want to mention that?}\etcomment{maybe better not to here}

\algrenewcommand\algorithmicrequire{\textbf{Input:}}

\begin{algorithm}
\caption{Dynamic Max-Min Fair Allocation Mechanism}
\label{alg:DMMF}
\begin{algorithmic}
\Require Agents $[n]$, fair shares of agents $\{\alpha_i\}_{i \in [n]}$
\State $\TotalWin_i[0] \gets 0$ for all $i \in [n]$
\For{$t = 1, 2, \cdots$}
    \State Collect a possible request $\Req_i[t] \in \{0, 1\}$ from each agent
    \State $S  = \left\{i \in [n]:  \Req_i[t] = 1\right\}$
    \State $w[t] = \arg\min_{i \in S} \left\{\frac{\TotalWin_i[t-1]}{\alpha_i}\right\}$ where, with ties broken arbitrarily.     
    \State $\TotalWin_i[t] \gets \TotalWin_i[t-1] + \mathbbm{1}(w[t] = i)$ for all $i \in [n]$
\EndFor
\end{algorithmic}
\end{algorithm}

\subsection{Agent Strategies under \dmmf}
\label{ssec:threshold_strategies}

We assume agents are fully aware of the allocation mechanism and which agents have requested on any of the previous rounds. On the other hand, an agent's actual value for the resource in any round is only known by the agent.
%\etcomment{in our strategy all the agents need to know is who is getting allocated and not the whole history, right?}
We use $\vec{\str} = (\str_1, \cdots, \str_n)$ to denote the set of agent strategies, and define $\U_i[t](\vec{\str})$ to be the utility realized by any agent $i$ and round $t$ when agents follow $\vec{\str}$ (i.e., each agent $j\in[n]$ uses strategy $\str_j$) under \dmmf. %Since the values are drawn from a distribution, $\U_i[t](\vec{\str})$ is a random variable.
Moreover, for any strategy vector $\vec{\str} = (\str_1, \cdots, \str_n)$ , agent $i$ and strategy $\str'$, we use the shorthand $\U_i[t](\str', \vec{\str}_{-i}) = \U_i[t](\str_1, \cdots, \str_{i-1}, \str',\str_{i+1}, \cdots, \str_n)$.

We define a \textit{dynamic threshold strategy} to be a strategy in which an agent requests only if her value is above a certain threshold. 
Specifically, in any round $t\inN$, agent $i$ selects a probability $p[t]$ (possibly history dependant), and requests only if their value $\val_i[t]$ is in the top $1-p[t]$ quantile (i.e., to maximize their expected value conditioned on their request probability being $p[t]$).
Formally, a dynamic threshold strategy $\str_i$ for agent $i$ is a sequence of maps $\str_i[t]: (\{0, 1\}^n)^{t-1} \rightarrow [0, 1]$ such that $p[t] = \str_i[t](\Hist[t-1])$ is the probability that agent $i$ requests  with, and $\lambda(p[t]) = \sup\{\lambda \in [0, 1]: \pr(\val_i > \lambda) \geq p[t]\}$ the resulting threshold that indicates when to request.
For continuous value distributions, the agent requests only if $\val_i[t] > \lambda(p[t])$; 
%\sbdelete{ and does not request if $\val_i[t] \leq \lambda(p[t])$. For distributions with atoms, we have to be careful when $\val_i[t] = \lambda(p[t])$ happens with positive probability; in which case, the agent requests with probability 
%$$\frac{\pr(\val_i[t] > \lambda(p[t])) - p[t]}{ \pr(\val_i = \lambda(p[t]))}.$$
%Hence} 
for distributions with atoms, under strategy $\str_i[t]$ an agent $i$ requests when their value is strictly greater than $\lambda(p[t])$, and when their value equals $\lambda(p[t])$, requests with a probability that ensures their probability of requesting overall is exactly $p[t]$.

Note that not all strategies are dynamic threshold strategies; for example, an agent can request the resource with a different probability for each value $\val_i[t]$. The following proposition however asserts that any strategy is dominated by a dynamic threshold strategy.
%Let $\U_i[t](\str_1, \cdots, \str_n)$ be the total realized utility in round $t$ received by agent $i$ when agent $j$ uses strategy $\str_j$ for all $j$. Since the values are drawn from a distribution, $\U_i[t](\str_1, \cdots, \str_n)$ is a random variable. For simplicity of notation, let $\U_i[t](\str_1, \cdots, \str_{i-1}, \str',\str_{i+1}, \cdots, \str_n) = \U_i[t](\str', \vec{\str}_{-i})$ where $\vec{\str} = (\str_1, \cdots, \str_n)$.

\begin{restatable}{proposition}{PropThresholdsDominate}
\label{prop:PropThresholdsDominate}
For any strategy $\str_i$, there exists a dynamic threshold strategy $\str'_i$ s.t. $\forall\,t\inN$
$$\U_i[t](\str'_i,\vec{\str}_{-i}) \geq \U_i[t](\str_i,\vec{\str}_{-i}) - o(t)$$
almost surely~\footnotemark{} for any strategy vector $\vec{\str}_{-i}$. 
\end{restatable}

\footnotetext{For notational convenience, we say a sequence of events indexed by $t \in \N$ happens \emph{almost surely} to mean that the $t$-th event occurs with probability at least $1-O\left(1/t^2\right)$. We also use the notation $\eqas$ to mean two time-indexed random variables are equal almost surely for all $t\inN$.}

%\sbreplace{We prove this proposition by noticing that any strategy can be improved by considering a corresponding dynamic threshold strategy. In particular, the agent can use the dynamic threshold strategy corresponding to identical request probabilities. This leads to identical requests, leaving the history unaltered in distribution, while each successful request results in utility that is not less.}
This follows by picking $\str'$ to have the the same request probability in each round as $\str_i$, resulting in each successful request to have (weakly) higher utility. Overall this results in (weakly) higher utility in expectation, and hence sub-linear fluctuations almost surely. The formal proof is given in~\cref{sec:appendixA}.

Henceforth, without loss of generality, we assume all strategies are dynamic threshold strategies.
Define $\ValueFunc_{i}(p)$ to be the expected value of agent $i$ conditioned on requesting via a dynamic threshold strategy with probability $p$: when $\valDist_i$ is continuous, $\ValueFunc_{i}(p) = \E_{\val_i \sim \valDist_i}[\val_i|\val_i > \lambda(p)]$; more generally, for distributions with atoms,
$$\ValueFunc_{i}(p) = \frac{1}{p}\left(\int_{(\lambda(p), 1]} v\; d\mu_{i}(v) 
 + \lambda(p) \cdot (\pr_{\val_i \sim \valDist_i}(\val_i \geq \lambda(p)) - p)\right)$$
where $\mu_i(v)$ is the probability measure of $\valDist_i$. Note that \emph{$\ValueFunc_{i}(p)$ is continuous for any distribution $\valDist_i$}; we rely on this in the coming results.

%\etdelete{$\ValueFunc_{i}(\cdot)$ now easily defines the standard benchmark in this setting, the \textit{ideal utility}. Specifically, agent $i$'s ideal utility is $\frac{1}{n} \ValueFunc_{i}(\frac{1}{n})$, i.e., the utility they get in expectation if they get allocated all the requests in the top $1/n$ percentile of their value distribution.}\etcomment{This is defined again in the next section, and there is motivated. Here it seems only confusing.}

%\subsubsection{Fixed Threshold Strategies} \label{ssec:threshold_strategies}

A special subclass of dynamic threshold strategies are those in which $\str_i[t](\Hist[t-1])$ is a static function (i.e., $\str_i[t](\Hist[t-1]) = p$ for some fixed $p \in [0, 1]$ depending on the agent's distribution $\valDist_i$ and fair share $\alpha_i$, but independent of $\Hist[t-1]$ or time $t$). We refer to strategies of this form as static threshold strategies or simply \emph{threshold strategies}; moreover, we use $p$-threshold strategy to indicate $\str_i[t](\Hist[t-1]) = p$.

Threshold strategies are appealing as a simple behavior policy for agents in such settings, as they correspond to them requesting at a fixed rate based on when they most need the resource, and regardless of the history and the behavior of other agents.
Consequently, they were used in prior work to define a natural utility benchmark called the \textit{ideal utility}~\cite{DBLP:conf/sigecom/GorokhBI21}, which captures how much utility agent $i$ should hope for under no competition, while observing their fair share constraint. Formally, in the symmetric setting, the ideal utility of agent $i$ is $\frac{1}{n} \ValueFunc_{i}(\frac{1}{n})$, i.e., the utility they get in expectation if they get allocated all the requests in the top $1/n$ percentile of their value distribution. (More generally, an agent with fair share $\alpha_i$ has ideal utility $\alpha_i \ValueFunc_{i}(\alpha_i)$).
More significantly, threshold strategies are also known to provide strong individual utility guarantees irrespective of the behavior of other agents~\cite{DBLP:conf/sigecom/GorokhBI21,DBLP:conf/sigecom/BanerjeeFT23}, and have also been shown to be rich enough to enable distribution-dependent robustness guarantees, as well as handle correlated values~\cite{fikioris2023online}.

Due to the strong guarantees offered by such simple strategies, it is reasonable first to study the \dmmf mechanism when everyone uses such a strategy.
In~\cref{sec:analysis_thrs}, we study the long-term effect when every agent is using a threshold strategy.
In~\cref{ssec:nonexistence}, we prove that even for very simple examples, there might be no equilibrium when agents only use threshold strategies.

\section{The Long-Run Behavior of Threshold Strategies Under \dmmf} 
\label{sec:analysis_thrs}

Our goal is to study equilibria under the \dmmf mechanism when all agents use threshold strategies, but we first need to understand the long-run utility of each agent.
To do so, we assume for this section that the agents use some given vector of threshold strategies $\vec{\str} = (\str^{\thr}_{p_1}, \cdots, \str^{\thr}_{p_n})$.
In this context, we provide a complete characterization of the steady-state dynamics of \dmmf under any such policy -- in particular, for each agent $i$, we exactly characterize her long-run average utility $\lim_{t\rightarrow\infty}\U_i[t](\vec{\str})/t$. This then allows us to study equilibria in the next section. In this section, we provide the main lemmas and theorems, and sketch the ideas of the proofs. Complete proof of all results of this section can be found in \cref{appsec:longrun}.

Recall for agent $i$, we have $\val_i[t] \sim \valDist_i$ i.i.d., and $\TotalWin_i[t] = \sum_{\tau=1}^t \Win_i[\tau]$ is the total allocations of agent $i$ by round $t$. The following proposition shows that tracking $\TotalWin_i[t]$ is sufficient for determining the agent's utility.

\begin{proposition}
\label{prop: expected utility expression}
Under any threshold strategy vector $\vec{\str} = (\str^{\thr}_{p_1}, \cdots, \str^{\thr}_{p_n})$, for all $t \inN$ we can express the utility of any agent $i$ up to time $t$ as
\begin{equation}
\label{eq: expected utility expression} 
\U_i[t](\vec{\str}) \stackrel{a.s.}{=} \ValueFunc_{i}(p) \cdot \TotalWin_i[t] + o(t) 
\end{equation}
\end{proposition}

This result is a consequence of the fact that conditional on requesting the resource, whether or not an agent wins the resource is independent of their value. Furthermore, the expectation of the value conditional on requesting the resource is given by $\ValueFunc_{i}(p)$ on each round. Hence, we achieve the expression almost surely by the law of large numbers. 

\begin{figure}
    \begin{minipage}[c]{0.5\textwidth}
        \includegraphics[width=\linewidth]{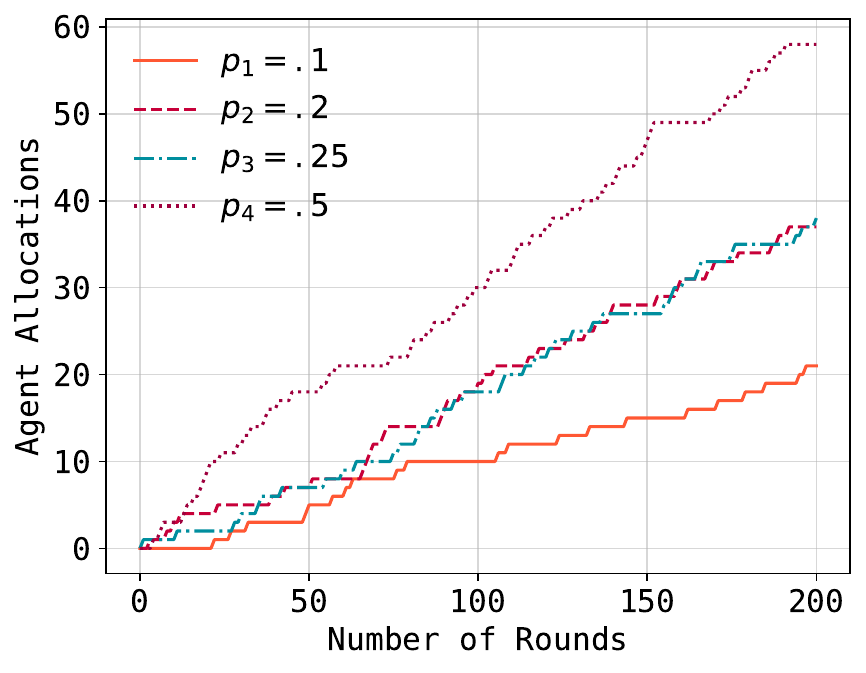}
    \end{minipage}
    \hfill
    \begin{minipage}[c]{\textwidth}
        %\vspace{-1cm}
        \caption{\em A sample path of allocations under \dmmf with $4$ agents in the symmetric setting (i.e., with $\alpha_i = 1/4$ for all $i$). The agents all use threshold strategies, with strategy vector $(\str^{\thr}_{0.1},\str^{\thr}_{0.2},\str^{\thr}_{0.25},\str^{\thr}_{0.5})$. Note that while for all agents $W_i[t]$ grows linearly, they split up into $3$ stable subgroups, each with a different average win rate. Note also that \dmmf gives agent $1$ the highest priority and agent $4$ the lowest priority in \emph{every} round after round $70$; in fact, we formally prove that the trajectories of agents in different subgroups almost surely never intersect after some finite time.
        } 
        \label{fig:dmmf_progression}
    \end{minipage}
\end{figure}

The main result of this section (\cref{thm: condition for stablity}) establishes conditions on the threshold strategies such that each agent wins exactly their fair share (up to sub-linear error) of the number of rounds where at least one agent requested the resource. 
More generally, in \cref{sec:decomposition}, we show that given fair share $\alpha_i$ and threshold strategy $\str^{\thr}_{p_i}$ for each agent $i$, we can partition the agents into subgroups, wherein agents in each subgroup get their fair share within the subgroup, but different subgroups win the resource at different rates. In \cref{fig:dmmf_progression}, we show a sample path of agents' allocations ${(\TotalWin_i[t])_{i \in\N}}$ in a setting with four agents who split up into three subgroups.

To derive the condition for all agents to get their fair share, we need to establish the following critical property of $(\TotalWin_i[t])_{i \in\N}$ we call \emph{global state-space collapse}: for all $t\inN$ and agents $i,j\in[n]$ we have $\left|\frac{\TotalWin_i[t]}{\alpha_i}-\frac{\TotalWin_j[t]}{\alpha_j}\right| \le o(t)$ almost surely. 
{In \cref{lemma: Sufficient condition}, we prove a sufficient condition we call \emph{subgroup stability} (\cref{def:subgpstable}) for the state-space collapse property to hold for all agents in some subset $S\subseteq[n]$, in terms of the request probabilities and fair shares of each agent.
By instantiating this for the complete set $[n]$, we get a sufficient condition for global state-space collapse.
Crucially, note that our condition does not need all request policies and fair shares to be symmetric, but rather, holds for a much larger set of thresholds.

On the other hand, in~\cref{thm:necessary condition 1}, we show that when set $[n]$ violates the subgroup stability condition, then we no longer have global state-space collapse, in that there is some subset of agents \coedit{$S \subsetneq[n]$} such that $\left|\frac{\sum_{k \in S}\TotalWin_k[t]}{\sum_{k \in S}\alpha_k}-\frac{\sum_{k \not\in S}\TotalWin_k[t]}{\sum_{k \not\in S}\alpha_k}\right| \ge \Omega(t)$ for all $t \in \N$ almost surely. } 
\coedit{In \cref{sec:decomposition}, we show that this further implies that there exists a set $U \subsetneq [n]$ such that $\frac{\TotalWin_i[t]}{\alpha_i}-\frac{\TotalWin_j[t]}{\alpha_j} \ge \Omega(t)$ for any $i \in U$ and $j \in V = [n] - U$. Hence, an agent in $U$ will (in the long run) always have less priority than an agent in $V$. Thus, if an agent in $V$ requests the resource on a round, no agent in $U$ can win the resource, which means agents in $V$ are essentially engaging in an instance of the DMMF mechanism where agents in $U$ do not exist. On the other hand, agents in $U$ are engaging with an instance of the DMMF mechanism where agents in $V$ do not exist but, each round has some probability (ie. the probability that an agent in $V$ requests on the round) that the resource is not allocated on that round regardless of how agents in $U$ request. By repeating this argument on the smaller instances of the DMMF mechanism, we show that we can partition agents such that the subgroup stability condition holds for each group of agents in the partition, agents in a part will get proportional allocations (up to sub-linear error) and $\left|\frac{\TotalWin_i[t]}{\alpha_i}-\frac{\TotalWin_j[t]}{\alpha_j}\right|\ge \Omega(t)$ for any $i$ and $j$ in different parts.}

% \begin{figure}[t]
%     \centering
%     \includegraphics[width=0.6\linewidth]{Paper Draft/Figures/dmmf_prog.png}
%     \caption{Allocations of $4$ agents. \gfcomment{To be made prettier and placed in the correct spot.}}
%     \label{fig:dmmf_progression}
% \end{figure}

\subsection{The \dmmf Process and Subgroup Stability Criterion}

For the rest of the section, we fix the threshold strategies used by each agent $\vec{\str} = (\str^{\thr}_{p_1}, \cdots, \str^{\thr}_{p_n})$. Recall that under the $\str^{\thr}_{p_i}$ strategy, agent $i$ requests the resource with probability $p_i$ independently in each round. Furthermore, whether an agent wins on round $t$ can be \coedit{determined} by observing $\TotalWin_i[t-1]$ and $\Req_i[t]$ for all $i \in [n]$. Hence, the vector of the agents' allocations, $\hat{\mChain}[t] = (\TotalWin_1[t], \cdots, \TotalWin_n[t])$, evolves as a Markov process. Clearly, if $p_i > 0$ for any $i$, $\norm{\hat{\mChain}[t]} \geq \Omega(t)$ almost surely. 

In order to better analyze the allocation process, we consider instead a transformed version with better convergence properties.
Define the total number of wins by round $t$, $\TotalAnyReq[t] = \sum_{i=1}^n \TotalWin_i[t]$, and use this to re-scale and center each agent's allocation as $\mChain_i[t] = \frac{\TotalWin_i[t]}{\alpha_i} - \TotalAnyReq[t]$.
%\etcomment{do we want this definition, or rather $\mChain_i[t] = \TotalWin_i[t] - \alpha_j\TotalAnyReq[t]$?} \gfcomment{I think this is better so that $X_i$ is close to $X_j$}\cocomment{I disagree here. I define $\mChain_i[t] = \frac{\TotalWin_i[t]}{\alpha_i} - \TotalAnyReq[t]$ so that the ordering of $\mChain_i[t]$ is also the priority ordering: $\mChain_i[t] < \mChain_j[t]$ if and only if $\frac{\TotalWin_i[t]}{\alpha_i} < \frac{\TotalWin_j[t]}{\alpha_i}$.} \gfcomment{The agree was meant for the $W/\alpha - K$ def.}
We now consider the process 
$\mChain[t] = (\mChain_1[t], \cdots, \mChain_n[t])$.

Given the above definition, we next define the \emph{global state-space collapse property} for the process as follows: we say $\mChain[t]$ {has the} global {state-space collapse property} if for all $i, j \in [n]$ and all $t\inN$ we have $|\mChain_i[t] - \mChain_j[t]| = \left|\frac{\TotalWin_i[t]}{\alpha_i} - \frac{\TotalWin_j[t]}{\alpha_j} \right| =  o(t)$ almost surely.
For sake of argument, suppose we know that the process $\mChain[t]$ satisfies global state-space collapse.
This would then give us
\begin{equation*}
\TotalAnyReq[t] = \sum_{k \in [n]}\TotalWin_k[t] \eqas \sum_{k \in [n]}\qty(\TotalWin_i[t]\frac{\alpha_k}{\alpha_i} + o(t)) =  \frac{\TotalWin_i[t]}{\alpha_i} \sum_{k \in [n]} \alpha_k + o(t) = \frac{\TotalWin_i[t]}{\alpha_i} + o(t).
\end{equation*}
Thus, we would have that $|\mChain_i[t]| = \TotalAnyReq[t] - \frac{\TotalWin_i[t]}{\alpha_i} \eqas o(t)$, making it easy to reason about its convergence. 
In addition, we could also conclude that $\TotalWin_i[t] \eqas \alpha_i \TotalAnyReq[t] + o(t)$ where $\TotalAnyReq[t]$ is the number of rounds in which at least one agent requests the resource (since on any such round some agent will win the resource), and is easily computed by the law of large numbers. Furthermore, this means that each agent would win exactly their fair share of all requested rounds (with some sublinear error due to stochastic fluctuations).

In order to relate global state-space collapse of the $\mChain[t]$ process to the underlying primitives of the setting, we first need the following definition:
\begin{definition}[\textbf{Subgroup Stability Condition}]
\label{def:subgpstable}
We say that a subgroup of agents $S \subseteq [n]$ satisfies the subgroup stability condition if, for all $R \subseteq S$, 
    \begin{equation}
    \label{eq:subgpstable}
        \frac{1 - \prod_{k\in R}(1-p_k)}{\sum_{k \in R}\alpha_k} \geq \frac{1 - \prod_{k\in S}(1-p_k)}{\sum_{k \in S}\alpha_k}.    
    \end{equation} 
    \coedit{We say $S$  satisfies the subgroup stability condition strictly if the above inequality is strict for all $R \subsetneq S$.} 
\end{definition}

The inequality in~\cref{eq:subgpstable} has a simple intuition: it ensures that if the agents in $R$ fall behind the agents in $S \setminus R$ (i.e., each has less total number of wins proportional to their share), then agents in $R$ win rounds at a faster rate (relative to their total fair share) compared to agents in $S \setminus R$.
To understand this in more detail, consider the case for $S = [n]$ and $\alpha_i = \frac{1}{n}$ for all $i \in [n]$: Now note that we can rewrite the inequality as
\begin{equation}
\label{eq: condition for stability}
    \frac{1 - \prod_{k\in R}(1-p_k)}{|R|}
    \ge
    \frac{1 - \prod_{k\in [n]}(1-p_k)}{n}
\end{equation}
Now consider a round $t$ where every agent in $R$ has a smaller total allocation than every agent in $[n] \setminus R$.
Then the LHS of \cref{eq: condition for stability} is the expected rate of increase of $\frac{1}{|R|} \sum_{k \in R} \TotalWin_i[t-1]$, the total number of wins of agents in $R$, averaged across those agents.
Specifically, $1 - \prod_{k\in R}(1-p_k)$ is the probability that at least one agent in $R$ will request, and since they are all behind, some agent in $R$ will get allocated.
On the other hand, the RHS of \cref{eq: condition for stability} is the corresponding quantity for all agents.
This implies that agents in $R$ have a higher average rate of wins than all agents and, therefore, will ``catch up'' to the other agents, thus ensuring that all normalized allocations stay close to each other.

\subsection{Necessary and Sufficient Conditions for Global State-Space Collapse}

Using the formalism introduced above, we can now state the main result of this section:
\begin{theorem}[\textbf{Global State-Space Collapse for the Allocation Process}] 
\label{thm: condition for stablity}
The process $\mChain[t]$ satisfies $|\mChain_i[t] - \mChain_j[t]| \eqas o(t)$ for all $i, j \in [n]$ and $t \in \N$ \emph{if and only if} the global set $[n]$ satisfies the stability criterion in \cref{def:subgpstable}. Moreover, if global state-space collapse is satisfied, then for all $i$, we have 
\begin{equation*}
\TotalWin_i[t] \eqas \alpha_i\left(1 - \prod_k(1-p_k)\right)\cdot t + o(t).
\end{equation*}
\end{theorem}
This formalizes the intuition above for the subgroup stability criterion. \cref{thm: condition for stablity} implies that when~\cref{def:subgpstable} does not hold for the set $[n]$, then we do not have global state-space collapse. In~\cref{sec:decomposition}, we show that in this case, we can partition $\mChain[t] = (\mChain_{1}[t], \cdots,\mChain_n[t])$ as $\left((\mChain_i[t])_{i \in S}, (\mChain_i[t])_{i \in [n] \setminus S}\right)$ such that $(\mChain_i[t])_{i \in S}$ and $(\mChain_i[t])_{i \in [n] \setminus S}$ depend on each other (in a non-trivial way) only finitely often. Hence, for large enough $t$, $(\mChain_i[t])_{i \in S}$ and $(\mChain_i[t])_{i \in [n] \setminus S}$ are Markov processes when considered in isolation. Hence, we can reduce the problem into instances of smaller dimensions.

To prove the theorem, we separately establish the necessary and sufficient conditions for the process $\mChain[t]$ to satisfy global state-space collapse.
We first provide our necessary condition, which states that when the set $[n]$ violates the subgroup stability condition, then we can find disjoint subsets $R,U \subseteq [n]$ whose agents have very different win rates. Subsequently, this proves one side of the if and only if the condition of \cref{thm: condition for stablity}.
First, we have the following:
\begin{restatable}{lemma}{ThmNecessaryCondition}
\label{thm:necessary condition 1}
%For the process $\mChain[t]$ to be globally stable, it must hold that for, any disjoint $R, U \subset [n]$,
Given $\{\alpha_i,p_i\}_{i\in[n]}$, suppose there are disjoint subsets $R, U \subseteq [n]$ such that:
$$\frac{1-\prod_{k \in U}(1-p_k)}{\sum_{j \in U}\alpha_i} < \prod_{k \not\in R}(1-p_k)\cdot \frac{1-\prod_{k \in R}(1-p_k)}{\sum_{j \in R}\alpha_i},$$ 
%In particular, if this does not hold for some $R, U$, 
then, there exists $C > 0$ such that, for all rounds $t\inN$, we have
$$\frac{\sum_{i\in R}\TotalWin_i[t]}{\sum_{i \in R}\alpha_i} - \frac{\sum_{i\in U}\TotalWin_i[t]}{\sum_{i \in U}\alpha_i} \stackrel{a.s.}{\geq} (C+o(1))t.$$  
\end{restatable}

\begin{proof}[Proof Sketch]
The intuition for this necessary condition is as follows: on any round where agent $i$ has the highest priority and agent $j$ has the lowest priority, if the expected drift in $\frac{\TotalWin_j[t]}{\alpha_j} - \frac{\TotalWin_i[t]}{\alpha_i}$ is positive then on every round (independent of the state), the expected drift in this quantity is also positive, and hence, by the law of large numbers, this gap goes to infinity at a rate of $\Omega(t)$ almost surely. %Hence, $\frac{\TotalWin_j[t]}{\alpha_j} - \frac{\TotalWin_i[t]}{\alpha_i} \geq \Omega(t)$ leading to the process being unstable. 
The same type of behavior (appropriately normalized) must also hold when we consider disjoint groups $U,R$ of agents, such that agents in $U$ have the highest priorities, and agents in $R$ have the lowest.
%\gfedit{which proves instability of the sets of agents $S,R$ conditioned on the set $R$ winning more than the set $S$ when $R$ is ahead of all other agents and $S$ is behind all other agents}\cocomment{I'm not sure what you are trying to say here that was not mentioned above.}.
Finally, taking $R = [n] \setminus U$, we can algebraically manipulate the above condition to obtain that the condition in~\cref{def:subgpstable} is necessary. The complete proof is provided in~\cref{appsec:longrun}.
\end{proof}

We next complement this with sufficient conditions for the process $(\mChain_i[t])_{i \in S}$ to satisfy global state-space collapse for some $S \subseteq[n]$.
For this, first define the sub-process with respect to the agents in $S \subseteq[n]$ as $\mChain^S[t] = (\mChain^S_1[t], \ldots, \mChain^S_n[t])$ such that
\begin{equation*}
    \mChain^S_i[t] =
    \begin{cases}
        \frac{\TotalWin_i[t]}{\alpha_i(S)} - \TotalAnyReq_S[t] , & \text{ if } i\in S
        \\
        0 , & \text{ otherwise}
    \end{cases}   
\end{equation*}
where $\alpha_i(S)$ is agent $i$'s share relative to the other agents in $S$, i.e., $\alpha_i(S) = \frac{\alpha_i}{\sum_{k \in S}\alpha_k}$ and $\TotalAnyReq_S[t] = \sum_{i\in S} \TotalWin_i[t]$ is the total allocations of agents in $S$. We can alternatively think of $\mChain^S_i[t]$ as a process with domain $\R^{|S|}$ by removing the indices $i \not\in S$.
%In essence, $\mChain^S[t]$ is version of $\mChain[t]$ where we only consider agents in $S$ then recenter and rescale the process appropriately. 
Observe that, for $S \subsetneq [n]$, $\mChain^S[t]$ may not be a Markov process since its evolution may depend on information in $\mChain^{[n] \setminus S}[t]$.
%, e.g., if an agent in $S$ has more allocations than an agent in $[n] - S$ (proportionally to each one's share).

The following sufficient condition now shows that the issue captured by the necessary condition is essentially the only way in which the process is not stable.
\begin{restatable}{lemma}{ThmSufficientCondition}
\label{lemma: Sufficient condition}
    For all $S \subseteq[n]$ such that $S$ satisfies the stability criterion, $\norm{\mChain^S[t]} = o(t)$ almost surely.
    Hence, for all $i, j \in S$, $\left|\frac{\TotalWin_i[t]}{\alpha_i} -  \frac{\TotalWin_j[t]}{\alpha_j}\right| = o(t)$ almost surely. 
    \coedit{Specifically, if $S$ satisfies the stability criterion strictly then $\left|\frac{\TotalWin_i[t]}{\alpha_i} -  \frac{\TotalWin_j[t]}{\alpha_j}\right| < O(\sqrt{t\log t})$.}
\end{restatable}

%\sbcomment{Add a line saying that this variance is better than that of an unbiased random walk (which would grow as $\Theta(t)$ - so there is some mean reversion, but weaker than OU - see \href{https://math.stackexchange.com/questions/2538376/calculating-variance-of-an-ornstein-uhlenbeck-process}{this MO article}}

\begin{proof}[Proof Sketch]
We first consider the case when the stability criterion holds with strict inequality for every $R \subsetneq S$. To establish the state-space collapse, we consider the {Lyapanov} function $G\left(\mChain^S[t]\right) = \sum_{i \in S} \alpha_i(S) \mChain^S_i[t]^2$, and establish that its drift satisfies $\E\left[\nabla G\left(\mChain^S[t]\right)\right] \leq C + 2\inner{\mChain^S[t]}{\vec{w}}$, where $w_i$ denotes the probability that agent $i$ wins the resource on any round when the state is $\mChain[t]$, and for some fixed $C > 0$. The main technical part of the proof is then to show that $\inner{\mChain^S[t]}{\vec{w}} < -C$ when $\max_{i \in S}\mChain^S_i[t]$ is sufficiently large. Hence, we get that $\E\left[\nabla G\left(\mChain^S[t]\right)\right] < -C$ whenever $G\left(\mChain^S[t]\right)$ sufficiently large.

Note that since $\mChain^S[t]$ is a lossy projection of a Markov chain, we can not directly apply a standard Foster-Lyapunov argument. Nevertheless, a result of \citet{pemantle1999moment} tells us that in any random process $Z_t \geq 0$ where $|Z_{t+1} - Z_{t}|$ is bounded by a constant and $\E[Z_{t+1} - Z_{t}|\mathcal{F}_t] < -\epsilon$ whenever $Z_t$ is sufficiently large, $\E[Z_{t}]$ is bounded by a constant almost surely. We show a modified version of the result that says that such a process is also bounded above by $o(t)$ almost surely. 
Now, since the process $G\left(\mChain^S[t]\right)$ satisfies the above conditions, we get  $\norm{\mChain^S[t]}^2 = \Theta(G\left(\mChain^S[t]\right))= o(t)$. We then apply a continuity argument to show that $\norm{\mChain^S[t]} = o(t)$ even when we allow the stability criterion to hold with equality for $R \subsetneq S$.
The complete proof is provided in~\cref{appsec:longrun}.
\end{proof}

Taking $S = [n]$, \cref{lemma: Sufficient condition} shows that $[n]$ satisfying the stability criterion is sufficient for the process to be stable. Moreover, if any $S \subseteq[n]$ satisfies the stability criterion, agents in that subset also have the stable property (ie. $|\mChain_i[t] - \mChain_j[t]| = o(t)$ for all $i, j \in S$). 
The two lemmas above, taken together, thus prove \cref{thm: condition for stablity}.
%The complete proofs of \cref{thm:necessary condition 1,lemma: Sufficient condition} can be found in \cref{appsec:longrun}.

Note that the necessary condition can be used to show that $\mChain[t] = \mChain^{[n]}[t]$ is unstable whereas the sufficient condition can show the stability of  $\mChain^{S}[t]$ for any $S \subseteq[n]$. 

In \cref{sec:decomposition}, we provide a more general characterization of the process, decomposing agents into a collection of disjoint subgroups, with the trajectories in each subgroup undergoing a `local' state-space collapse. Before discussing this general case, however, we turn first to our primary motivation of studying strategic behavior in such settings (in particular, as our results in this section are essentially sufficient for this purpose). In~\cref{sec:equilibrium}, we show that the existence of fixed threshold strategy equilibrium is not guaranteed (\cref{thm: No Pure Nash Eq}) and propose a data-driven variant of a threshold strategy and use \cref{thm: condition for stablity} to prove that the proposed strategy offers an approximate equilibrium.

%Hence, the only undecided case is when we have equality in some of these expressions. We will somewhat circumvent this issue using a continuity argument later in the paper. 

\section{Characterizing Equilibria under Threshold Strategies in \dmmf}
\label{sec:equilibrium}

In this section, we will consider the structure of equilibrium strategies in the \dmmf mechanism. The strategy set we will first consider for the agents is the set of threshold strategies $\{\strThresh_{p}: p \in [0, 1]\}$. As previously mentioned, threshold strategies have strong robustness properties: under the right threshold, any agent can guarantee half her ideal utility, even under adversarial behavior of the other agents.
Under such strong guarantees, one might wonder if threshold strategies (possibly different than the ones in the robust guarantee) form an equilibrium among the agents.
In \cref{ssec:nonexistence}, we prove this is not the case.
Even under simple symmetric distributions, there are no threshold strategies that form an equilibrium.
In \cref{ssec:followthedeviator}, we show how a simple data-driven modification can fix this and how this leads to the agents choosing the threshold strategies to maximize the total utility.

\subsection{Non-Existence of Equilibria under Fixed Thresholds}
\label{ssec:nonexistence}

Using our results from \cref{sec:analysis_thrs}, it is easy to analyze the agents' utilities when they all use threshold strategies.
In \cref{thm: characterize number of wins}, we will show that when all agents commit to threshold strategies, the average per round utility for agent $i$, $\lim_{t\rightarrow \infty}\frac{\U_i[t](\str^{\thr}_{p_1}, \cdots, \str^{\thr}_{p_n})}{t}$ converges almost surely to a fixed value, $\U_i(p_1, \cdots, p_n)$. Hence, we can consider an $n$-player game where each player $i$ selects an action $p_i \in [0, 1]$ and receives payoff $\U_i(p_1, \cdots, p_n)$ from that action. We will call this the \textit{Threshold Game}. Lemma \ref{lemma: U is cont} tells us that $\U_i(p_1, \cdots, p_n)$ is continuous with respect to $p_j$ for all $j \in [n]$. Thus, the payoff functions in this game are continuous. Furthermore, the action set of the agents is compact. \cite{glicksberg1952further} showed that any such game must have mixed Nash equilibrium.

However, the existence of a mixed Nash equilibrium in the Threshold Game is unsatisfying.
Specifically, a mixed Nash equilibrium only guarantees that agents are best responding \textit{before} they know the realization of the other agents' thresholds.
However, when running the \dmmf mechanism, agents can estimate the realized thresholds and best respond differently than the equilibrium implies.

The above scenario would not be problematic if the equilibrium was pure.
In that case, the agents' thresholds would best respond to the predictable progression of the Threshold game.
Unfortunately, the next theorem proves that pure Nash equilibria might not exist, even in simple examples.

\begin{restatable}[\textbf{Non-Existence of Equilibria in Fixed Thresholds}]{theorem}{ThmNoPureNash}
\label{thm: No Pure Nash Eq}
The Threshold Game may not have a pure Nash equilibrium. This, in particular, holds when $n=2$, $\alpha_i = \frac{1}{2}$ and $\valDist_1 = \valDist_2$ is a single distribution supported on $1$ and $\epsilon$ for well chosen $\epsilon \in [0, 1]$ for all $i$.
\end{restatable}
\begin{proof}[Proof Sketch]
We assume we have an equilibrium where the profile of request probabilities is given by $(p_1, p_2)$. If the profile $(p_1, p_2)$ is not stable then, in the long run, the agent winning more can only win when the other agent does not request. Hence, that agent might as well request as much as possible. Hence, it must be the case that the agent asking more is requesting every round. On the other hand, when $(p_1, p_2)$ is stable, we show that  $\U_i(p_1, p_2)$ is convex in $p_i$. Hence, each agent maximizes their utility at a boundary of stability. However, being at a boundary of stability is equivalent to being unstable. Hence, one agent must request every round in all cases. Assume this is agent 1. Let $\hat{p}$ be agent 2's best response to agent 1 requesting in every round. We show that  $(1, \hat{p})$ is not a Nash because agent 1's best response to $\hat{p}$ is to not request all the time; in fact, agent 1 should request with a probability smaller than $\hat{p}$. (A detailed proof of this result can be found in \cref{appsec: proof of no nash}.)
\end{proof}

\cref{thm: No Pure Nash Eq} proves that the agents' behavior when using the \dmmf mechanism is not as simple as using a fixed threshold to request.
Ideally, we would want the agents to use symmetric threshold strategies, so that the total allocated value across all agents (i.e., social welfare) is maximized.
In the next section, we show how the agents can use a simple data-driven strategy to achieve the desired behavior.

\subsection{Inducing Equilibria via Dynamic Thresholds: The \algorithmName Strategy}
\label{ssec:followthedeviator}

In this section, we focus on the case where agents are symmetric, that is, all agents draw values from the same distribution (i.e. $\valDist_i = \valDist$) and all agents have equal fair shares, which we will assume for this section.
As a remedy for the non-existence of pure Nash equilibrium in the Threshold Game, we propose an alternative strategy for this case such that no agent, when other agents are using this strategy, had any reason to play any other threshold strategy in the long run. In the case of symmetric agents it is meaningful to also consider the social welfare of the resulting equilibrium outcome, which we will show to be high compared to what the algorithms offering robustness guarantees achieve.

The objective of the strategy we shall propose in this section is to steer the agents to the optimal fixed threshold strategy for welfare.
If an agent deviates from this strategy to a different fixed threshold strategy, the other agents following the equilibrium strategy will steer the request rate to guarantee that this is not a profitable deviation. In particular, this can be done without explicit knowledge of which strategy each agent chooses to use. %\etcomment{is the next sentence worth including? Maybe at the end, but not here before even defining the strategy.} \etdelete{Furthermore, the property of the strategy profile being steered to the restricted class will hold even when agents decide to use an arbitrarily strategy from some finite number of rounds before choosing the threshold strategy to use indefinitely from that round on.}

Let $\Strats^{\text{Sym}}$ be the set of all strategy profiles where all agents use Threshold strategies with the same Threshold (i.e. $\Strats^{\text{Sym}} = \{(\strThresh_{p}, \cdots, \strThresh_{p}): p \in [0, 1]\}$). An important property of this set of strategies is that if the only allowed strategy profiles where profiles from $\Strats^{\text{Sym}}$, all agents would be aligned on which profile to select due to the symmetry among agents. Explicitly, the following proposition holds.

\begin{proposition} \label{prop:symmetricUtilMax}
    $\arg\max_{p \in [0, 1]}\U_i(p, \cdots, p) = \arg\max_{p \in [0, 1]}\U_j(p, \cdots, p) = p^*\,\forall\,i, j \in [n]$.
\end{proposition}

Suppose agent $i$ was forced to announce a threshold strategy $p$, and all other agents announced ahead of time that they will all follow the strategy agent $i$ announced. Given this fact, the best agent $i$ can do is to announce using the optimal threshold strategy $p^*$.
Of course, in reality, agents do not announce their strategies. However, we can derive information about an agent's strategy using the history of the mechanism. Based on this principle, we now describe a strategy we, henceforth, refer to as \algorithmName. The idea of our \algorithmName strategy is for players to aim to match the total win-rate in the mechanism with a small drift towards the optimal request rate $p^*$ for the value distribution $\valDist$. The time-average total win-rate till time $t$ is $\frac{K[t]}{t}$. If all players use the $p$-threshold strategy, the expected total win-rate would be $\Phi(p)=1-(1-p)^n$, so to match the observed win-rate the agents have to request with probability $\Phi^{-1}\left(\frac{\TotalAnyReq[t]}{t}\right)$. 

We present the full \algorithmName strategy in \cref{alg:WinRateMatching}.
\coedit{To implement the \algorithmName strategy, the only information agents need to have is the proportion of rounds in which at least 1 agent requested the resource, $\frac{K[t]}{t}$ and the number of agents in the mechanism. Of course, if the full request history is public information, $K[t]$ can be deduced. However, it would be sufficient for the principal to make whether the resource was allocated in a round public for agent to implement the algorithm. Furthermore, since $p^*$ does not depend on what other agents are doing, each agent can compute $M_{\eta,\zeta}[t]$ independently without the need for coordination from the principal.}

% \sbcomment{Can we actually make this an algorithm? Makes it more visible then}
% \gfcomment{Made the algo but didn't delete the definition to check that we have moved everyting}

% \begin{definition}[\textbf{The \algorithmName Strategy}]
% \label{alg:winratematch}
%     Fix $\eta, \zeta:\N \rightarrow [0, 1]$ such that $\zeta(t) = 1 - o(1)$. \etdelete{Let $\phi(x) = 1-(1-x)^{n}$.} Then,
    
%     $$M_{\eta,\zeta}[t+1] = (1 - \eta(t+1)) \phi^{-1}\left(\frac{\zeta(t+1)\TotalAnyReq[t]}{t}\right) + \eta(t+1) \cdot p^*$$ where $\TotalAnyReq[t] = \sum_{s=1}^t \sum_{i=1}^n\Win[t]$. The \algorithmName\ strategy with drift rate $\eta(t)$ and gap $\zeta(t)$, $\str^{\ftd}_{\eta, \zeta}$, is the strategy in which $\str^{\ftd}_{\eta, \zeta}[t](\Hist[t-1])= M_{\eta,\zeta}[t]$. Hence, in round $t$, the agent requests the resource with probability $M_{\eta,\zeta}[t]$.
% \end{definition}

We now show that, with the right choice of parameters, this strategy will ensure the desired guarantee about the behaviour of the agents. The \algorithmName strategy offers two parameters: $\eta$ guarantees a drift towards the optimal probability $p^*$, while $\zeta$ helps keep the process away from requesting all the time, which is a sink state of the process resulting in the round-robin allocation with low welfare. 

\algrenewcommand\algorithmicrequire{\textbf{Input:}}

\begin{algorithm}[t]
\caption{The \algorithmName Strategy for any agent}
\label{alg:WinRateMatching}
\begin{algorithmic}
\Require Optimal probability $p^*$, functions $\eta, \zeta:\N \rightarrow [0, 1]$
\State $\TotalAnyReq[0] \gets 0$
\For{$t = 1, 2, \cdots$}
    \State Set
    $$M_{\eta,\zeta}[t] = \qty\big(1 - \eta(t)) \Phi^{-1}\left(\frac{\zeta(t)\TotalAnyReq[t-1]}{t-1}\right) + \eta(t) \cdot p^*$$
    where $\Phi(p) = 1 - (1 - p)^n$ and $p^* = \arg\max_{p \in [0, 1]}\U_i(p, \cdots, p)$
    \State Request with probability $M_{\eta,\zeta}[t]$
    \State Set $\TotalAnyReq[t] \gets \TotalAnyReq[t-1] + \ind{\exists j : \Req_j[t] = 1}$
\EndFor
\end{algorithmic}
\end{algorithm}

%  \begin{definition}[\textbf{The \algorithmName Strategy}]
% \label{alg:winratematch}
%     Fix $\eta, \zeta:\N \rightarrow [0, 1]$ such that $\zeta(t) = 1 + o(1)$. \etdelete{Let $\phi(x) = 1-(1-x)^{n}$.} Then,
    
%     $$M_{\eta,\zeta}[t+1] = (1 - \eta(t+1)) \phi^{-1}\left(\frac{\zeta(t+1)\TotalAnyReq[t]}{t}\right) + \eta(t+1) \cdot p^*$$ where $\TotalAnyReq[t] = \sum_{s=1}^t \sum_{i=1}^n\Win[t]$. The \algorithmName\ strategy with drift rate $\eta(t)$ and gap $\zeta(t)$, $\str^{\ftd}_{\eta, \zeta}$, is the strategy in which $\str^{\ftd}_{\eta, \zeta}[t](\Hist[t-1])= M_{\eta,\zeta}[t]$. Hence, in round $t$, the agent requests the resource with probability $M_{\eta,\zeta}[t]$.
% \end{definition}

%\gfcomment{I am a bit confused. The first theorem we have here is \cref{thm:winratematchequilibrium} which says what happens when agents deviate from \algorithmName. I would expect that we first say what happens when agents follow \algorithmName (\cref{Thm:ThresholdConvergesWhenAllRequest}), since that is not obvious. It seems like whoever was writing the text after \cref{thm:winratematchequilibrium} had the same idea as me, because the text describes \cref{Thm:ThresholdConvergesWhenAllRequest}.}

For ease of notation, we use $\str^{\ftd}_{\eta, \zeta}$ to denote the \algorithmName strategy. In the language previously used, $\str^{\ftd}_{\eta, \zeta}$ is a strategy in which $\str^{\ftd}_{\eta, \zeta}[t](\Hist[t-1]) = M_{\eta,\zeta}[t]$; hence, given history, $\Hist[t-1]$, an agent using this strategy requests with probability $M_{\eta,\zeta}[t]$ on round $t$.

We now consider the behaviour of this strategy. Consider first the case when all agents use \algorithmName, that is requesting with probability $M_{\eta,\zeta}[t]$ on each round $t$, and assume they use drift rate $\eta(t) = \eta > 0$ and gap $\zeta(t) = 1$. %Assume all agents request with probability $M_{\eta,\zeta}[t]$ on \etedit{each} round $t$ and, 
Assume further that $M_{\eta,\zeta}[t]$ converge to some value $p$. Hence, agents playing the \algorithmName strategy end up essentially playing the $p$-threshold strategy. Thus, almost surely, $\frac{\TotalAnyReq[t]}{t} = 1- (1-p)^n = \Phi(p)$. Hence, we must have that $p = (1-\eta)\Phi^{-1}(\Phi(p)) + \eta\cdot p^*$. Hence, $p = p^*$. Note that when $\eta = 0$, we no longer have that $p$ must be any specific value. This shows that with the right parameter choices when all agents use \algorithmName, if the $M_{\eta,\zeta}[t]$ (the request probability) converges, it goes to $p^*$, the desired request rate.

On the other hand, when one agent requests with probability $\hat{p}$ instead, the total win-rate becomes $\frac{\TotalAnyReq[t]}{t} = 1- (1-p)^{n-1}(1-\hat{p})$ almost surely. Hence, $p = (1-\eta)[1-(1-p)^{\frac{n-1}{n}}(1-\hat{p})^{\frac{1}{n}}] + \eta\cdot p^*$. When $\eta = 0$, we have that $p = \hat{p}$ and, as $\eta$ increases, $p$ moves from $\hat{p}$ to $p^*$. In all cases, we see that the random process $M_{\eta,\zeta}[t]$ has a unique value to which it can converge. We would, however, like to select $\eta$ so as to ensure two seemingly contradicting statements.
On the one hand, when all agents follow \algorithmName, we want $M_{\eta,\zeta}[t]$ to converge as if $\eta > 0$.
On the other hand, when one agent deviates and requests with probability $\hat p$, we want $M_{\eta,\zeta}[t]$ to converge as if $\eta = 0$.
In particular, we show that there exists $\eta(t) = o(1)$ that will provide this guarantee.

For $\eta(t) = o(1)$, we have a new issue. Specifically, if $M_{\eta,\zeta}[t]$ is close to 1 even via random fluctuation, $\frac{\TotalAnyReq[t]}{t}$ will be close to 1. More problematically, $M_{\eta,\zeta}[t] = 1$ is a sink state if any agent follows this strategy because this would lead to that agent requesting every round. Mathematically, this issue arises from the fact that $(\Phi^{-1})'(1)  = \infty$. We remedy this issue with the gap parameter $\zeta(t)$, that provides a strict bound on how close the argument of $\Phi^{-1}(x)$ can get to 1 on any round. We want $\zeta(t) \rightarrow 1$ so that $M_{\eta,\zeta}[t]$ is able to achieve any value eventually. Our task is then to show that there exists drift rate $\eta(t) = o(1)$ and gap $\zeta(t) = 1 - o(1)$ such that $M_{\eta,\zeta}[t]$ achieves the desired properties. In particular, we show the following lemmas.
\begin{restatable}{lemma}{ThmThresholdConvergesWhenAllRequest} \label{Thm:ThresholdConvergesWhenAllRequest}
Assume all agent use the \algorithmName strategy with drift rate $\eta(t) = \frac{1}{\log(t)^{\frac{1}{2}-\epsilon}}$ for some $0 < \epsilon < \frac{1}{4}$ and gap $\zeta(t) = 1 - t^{-\frac{1}{4}}$. Then $M_{\eta,\zeta}[t] \rightarrow p^*$ almost surely.
\end{restatable}

\begin{restatable}{lemma}{ThmThresholdConvergesWhenDeviator}
\label{Thm:ThresholdConvergesWhenDeviator}
Assume all agents except agent $i$ use the \algorithmName strategy with drift rate $\eta(t) = \frac{1}{\log(t)^{\frac{1}{2}-\epsilon}}$ for some $0 < \epsilon < \frac{1}{4}$ and gap $\zeta(t) = 1 - t^{-\frac{1}{4}}$ and agent $i$ uses the $\hat{p}$-Threshold strategy, for any $\hat p$. Then $M_{\eta,\zeta}[t] \rightarrow \hat p$ almost surely.
\end{restatable}

\begin{proof}[Proof Sketch]
    We now give a sketch of the proof of \cref{Thm:ThresholdConvergesWhenAllRequest}. The proof of \cref{Thm:ThresholdConvergesWhenDeviator} follows the same outline with $\hat p$ in place of $p^*$.
    
    At a high level, the result follows from studying the random process $Z[t] = \frac{M_{\eta,\zeta}[t] - p^*}{1-\eta(t)}$, and establishing that it almost surely converges to $0$. 
    To show this, we perform a Taylor expansion on $\Phi^{-1}(x)$ to get that, for all $t$, we have 
    $$Z[t+1] = Z[t]\left(1-\Theta(\eta(t)/t)\right) + L[t] + O\left(t^{-1-\beta}\right).$$ 
    where 
    $L[t]$ is a martingale correction term which  satisfies $\E[L[t]|\mathcal{F}_{t}] = 0$ and $|L[t]| \leq O(t^{-1+\beta})$ for some $0 < \beta < \frac{1}{2}$. % (that depends on $\kappa$). 
    Telescoping, we get that for any $k<t$ we have
    $$|Z[t+1]| \leq |Z[k]|\cdot \exp\left(-\sum_{\tau = k}^t\Theta\left(\frac{\eta(\tau)}{\tau}\right) + \sum_{\tau = k}^t\frac{L[\tau]}{Z[\tau]} + \sum_{\tau = k}^t O\left(\frac{\tau^{-1-\beta}}{Z[\tau]}\right)\right).$$ 
    Bounding the term in the exponent is challenging, as it may become large if $Z[\tau]$ gets too small. In order to avoid this, we need to carefully stop the telescoping at some intermediate $k>0$ such that we can guarantee that either $Z[k]$ is sufficiently small, or the terms in the bracket can all be controlled to be small. The technical part of our argument establishes that it is always possible to choose such a $k$; for complete details, refer to~\cref{sec:app:followthedeviator}.
\end{proof}

Combining the two Lemmas, we get that the \algorithmName offers an approximate equilibrium of the \dmmf process.

\begin{restatable}[\textbf{Equilibrium Under \algorithmName}]{theorem}{ThmFTDIsEquilibrium}
\label{thm:winratematchequilibrium}
Fix drift rate $\eta(t) = \frac{1}{\log(t)^{\frac{1}{2}-\epsilon}}$ for some $0 < \epsilon < \frac{1}{4}$ and gap $\zeta(t) = 1 - t^{-\frac{1}{4}}$ . Then, for all $i \in [n]$,
$$\U_i(\str^{\ftd}_{\eta, \zeta}, \cdots, \str^{\ftd}_{\eta, \zeta})[t] \geq \U_i(\str^{\ftd}_{\eta, \zeta}, \cdots,\strThresh_{p}, \cdots, \str^{\ftd}_{\eta, \zeta})[t] -  o(t)$$ for any $p \in [0, 1]$.
\end{restatable}

%\etdelete{Note that the \algorithmName strategy (and all threshold strategies) are allocation independent, which is to say the behaviour of an agent, using such a strategy, in round $t$ is not a function of the current resource allocation in the mechanism.}
%\etcomment{I moved this comment here from the top. Not sure if we want it, but here is better than the top.}
%\gfcomment{At first read this paragraph sounds wrong because \algorithmName does depend on the allocation. I guess what you mean is that it does not depend on individual allocations and just the overall allocation? Or am I misunderstanding completely and by current resource allocation you mean what is happening on this round? In any case, if we want this, I think we should have some statement like 'this means that our results could extend to other mechanisms' (which I assume was the point of this).}\etcomment{maybe best to delete}

\begin{figure}[t]
    \centering
    
    \includegraphics[width=0.5\linewidth]{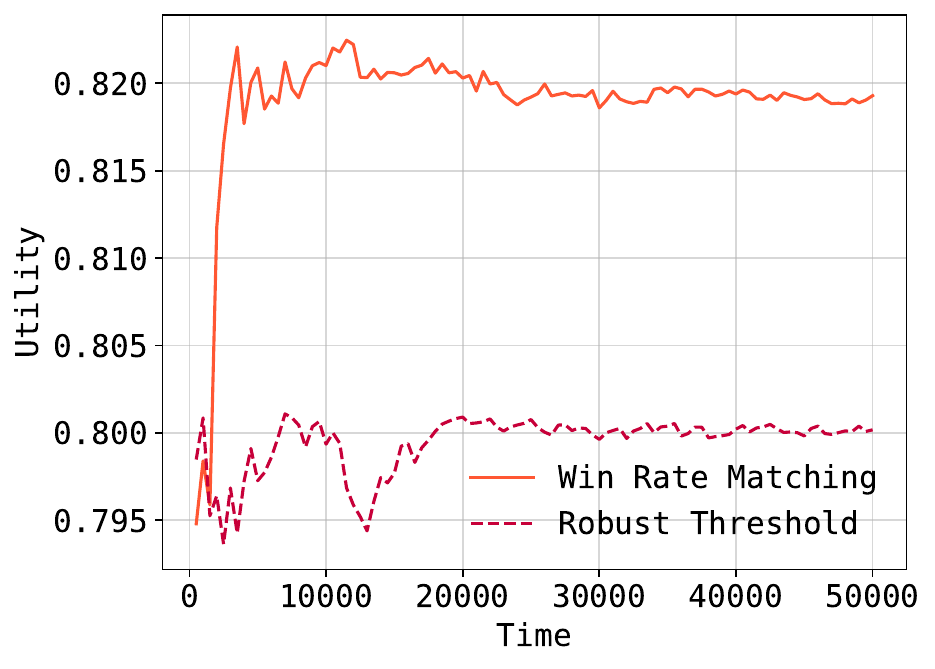}
    
    \caption{\gfedit{Utility of an agent who follows \algorithmName or the robust threshold of \cite{fikioris2023online}, when every other agent follows \algorithmName. The utility is normalized using the total ideal utility up to that round. We assume there are $5$ agents with value distributions uniform in $[0, 1]$. In addition, to get faster empirical convergence, we use $\zeta(t) = 1$ and $\eta(t)$ that quickly decreases linearly until $0.05$.}}
    \label{fig:utility}
\end{figure}

\gfedit{We showcase the above theorem in an experimental setting by comparing the following two scenarios.
First, when every agent is using \algorithmName.
Second, when one of these agents deviates to a fixed threshold, namely the one suggested by \cite{fikioris2023online}, that achieves the robust ideal utility guarantee.
We consider this when there are $n = 5$ agents with value distributions uniform in the interval $[0, 1]$.
\cref{fig:utility} shows that an agent would lose utility when deviating to the other threshold.}

Next, we show how much utility each agent gets when everyone follows the \algorithmName strategy.
First, we give a bound for arbitrary distributions, where we prove that every agent gets a $1 - \frac{1}{e} \approx 0.63$ fraction of her ideal utility (recall the ideal utility benchmark from \cref{ssec:threshold_strategies}).
This bound significantly improves the $0.5$ fraction of ideal utility that follows from the robust guarantees of \cite{DBLP:conf/sigecom/GorokhBI21,DBLP:conf/sigecom/BanerjeeFT23,fikioris2023online}.
For agents with uniformly distributed values, we prove that every agent gets almost all of her ideal utility.
This fraction is $1 - O\left(\log n/n\right)$, which improves over the $1 - O\left(1/\sqrt n\right)$ fraction of \cite{fikioris2023online}.

\begin{restatable}[\textbf{Improved Outcomes Under Equilibrium Strategies}]{theorem}{ThmUtilityComparison} \label{thm:utility_comparison}
    Assume that the agents use the \algorithmName strategy (\cref{alg:WinRateMatching}) with parameters as specified in \cref{thm:winratematchequilibrium}.
    Then, the following hold
    \begin{itemize}
        \item If agents have arbitrary (but identical distributions), then every agent receives at least a $\left(1 - \frac{1}{e}\right)$ fraction of her ideal utility almost surely as $T \to \infty$.
    
        \item If agents have values sampled from uniform distributions, then every agent receives at least a $1 - O\left(\frac{\log n}{n}\right)$ fraction of her ideal utility almost surely as $T \to \infty$.
    \end{itemize}
\end{restatable}

The theorem's proof is based on the combination of all our above results, which prove that agents behave as they would have if they were using the $p^*$-threshold strategy.
To get the utility bounds, we compare the agent's utility with the utility they would have gotten if everyone had used the $p$-threshold strategy for a well-chosen $p$.
Note that this $p$ can only lead to less utility than $p^*$, because of \cref{prop:symmetricUtilMax}.
For arbitrary distributions we use $p = \frac{1}{n}$ and for uniform distributions we use $p = \frac{\log n}{n}$.
The full proof can be found in \cref{sec:app:followthedeviator}.

\section{Decomposition of \texorpdfstring{$\mChain[t]$}{the Markov Chain} into stable groups}
\label{sec:decomposition}

In this section, we will provide a general characterization $\mChain[t]$. As in~\cref{sec:analysis_thrs}, we provide the lemmas and theorems, and sketch the ideas of the proofs. Complete proof of all results in this section can be found in \cref{appsec:longrun}.

Recall for every agent $i$, $\mChain_i[t] = \frac{\TotalWin_i[t]}{\alpha_i} - \TotalAnyReq[t]$. In particular, we will show that when $\mChain[t]$ is not globally state-space collapsing, we can partition the agents into disjoint sets $(C_1, \cdots C_m)$ such that $\mChain^{C_i}[t]$ will evolve like a lower-dimensional instance of $\mChain[t]$ where agents in $[n] \setminus C_i$ do not exist and will undergo global state-space collapse.
For brevity, we abuse notation and henceforth say a subgroup is `stable' (or satisfies the `stability criterion') if the trajectories in the subgroup undergo state-space collapse. 
The following condition will be key in our analysis of $\mChain[t]$:
\begin{definition}[\textbf{Splitting Property}]
    We say a set $S \subsetneq [n]$ has the \textit{splitting property} if, almost surely, for all $i \in S$ and $j \not\in S$, $ \frac{\TotalWin_j[t]}{\alpha_j} - \frac{\TotalWin_i[t]}{\alpha_i} \geq \Omega(t)$ for all $t \in \N$.
\end{definition}

Assume $S \subseteq[n]$ has the splitting property. This implies that, for $t \geq T$ (for some $T \in \N$), $\frac{\TotalWin_j[t]}{\alpha_j} > \frac{\TotalWin_i[t]}{\alpha_i}$ for all $i \in S$ and $j \not\in S$. After this happens, on a round where an agent in $S$ requests the resource, no agent in $[n] \setminus S$ can win the resource (regardless of if they request) as agents in $S$ always have priority over agents in $[n] \setminus S$. 

From this point, whether an agent in $S$ wins the resource is not affected by how agents in $[n] \setminus S$ request. Thus, agents in $S$ are essentially engaging in a copy of the \dmmf mechanism where agents in $[n] \setminus S$ do not exist. In particular, this means that the process $\mChain^{S}[t]$ is well and truly Markovian from this point on, and we can analyze this subgroup separately.

Let $\AnyReq_S[t]$ be the indicator variable for whether some agent in $S$ wins the resource in round $t$. As we have argued above, $\AnyReq_S[t] = \ind{\exists j \in S: \Req_j[t] = 1}$ for $t \geq T$. We now consider the behaviour of $\mChain^{[n] \setminus S}[t]$. For $t \geq T$, agents in $[n] \setminus S$ only win the resource when no agent in $S$ requests the resource. This can be conceptualized as the process $\mChain^{[n] \setminus S}[t]$ being forced to stay fixed on any round where any agent in $S$ requests. Alternatively, we can think of the process $\mChain^{[n] \setminus S}[t]$ as only transitioning on time steps where $\AnyReq_S[t] = 0$. This is the only dependence between $\mChain^{S}[t]$ and $\mChain^{[n] \setminus S}[t]$. Hence, conditioned on $\AnyReq_S[t] = 0$, $\mChain^{[n] \setminus S}[t]$ is also Markovian. Notice that the condition, $\AnyReq_S[t] = 0$ is equivalent to the condition $\TotalAnyReq_S[t] = \sum_{\tau=1}^t \AnyReq_S[\tau] = \TotalAnyReq_S[t-1]$.

In both cases $\mChain^{S}[t]$ and $\mChain^{[n] \setminus S}[t]$ evolve like independent copies of $\mChain[t]$ with fewer agents. Towards formalizing this intuition, let $\States(S)$ be the state space of $\mChain^S$ for any $S \subseteq [n]$. We now make the following definition:

\begin{definition} \label{def: Y subprocess}
    For a state of the system $\omega \in \States(S)$, let $\mChainY^S(\omega)[t] \in \States(S)$ be a random process that evolves like $\mChain[t]$ if only agents in $S$ existed and $\mChainY^S(\omega)[0] = \omega$. Hence, $\mChainY^S(\omega)$ has the same transition matrix as an instance of $\mChain[t]$ with only agents in $S$ but the initial state of $\mChainY^S(\omega)[0]$ is $\omega$.
    For a state of the system $\omega' \in \States([n] \setminus S)$, we define $\mChainY^{[n] \setminus S}(\omega')[t] \in \States([n] \setminus S)$ identically.
\end{definition}

For the sake of clarity, note that the difference between the state transitions of $\mChain^S[t]$ and $\mChainY^S(\omega)[t]$ is that the transition in $\mChain^S[t]$ may depend on the behaviour of agent in $[n] \setminus S$ where as the transition in $\mChainY^S(\omega)[t]$ behave as if agents in $[n] \setminus S$ did not exist. Hence, though $\mChain^S[t]$ may not be Markovian, $\mChainY^S(\omega)[t]$ is. This said, the following lemma tells us that eventually $\mChain^S[t]$ will behave like a time shifted copy of $\mChainY^S(\omega)[t]$ and $\mChain^{[n] \setminus S}[t]$ will behave like a time shifted copy of $\mChainY^{[n] \setminus S}(\omega')[t]$.

\begin{restatable}{lemma}{SplitChainLemma}
    \label{lemma: process splittng lemma}
    Let $S \subseteq[n]$ such that $S$ has the splitting property. Then, almost surely, there exist $T \inN$ such that, for an appropriate coupling of $\mChain[t]$, $\mChainY^S(\omega)[t]$ and $\mChainY^{[n] \setminus S}(\omega')[t]$ for $\omega = \mChain^{S}[T]$ and $\omega' = \mChain^{[n] \setminus S}[T]$, 
    
    $$\mChain^{S}[t] = \mChainY^S\left(\mChain^{S}[T]\right)[t - T]$$ 
    and 
    $$\mChain^{[n] \setminus S}[t] = \mChainY^{[n] \setminus S}\left(\mChain^{[n] \setminus S}[T]\right)\left[t - \TotalAnyReq_S[t] - (T-\TotalAnyReq_S[T])\right]$$ for all $t > T$.
    
\end{restatable}

This result essentially says that, if the splitting property holds then, for an appropriate coupling, if the processes 
$\mChain^S$ and  $\mChainY^S$ are in the same state then they will continue to be in the same state in all future rounds. Similarly, if the process $\mChain^{[n] \setminus S}$ and $\mChainY^{[n] \setminus S}$ are in the same state, they will continue to be in the same state in all future rounds after changing the time scales of the processes appropriately.

The following corollary of \cref{thm: condition for stablity,lemma: process splittng lemma} now gives us an almost sure expression for $\TotalWin_i$ for all $i \in S$ if $S$ satisfies the stability criterion or for all $i \in [n] \setminus S$ if $[n] \setminus S$ satisfies the stability criterion.

\begin{restatable}{corollary}{UtilityAfterSplit} \label{cor: utility after split}
     Let $S \subseteq[n]$ such that $S$ has the splitting property. Then, if $S$ satisfies the stability criterion,
     $$\TotalWin_i[t] \eqas \frac{\alpha_i}{\sum_{k \in S}\alpha_k}\left(1-\prod_{i \in S}(1-p_i)\right)\cdot t + o(t)$$ for all $i \in S$. On the other hand, if $[n] \setminus S$ satisfies the stability criterion, 
     $$\TotalWin_i[t] \eqas \frac{\alpha_i}{\sum_{k \in [n] \setminus S}\alpha_k}\left(1-\prod_{i \in [n] \setminus S}(1-p_i)\right)\prod_{i \in S}(1-p_i)\cdot t + o(t)$$ for all $i \in [n] \setminus S$.
\end{restatable}

Up to this point, we have not said anything about whether the splitting property is achievable. We now show that if the global state-space collapse property does not hold, then there must exist $S \subset [n]$ that has the splitting property. In particular, we have the following.

\begin{restatable}{lemma}{SplittingSetsExist} \label{lemma: splitting sets exist}
    Assume $[n]$ does not satisfy the stability criterion. Let $i^* = \arg\min_{i \in [n]} \frac{p_i}{\alpha_i}$. Let $S^* \subseteq[n]$ be such that $S^*$ satisfies the stability criterion, contains $i^*$ and is maximal with this property. Then $S^*$ has the splitting property. Note that such an $S^*$ must exist and is unique.
\end{restatable} 
\begin{proof}[Proof Sketch]
    We show this result inductively. Assume we have that $U \subseteq[n]$ with the property that $\frac{\TotalWin_j[t]}{\alpha_j} - \frac{\TotalWin_{i^*}[t]}{\alpha_{i^*}} \geq \Omega(t)$ for all $j \in U$. If $[n] \setminus U$ is not stable, we show that there must exist a set $V \subseteq[n] \setminus U$ such that $\frac{\sum_{k \in V}\TotalWin_k[t]}{\sum_{k \in V}\alpha_k} - \frac{\TotalWin_{i^*}[t]}{\alpha_{i^*}}  \geq \Omega(t)$. Furthermore, we can select $V$ to be stable. Hence, $\frac{\TotalWin_j[t]}{\alpha_j} - \frac{\TotalWin_{i^*}[t]}{\alpha_{i^*}} = \frac{\sum_{k \in V}\TotalWin_k[t]}{\sum_{k \in V}\alpha_k} - \frac{\TotalWin_{i^*}[t]}{\alpha_{i^*}} + o(t)  \geq \Omega(t)$ for all $j \in V$. Hence, $U \cup V$ has the same property as $U$. We repeat this process until we have $U$ with the property and $[n] \setminus U$ is stable. We then show that $S^* = [n] \setminus U$ and, by stability of $S^*$, $S^*$ has the splitting property. 
\end{proof}

\cref{lemma: splitting sets exist} tells us that $S^*$ has the splitting property and is stable. \cref{lemma: process splittng lemma} then tells us that we can split $\mChain[t]$ into $\mChainY^S(\omega)[t]$ and $\mChainY^{[n] \setminus S}(\omega')[t]$ for some $\omega \in \States(S)$ and $\omega' \in \States([n] \setminus S)$ where $\mChainY^S(\omega)[t]$ is stable. Hence, using \cref{cor: utility after split}, we can characterize $\TotalWin_i[t]$ for all $i \in S^*$. To characterize the utility of agents not in $S^*$, it is sufficient to analyse the process $\mChainY^{[n] \setminus S^*}(\omega')[t]$. However, this is just an instance of DMMF where only agents not in $S^*$ exist. Hence, we can repeat all the analysis we have done on this smaller instance of the problem. Hence, either $[n] \setminus S^*$ is stable or there exists a set $S^*_2 \subseteq [n] \setminus S^*$ that is stable and has a version of the splitting property restricted to agents in $[n] \setminus S^*$. We can then repeat this reasoning. To formalize this argument, we make the following generalization of the splitting property:

\begin{definition}
    Fix $S \subseteq[n]$. We say a set $R \subseteq S$ has the splitting property wrt. $S$ if, almost surely, for all $i \in R$ and $j \not\in S\setminus R$, $ \frac{\TotalWin_j[t]}{\alpha_j} - \frac{\TotalWin_i[t]}{\alpha_i} \geq \Omega(t)$.
\end{definition}

The following lemma now tells us that we can partition $[n]$ into subsets such that each subset is stable and these sets are well ordered with respect to the splitting property, making concrete the idea of applying the splitting property iteratively.

\begin{restatable}{lemma}{SplittingPartitionsExist} \label{lemma: splitting partitions exist}
    We will say an ordered partition of $[n]$, $(C_1, \cdots, C_m)$, is called a splitting partition wrt. the threshold strategy profile if $C_i$ is stable and, for all $i$, $C_i$ has the splitting property with respect to $\bigcup_{k=i}^mC_k$. Then, for any threshold strategy probabilities, there exists a (unique) splitting partition wrt. the threshold strategy probabilities.
\end{restatable}
Using this result and repeated application of~\cref{cor: utility after split}, we show the following theorem that characterizes the number of rounds won by each agent.
\begin{restatable}[\textbf{Characterization of Win Rate}]{theorem}{FinalWinRateCharaterization} 
\label{thm: characterize number of wins}
    Let $(C_1, \cdots, C_m)$ be the unique splitting partition wrt. the threshold strategy probabilities. Then, $$\TotalWin_i[t] \eqas \frac{\alpha_i}{\sum_{k\in C_u} \alpha_k} \left(1 - \prod_{k \in C_{u}}(1-p_k)\right)\prod_{v < u}\prod_{k \in C_{v}}(1-p_k) \cdot t + o(t),$$ 
    for all $i \in C_{u}$ \coedit{where $o(t) \leq O(\sqrt{t\log t})$ if $C_i$ satisfies the stability criterion strictly}
\end{restatable}

This result can be combined with Equation \eqref{eq: expected utility expression} to construct an almost sure expression for $\U_i[t](\str^{\thr}_{p_1}, \cdots, \str^{\thr}_{p_n})$. We can now observe that $\U_i(p_1, \cdots, p_n) = \lim_{t\rightarrow \infty}\frac{\U_i[t](\str^{\thr}_{p_1}, \cdots, \str^{\thr}_{p_n})}{t}$, the average per round utility achieved by agent $i$, almost surely exists and is a unique value. Furthermore, in~\cref{lemma: U is cont}, we show that, $\U_i(p_1, \cdots, p_n)$ is continuous in $p_j$ for all $j \in [n]$.

\section{Discussion}
In this work, we have demonstrated that, even though the DMMF mechanism does not have equilibrium under the simple and natural class of threshold strategies, a simple data-driven policy can be used to yield an approximate equilibrium on the agents. Furthermore, we show that this equilibrium leads to significantly better outcomes than the robust policy of \cite{fikioris2023online}. 
In the process, we also obtain a complete description (in terms of a subgroup state-space collapse characterization) of long-term outcomes of a system under \dmmf.
We believe that the techniques we have used to achieve these results may be of broader interest in the analysis of equilibria in other mechanisms for tackling allocation problems. 

The DMMF mechanism is an instance of the more general dynamic allocation mechanism framework that incorporates the constraint of ensuring each agent is able to receive at least their ``fair share" of the resource being allocated. However, one may consider different allocation functions designed to incorporate other goals or constraints that a principal might want to enforce on a dynamic allocation. Our hope is that the approaches we have introduced in this work will be useful in reasoning about the behaviour of random processes induced by other such dynamic allocation mechanisms. 

We would also like to highlight the general design principle by which we developed the \algorithmName strategy that may be instructive in inducing equilibrium in other repeated games. The lack of pure Nash equilibria demonstrated by \cref{thm: No Pure Nash Eq} suggests that the agents have too many options for possible deviations. 
We may, instead, like to consider a ``good'' subset of strategy profiles under which an equilibrium does exist, and hope to make sure that all users only choose between these options. In our case, the ``good'' set of strategies we considered is the set of symmetric threshold strategy profiles, but one could imagine that this principle could be applied to a more complicated subset set of strategies. If one can then construct a data-driven strategy for all agents that forces the strategy profile of the agents onto this ``good'' set of strategies when most agents use the data-driven strategy, then one may be able to induce such an equilibrium.

\section{Acknowledgments}
Sid Banerjee acknowledges support from AFOSR grant FA9550-23-1-0068, ARO MURI grant W911NF-19-1-0217, NSF grants ECCS-1847393 and CNS-195599, and the Simons Institute for the Theory of Computing.
Giannis Fikioris acknowledges support from the Department of Defense (DoD) through the National Defense Science \& Engineering Graduate (NDSEG) Fellowship, the Onassis Foundation -- Scholarship ID: F ZS 068-1/2022-2023, and ONR MURI grant N000142412742.
Chido Onyeze acknowledges support from AFOSR grant FA9550-23-1-0068.
Eva Tardos acknowledges support from AFOSR grants FA9550-19-1-0183 and FA9550-23-1-0068, and  ONR MURI grant N000142412742.

      \bibliographystyle{ACM-Reference-Format}
      
      \bibliography{bibliography}


\begin{thebibliography}{35}

%%% the Web address.

\ifx \showCODEN    \undefined \def \showCODEN     #1{\unskip}     \fi
\ifx \showDOI      \undefined \def \showDOI       #1{#1}\fi
\ifx \showISBNx    \undefined \def \showISBNx     #1{\unskip}     \fi
\ifx \showISBNxiii \undefined \def \showISBNxiii  #1{\unskip}     \fi
\ifx \showISSN     \undefined \def \showISSN      #1{\unskip}     \fi
\ifx \showLCCN     \undefined \def \showLCCN      #1{\unskip}     \fi
\ifx \shownote     \undefined \def \shownote      #1{#1}          \fi
\ifx \showarticletitle \undefined \def \showarticletitle #1{#1}   \fi
\ifx \showURL      \undefined \def \showURL       {\relax}        \fi
% invisible to TeX
\providecommand\bibfield[2]{#2}
\providecommand\bibinfo[2]{#2}
\providecommand\natexlab[1]{#1}
\providecommand\showeprint[2][]{arXiv:#2}

\bibitem[Balseiro et~al\mbox{.}(2019)]%
        {balseiro2017}
\bibfield{author}{\bibinfo{person}{Santiago~R Balseiro}, \bibinfo{person}{Huseyin Gurkan}, {and} \bibinfo{person}{Peng Sun}.} \bibinfo{year}{2019}\natexlab{}.
\newblock \showarticletitle{Multiagent mechanism design without money}.
\newblock \bibinfo{journal}{\emph{Operations Research}} \bibinfo{volume}{67}, \bibinfo{number}{5} (\bibinfo{year}{2019}), \bibinfo{pages}{1417--1436}.
\newblock


\bibitem[Banerjee et~al\mbox{.}(2023a)]%
        {DBLP:conf/sigecom/BanerjeeFT23}
\bibfield{author}{\bibinfo{person}{Siddhartha Banerjee}, \bibinfo{person}{Giannis Fikioris}, {and} \bibinfo{person}{{\'{E}}va Tardos}.} \bibinfo{year}{2023}\natexlab{a}.
\newblock \showarticletitle{Robust Pseudo-Markets for Reusable Public Resources}. In \bibinfo{booktitle}{\emph{Proceedings of the 24th {ACM} Conference on Economics and Computation, {EC} 2023, London, United Kingdom, July 9-12, 2023}}. \bibinfo{publisher}{{ACM}}, \bibinfo{address}{London, United Kingdom}, \bibinfo{pages}{241}.
\newblock


\bibitem[Banerjee et~al\mbox{.}(2023b)]%
        {banerjee2023online}
\bibfield{author}{\bibinfo{person}{Siddhartha Banerjee}, \bibinfo{person}{Chamsi Hssaine}, {and} \bibinfo{person}{Sean~R Sinclair}.} \bibinfo{year}{2023}\natexlab{b}.
\newblock \showarticletitle{Online Fair Allocation of Perishable Resources}.
\newblock \bibinfo{journal}{\emph{ACM SIGMETRICS}} \bibinfo{volume}{51}, \bibinfo{number}{1} (\bibinfo{year}{2023}), \bibinfo{pages}{55--56}.
\newblock


\bibitem[Blanchard and Jaillet(2024)]%
        {blanchard2024near}
\bibfield{author}{\bibinfo{person}{Moise Blanchard} {and} \bibinfo{person}{Patrick Jaillet}.} \bibinfo{year}{2024}\natexlab{}.
\newblock \showarticletitle{Near-Optimal Mechanisms for Resource Allocation Without Monetary Transfers}.
\newblock \bibinfo{journal}{\emph{arXiv preprint arXiv:2408.10066}} (\bibinfo{year}{2024}).
\newblock


\bibitem[Bonald and Massouli{\'{e}}(2001)]%
        {bonald2001impact}
\bibfield{author}{\bibinfo{person}{Thomas Bonald} {and} \bibinfo{person}{Laurent Massouli{\'{e}}}.} \bibinfo{year}{2001}\natexlab{}.
\newblock \showarticletitle{Impact of fairness on Internet performance}. In \bibinfo{booktitle}{\emph{Proceedings of the Joint International Conference on Measurements and Modeling of Computer Systems, SIGMETRICS/Performance 2001, June 16-20, 2001, Cambridge, MA, {USA}}}, \bibfield{editor}{\bibinfo{person}{Mary~K. Vernon}} (Ed.). \bibinfo{publisher}{{ACM}}, \bibinfo{address}{Cambridge, MA, USA}, \bibinfo{pages}{82--91}.
\newblock


\bibitem[Bonald et~al\mbox{.}(2006)]%
        {bonald2006queueing}
\bibfield{author}{\bibinfo{person}{Thomas Bonald}, \bibinfo{person}{Laurent Massouli{\'e}}, \bibinfo{person}{Alexandre Proutiere}, {and} \bibinfo{person}{Jorma Virtamo}.} \bibinfo{year}{2006}\natexlab{}.
\newblock \showarticletitle{A queueing analysis of max-min fairness, proportional fairness and balanced fairness}.
\newblock \bibinfo{journal}{\emph{Queueing systems}}  \bibinfo{volume}{53} (\bibinfo{year}{2006}), \bibinfo{pages}{65--84}.
\newblock


\bibitem[Bonald and Roberts(2015)]%
        {bonald2015multi}
\bibfield{author}{\bibinfo{person}{Thomas Bonald} {and} \bibinfo{person}{James~W. Roberts}.} \bibinfo{year}{2015}\natexlab{}.
\newblock \showarticletitle{Multi-Resource Fairness: Objectives, Algorithms and Performance}. In \bibinfo{booktitle}{\emph{Proceedings of the 2015 {ACM} {SIGMETRICS} International Conference on Measurement and Modeling of Computer Systems, Portland, OR, USA, June 15-19, 2015}}, \bibfield{editor}{\bibinfo{person}{Bill Lin}, \bibinfo{person}{Jun~(Jim) Xu}, \bibinfo{person}{Sudipta Sengupta}, {and} \bibinfo{person}{Devavrat Shah}} (Eds.). \bibinfo{publisher}{{ACM}}, \bibinfo{address}{Portland, OR, USA}, \bibinfo{pages}{31--42}.
\newblock
\urldef\tempurl%
\url{https://doi.org/10.1145/2745844.2745869}
\showDOI{\tempurl}


\bibitem[Budish et~al\mbox{.}(2017)]%
        {budish2017course}
\bibfield{author}{\bibinfo{person}{Eric Budish}, \bibinfo{person}{G{\'e}rard~P Cachon}, \bibinfo{person}{Judd~B Kessler}, {and} \bibinfo{person}{Abraham Othman}.} \bibinfo{year}{2017}\natexlab{}.
\newblock \showarticletitle{Course match: A large-scale implementation of approximate competitive equilibrium from equal incomes for combinatorial allocation}.
\newblock \bibinfo{journal}{\emph{Operations Research}} \bibinfo{volume}{65}, \bibinfo{number}{2} (\bibinfo{year}{2017}), \bibinfo{pages}{314--336}.
\newblock


\bibitem[Cavallo(2014)]%
        {cavallo2014incentive}
\bibfield{author}{\bibinfo{person}{Ruggiero Cavallo}.} \bibinfo{year}{2014}\natexlab{}.
\newblock \showarticletitle{Incentive compatible two-tiered resource allocation without money}. In \bibinfo{booktitle}{\emph{International conference on Autonomous Agents and Multi-Agent Systems, {AAMAS} '14, Paris, France, May 5-9, 2014}}, \bibfield{editor}{\bibinfo{person}{Ana L.~C. Bazzan}, \bibinfo{person}{Michael~N. Huhns}, \bibinfo{person}{Alessio Lomuscio}, {and} \bibinfo{person}{Paul Scerri}} (Eds.). \bibinfo{publisher}{{IFAAMAS/ACM}}, \bibinfo{address}{Paris, France}, \bibinfo{pages}{1313--1320}.
\newblock
\urldef\tempurl%
\url{http://dl.acm.org/citation.cfm?id=2617457}
\showURL{%
\tempurl}


\bibitem[Elokda et~al\mbox{.}(2023)]%
        {elokda2023self}
\bibfield{author}{\bibinfo{person}{Ezzat Elokda}, \bibinfo{person}{Saverio Bolognani}, \bibinfo{person}{Andrea Censi}, \bibinfo{person}{Florian D{\"o}rfler}, {and} \bibinfo{person}{Emilio Frazzoli}.} \bibinfo{year}{2023}\natexlab{}.
\newblock \showarticletitle{A Self-Contained Karma Economy for the Dynamic Allocation of Common Resources}.
\newblock \bibinfo{journal}{\emph{Dynamic Games and Applications}}  \bibinfo{volume}{13} (\bibinfo{date}{25 4} \bibinfo{year}{2023}), \bibinfo{pages}{1--33}.
\newblock
\showISSN{2153-0793}
\urldef\tempurl%
\url{https://doi.org/10.1007/s13235-023-00503-0}
\showDOI{\tempurl}


\bibitem[Elokda et~al\mbox{.}(2022)]%
        {elokda2022carma}
\bibfield{author}{\bibinfo{person}{Ezzat Elokda}, \bibinfo{person}{Carlo Cenedese}, \bibinfo{person}{Kenan Zhang}, \bibinfo{person}{John Lygeros}, {and} \bibinfo{person}{Florian D{\"o}rfler}.} \bibinfo{year}{2022}\natexlab{}.
\newblock \showarticletitle{CARMA: Fair and efficient bottleneck congestion management with karma}.
\newblock \bibinfo{journal}{\emph{arXiv preprint arXiv:2208.07113}} (\bibinfo{year}{2022}).
\newblock


\bibitem[Fikioris et~al\mbox{.}(2024)]%
        {fikioris2024incentives}
\bibfield{author}{\bibinfo{person}{Giannis Fikioris}, \bibinfo{person}{Rachit Agarwal}, {and} \bibinfo{person}{{\'E}va Tardos}.} \bibinfo{year}{2024}\natexlab{}.
\newblock \showarticletitle{Incentives in dominant resource fair allocation under dynamic demands}. In \bibinfo{booktitle}{\emph{International Symposium on Algorithmic Game Theory}}. Springer, \bibinfo{pages}{108--125}.
\newblock


\bibitem[Fikioris et~al\mbox{.}(2023)]%
        {fikioris2023online}
\bibfield{author}{\bibinfo{person}{Giannis Fikioris}, \bibinfo{person}{Siddhartha Banerjee}, {and} \bibinfo{person}{{\'E}va Tardos}.} \bibinfo{year}{2023}\natexlab{}.
\newblock \showarticletitle{Online resource sharing via dynamic max-min fairness: efficiency, robustness and non-stationarity}.
\newblock \bibinfo{journal}{\emph{arXiv preprint arXiv:2310.08881}} (\bibinfo{year}{2023}).
\newblock


\bibitem[Freeman et~al\mbox{.}(2018)]%
        {DBLP:conf/sigmetrics/FreemanZCL18}
\bibfield{author}{\bibinfo{person}{Rupert Freeman}, \bibinfo{person}{Seyed~Majid Zahedi}, \bibinfo{person}{Vincent Conitzer}, {and} \bibinfo{person}{Benjamin~C. Lee}.} \bibinfo{year}{2018}\natexlab{}.
\newblock \showarticletitle{Dynamic Proportional Sharing: {A} Game-Theoretic Approach}. In \bibinfo{booktitle}{\emph{Abstracts of the 2018 {ACM} International Conference on Measurement and Modeling of Computer Systems, {SIGMETRICS} 2018, Irvine, CA, USA, June 18-22, 2018}}. \bibinfo{publisher}{{ACM}}, \bibinfo{address}{Irvine, CA, USA}, \bibinfo{pages}{33--35}.
\newblock


\bibitem[Gaitonde and Tardos(2023)]%
        {gaitonde2023price}
\bibfield{author}{\bibinfo{person}{Jason Gaitonde} {and} \bibinfo{person}{{\'E}va Tardos}.} \bibinfo{year}{2023}\natexlab{}.
\newblock \showarticletitle{The price of anarchy of strategic queuing systems}.
\newblock \bibinfo{journal}{\emph{J. ACM}} \bibinfo{volume}{70}, \bibinfo{number}{3} (\bibinfo{year}{2023}), \bibinfo{pages}{1--63}.
\newblock


\bibitem[Ghodsi et~al\mbox{.}(2011)]%
        {DBLP:conf/nsdi/GhodsiZHKSS10}
\bibfield{author}{\bibinfo{person}{Ali Ghodsi}, \bibinfo{person}{Matei Zaharia}, \bibinfo{person}{Benjamin Hindman}, \bibinfo{person}{Andy Konwinski}, \bibinfo{person}{Scott Shenker}, {and} \bibinfo{person}{Ion Stoica}.} \bibinfo{year}{2011}\natexlab{}.
\newblock \showarticletitle{Dominant Resource Fairness: Fair Allocation of Multiple Resource Types}. In \bibinfo{booktitle}{\emph{Proceedings of the 8th {USENIX} Symposium on Networked Systems Design and Implementation, {NSDI} 2011, Boston, MA, USA, March 30 - April 1, 2011}}, \bibfield{editor}{\bibinfo{person}{David~G. Andersen} {and} \bibinfo{person}{Sylvia Ratnasamy}} (Eds.). \bibinfo{publisher}{{USENIX} Association}, \bibinfo{address}{Boston, MA, USA}.
\newblock


\bibitem[Glicksberg(1952)]%
        {glicksberg1952further}
\bibfield{author}{\bibinfo{person}{Irving~L Glicksberg}.} \bibinfo{year}{1952}\natexlab{}.
\newblock \showarticletitle{A further generalization of the Kakutani fixed point theorem, with application to Nash equilibrium points}.
\newblock \bibinfo{journal}{\emph{Proc. Amer. Math. Soc.}} \bibinfo{volume}{3}, \bibinfo{number}{1} (\bibinfo{year}{1952}), \bibinfo{pages}{170--174}.
\newblock


\bibitem[Gorokh et~al\mbox{.}(2017)]%
        {gorokh2017}
\bibfield{author}{\bibinfo{person}{Artur Gorokh}, \bibinfo{person}{Siddhartha Banerjee}, {and} \bibinfo{person}{Krishnamurthy Iyer}.} \bibinfo{year}{2017}\natexlab{}.
\newblock \showarticletitle{From Monetary to Non-Monetary Mechanism Design via Artificial Currencies}. In \bibinfo{booktitle}{\emph{Proceedings of the 2017 {ACM} Conference on Economics and Computation, {EC} '17, Cambridge, MA, USA, June 26-30, 2017}}, \bibfield{editor}{\bibinfo{person}{Constantinos Daskalakis}, \bibinfo{person}{Moshe Babaioff}, {and} \bibinfo{person}{Herv{\'{e}} Moulin}} (Eds.). \bibinfo{publisher}{{ACM}}, \bibinfo{address}{Cambridge, MA, USA}, \bibinfo{pages}{563--564}.
\newblock
\urldef\tempurl%
\url{https://doi.org/10.1145/3033274.3085140}
\showDOI{\tempurl}


\bibitem[Gorokh et~al\mbox{.}(2021)]%
        {DBLP:conf/sigecom/GorokhBI21}
\bibfield{author}{\bibinfo{person}{Artur Gorokh}, \bibinfo{person}{Siddhartha Banerjee}, {and} \bibinfo{person}{Krishnamurthy Iyer}.} \bibinfo{year}{2021}\natexlab{}.
\newblock \showarticletitle{The Remarkable Robustness of the Repeated Fisher Market}. In \bibinfo{booktitle}{\emph{{EC} '21: The 22nd {ACM} Conference on Economics and Computation, Budapest, Hungary, July 18-23, 2021}}, \bibfield{editor}{\bibinfo{person}{P{\'{e}}ter Bir{\'{o}}}, \bibinfo{person}{Shuchi Chawla}, {and} \bibinfo{person}{Federico Echenique}} (Eds.). \bibinfo{publisher}{{ACM}}, \bibinfo{pages}{562}.
\newblock
\urldef\tempurl%
\url{https://doi.org/10.1145/3465456.3467560}
\showDOI{\tempurl}


\bibitem[Grandl et~al\mbox{.}(2014)]%
        {DBLP:conf/sigcomm/GrandlAKRA14}
\bibfield{author}{\bibinfo{person}{Robert Grandl}, \bibinfo{person}{Ganesh Ananthanarayanan}, \bibinfo{person}{Srikanth Kandula}, \bibinfo{person}{Sriram Rao}, {and} \bibinfo{person}{Aditya Akella}.} \bibinfo{year}{2014}\natexlab{}.
\newblock \showarticletitle{Multi-resource packing for cluster schedulers}. In \bibinfo{booktitle}{\emph{{ACM} {SIGCOMM} 2014 Conference, SIGCOMM'14, Chicago, IL, USA, August 17-22, 2014}}, \bibfield{editor}{\bibinfo{person}{Fabi{\'{a}}n~E. Bustamante}, \bibinfo{person}{Y.~Charlie Hu}, \bibinfo{person}{Arvind Krishnamurthy}, {and} \bibinfo{person}{Sylvia Ratnasamy}} (Eds.). \bibinfo{publisher}{{ACM}}, \bibinfo{address}{Chicago, IL, USA}, \bibinfo{pages}{455--466}.
\newblock
\urldef\tempurl%
\url{https://doi.org/10.1145/2619239.2626334}
\showDOI{\tempurl}


\bibitem[Grandl et~al\mbox{.}(2016)]%
        {DBLP:conf/osdi/GrandlKRAK16}
\bibfield{author}{\bibinfo{person}{Robert Grandl}, \bibinfo{person}{Srikanth Kandula}, \bibinfo{person}{Sriram Rao}, \bibinfo{person}{Aditya Akella}, {and} \bibinfo{person}{Janardhan Kulkarni}.} \bibinfo{year}{2016}\natexlab{}.
\newblock \showarticletitle{{GRAPHENE:} Packing and Dependency-Aware Scheduling for Data-Parallel Clusters}. In \bibinfo{booktitle}{\emph{12th {USENIX} Symposium on Operating Systems Design and Implementation, {OSDI} 2016, Savannah, GA, USA, November 2-4, 2016}}. \bibinfo{publisher}{{USENIX} Association}, \bibinfo{address}{Savannah, GA, USA}, \bibinfo{pages}{81--97}.
\newblock
\urldef\tempurl%
\url{https://www.usenix.org/conference/osdi16/technical-sessions/presentation/grandl\_graphene}
\showURL{%
\tempurl}


\bibitem[Guo and Conitzer(2010)]%
        {guo2010}
\bibfield{author}{\bibinfo{person}{Mingyu Guo} {and} \bibinfo{person}{Vincent Conitzer}.} \bibinfo{year}{2010}\natexlab{}.
\newblock \showarticletitle{Strategy-proof allocation of multiple items between two agents without payments or priors}. In \bibinfo{booktitle}{\emph{9th International Conference on Autonomous Agents and Multiagent Systems {(AAMAS} 2010), Toronto, Canada, May 10-14, 2010, Volume 1-3}}. \bibinfo{publisher}{{IFAAMAS}}, \bibinfo{address}{Toronto, Canada}, \bibinfo{pages}{881--888}.
\newblock


\bibitem[Hassin(2016)]%
        {hassin2016rational}
\bibfield{author}{\bibinfo{person}{Refael Hassin}.} \bibinfo{year}{2016}\natexlab{}.
\newblock \bibinfo{booktitle}{\emph{Rational queueing}}.
\newblock \bibinfo{publisher}{CRC press}.
\newblock


\bibitem[Hassin and Haviv(2003)]%
        {hassin2003queue}
\bibfield{author}{\bibinfo{person}{Refael Hassin} {and} \bibinfo{person}{Moshe Haviv}.} \bibinfo{year}{2003}\natexlab{}.
\newblock \bibinfo{booktitle}{\emph{To queue or not to queue: Equilibrium behavior in queueing systems}}. Vol.~\bibinfo{volume}{59}.
\newblock \bibinfo{publisher}{Springer Science \& Business Media}.
\newblock


\bibitem[Jackson and Sonnenschein(2007)]%
        {jackson}
\bibfield{author}{\bibinfo{person}{Matthew~O Jackson} {and} \bibinfo{person}{Hugo~F Sonnenschein}.} \bibinfo{year}{2007}\natexlab{}.
\newblock \showarticletitle{Overcoming incentive constraints by linking decisions}.
\newblock \bibinfo{journal}{\emph{Econometrica}} \bibinfo{volume}{75}, \bibinfo{number}{1} (\bibinfo{year}{2007}), \bibinfo{pages}{241--257}.
\newblock


\bibitem[Joe-Wong et~al\mbox{.}(2013)]%
        {joe2013multiresource}
\bibfield{author}{\bibinfo{person}{Carlee Joe-Wong}, \bibinfo{person}{Soumya Sen}, \bibinfo{person}{Tian Lan}, {and} \bibinfo{person}{Mung Chiang}.} \bibinfo{year}{2013}\natexlab{}.
\newblock \showarticletitle{Multiresource allocation: Fairness--efficiency tradeoffs in a unifying framework}.
\newblock \bibinfo{journal}{\emph{IEEE/ACM Transactions on Networking}} \bibinfo{volume}{21}, \bibinfo{number}{6} (\bibinfo{year}{2013}), \bibinfo{pages}{1785--1798}.
\newblock


\bibitem[Kelly et~al\mbox{.}(1998)]%
        {kelly1998rate}
\bibfield{author}{\bibinfo{person}{Frank~P Kelly}, \bibinfo{person}{Aman~K Maulloo}, {and} \bibinfo{person}{David Kim~Hong Tan}.} \bibinfo{year}{1998}\natexlab{}.
\newblock \showarticletitle{Rate control for communication networks: shadow prices, proportional fairness and stability}.
\newblock \bibinfo{journal}{\emph{Journal of the Operational Research society}}  \bibinfo{volume}{49} (\bibinfo{year}{1998}), \bibinfo{pages}{237--252}.
\newblock


\bibitem[Parkes et~al\mbox{.}(2012)]%
        {DBLP:conf/sigecom/ParkesPS12}
\bibfield{author}{\bibinfo{person}{David~C. Parkes}, \bibinfo{person}{Ariel~D. Procaccia}, {and} \bibinfo{person}{Nisarg Shah}.} \bibinfo{year}{2012}\natexlab{}.
\newblock \showarticletitle{Beyond dominant resource fairness: extensions, limitations, and indivisibilities}. In \bibinfo{booktitle}{\emph{Proceedings of the 13th {ACM} Conference on Electronic Commerce, {EC} 2012, Valencia, Spain, June 4-8, 2012}}. \bibinfo{publisher}{{ACM}}, \bibinfo{address}{Valencia, Spain}, \bibinfo{pages}{808--825}.
\newblock


\bibitem[Pemantle and Rosenthal(1999)]%
        {pemantle1999moment}
\bibfield{author}{\bibinfo{person}{Robin Pemantle} {and} \bibinfo{person}{Jeffrey~S Rosenthal}.} \bibinfo{year}{1999}\natexlab{}.
\newblock \showarticletitle{Moment conditions for a sequence with negative drift to be uniformly bounded in Lr}.
\newblock \bibinfo{journal}{\emph{Stochastic Processes and their Applications}} \bibinfo{volume}{82}, \bibinfo{number}{1} (\bibinfo{year}{1999}), \bibinfo{pages}{143--155}.
\newblock


\bibitem[Prendergast(2022)]%
        {prendergast2022allocation}
\bibfield{author}{\bibinfo{person}{Canice Prendergast}.} \bibinfo{year}{2022}\natexlab{}.
\newblock \showarticletitle{The allocation of food to food banks}.
\newblock \bibinfo{journal}{\emph{Journal of Political Economy}} \bibinfo{volume}{130}, \bibinfo{number}{8} (\bibinfo{year}{2022}), \bibinfo{pages}{1993--2017}.
\newblock


\bibitem[Shakkottai et~al\mbox{.}(2008)]%
        {shakkottai2008network}
\bibfield{author}{\bibinfo{person}{Srinivas Shakkottai}, \bibinfo{person}{Rayadurgam Srikant}, {et~al\mbox{.}}} \bibinfo{year}{2008}\natexlab{}.
\newblock \showarticletitle{Network optimization and control}.
\newblock \bibinfo{journal}{\emph{Foundations and Trends{\textregistered} in Networking}} \bibinfo{volume}{2}, \bibinfo{number}{3} (\bibinfo{year}{2008}), \bibinfo{pages}{271--379}.
\newblock


\bibitem[Shue et~al\mbox{.}(2012)]%
        {DBLP:conf/osdi/ShueFS12}
\bibfield{author}{\bibinfo{person}{David Shue}, \bibinfo{person}{Michael~J. Freedman}, {and} \bibinfo{person}{Anees Shaikh}.} \bibinfo{year}{2012}\natexlab{}.
\newblock \showarticletitle{Performance Isolation and Fairness for Multi-Tenant Cloud Storage}. In \bibinfo{booktitle}{\emph{10th {USENIX} Symposium on Operating Systems Design and Implementation, {OSDI} 2012, Hollywood, CA, USA, October 8-10, 2012}}, \bibfield{editor}{\bibinfo{person}{Chandu Thekkath} {and} \bibinfo{person}{Amin Vahdat}} (Eds.). \bibinfo{publisher}{{USENIX} Association}, \bibinfo{address}{Hollywood, CA, USA}, \bibinfo{pages}{349--362}.
\newblock
\urldef\tempurl%
\url{https://www.usenix.org/conference/osdi12/technical-sessions/presentation/shue}
\showURL{%
\tempurl}


\bibitem[Sinclair et~al\mbox{.}(2022)]%
        {sinclair2022sequential}
\bibfield{author}{\bibinfo{person}{Sean~R Sinclair}, \bibinfo{person}{Siddhartha Banerjee}, {and} \bibinfo{person}{Christina~Lee Yu}.} \bibinfo{year}{2022}\natexlab{}.
\newblock \showarticletitle{Sequential fair allocation: Achieving the optimal envy-efficiency tradeoff curve}.
\newblock \bibinfo{journal}{\emph{ACM SIGMETRICS}} \bibinfo{volume}{50}, \bibinfo{number}{1} (\bibinfo{year}{2022}), \bibinfo{pages}{95--96}.
\newblock


\bibitem[Vuppalapati et~al\mbox{.}(2023)]%
        {DBLP:conf/osdi/VuppalapatiF0CK23}
\bibfield{author}{\bibinfo{person}{Midhul Vuppalapati}, \bibinfo{person}{Giannis Fikioris}, \bibinfo{person}{Rachit Agarwal}, \bibinfo{person}{Asaf Cidon}, \bibinfo{person}{Anurag Khandelwal}, {and} \bibinfo{person}{{\'{E}}va Tardos}.} \bibinfo{year}{2023}\natexlab{}.
\newblock \showarticletitle{Karma: Resource Allocation for Dynamic Demands}. In \bibinfo{booktitle}{\emph{17th {USENIX} Symposium on Operating Systems Design and Implementation, {OSDI} 2023, Boston, MA, USA, July 10-12, 2023}}. \bibinfo{publisher}{{USENIX} Association}, \bibinfo{address}{Boston, MA, USA}, \bibinfo{pages}{645--662}.
\newblock


\bibitem[Yin and Kroer(2022)]%
        {yin2022optimal}
\bibfield{author}{\bibinfo{person}{Steven Yin} {and} \bibinfo{person}{Christian Kroer}.} \bibinfo{year}{2022}\natexlab{}.
\newblock \showarticletitle{Optimal Efficiency-Envy Trade-Off via Optimal Transport}.
\newblock \bibinfo{journal}{\emph{Advances in Neural Information Processing Systems (NeurIPS)}}  \bibinfo{volume}{35} (\bibinfo{year}{2022}), \bibinfo{pages}{25644--25654}.
\newblock


\end{thebibliography}

% %   \begin{acks}
% %   ...
% %   \end{acks}
        \newpage
        \appendix

\section{Appendix A}
\label{sec:appendixA}
\PropThresholdsDominate*
\begin{proof}
    Fix $\Hist[t-1] \in (\{0, 1\}^n)^{t-1}$. Let $p(\Hist[t-1]) = \pr(\Req_i[t] = 1)$ under the strategy $\str$. Hence, $p(\Hist[t-1]) = \int_{0}^1 f_i(v)\cdot \str(\Hist[t-1], v) \;d\mu(v)$ where $f_i$ is the PDF for $\valDist_i$. 
    
    Let $\lambda(\Hist[t-1]) = \sup_{x \in [0, 1]} \{\pr(\val_i \geq x) \geq p(\Hist[t-1])\}$. We now define $\str'[t]$ to be the strategy where $\str(\Hist[t-1], v) = 1$ for $v > \lambda(\Hist[t-1])$ and $\str(\Hist[t-1], v) = 0$ for $v < \lambda(\Hist[t-1])$. Finally, we set $\str(\Hist[t-1], \lambda(\Hist[t-1]))$ such that $\pr(\Req'_i[t] = 1) = p(\Hist[t-1])$ where $\Req'_i[t]$ is the indicator for the agent requesting under the strategy $\str'$.

    Notice that whether or not an agent wins a round is a function of whether or not they request in that round (and is independent of their value conditional on them requesting). Since in both strategy $S$ and strategy $S'$ the agent requests with probability $p(\Hist[t-1])$ independent of all other randomness in the system, they win with the same probability in both cases. For simplicity, let $\lambda = \lambda(\Hist[t-1])$. We, hence, see that
    \begin{eqnarray*}
        &&\E\left[\left.\val_i[t]\cdot \ind{\Req'_i[t] = 1}\right|\Hist[t-1]\right] - \E\left[\left.\val_i[t]\cdot \ind{\Req_i[t] = 1}\right|\Hist[t-1]\right]\\
        &=& \int_{\lambda}^1 v \cdot f_i(v)\cdot \str'(\Hist[t-1], \val_i[t]) \;d\mu(v) - \int_{0}^1 v \cdot f_i(v)\cdot \str(\Hist[t-1], v) \;d\mu(v) \\
         &=& \int_{(\lambda, 1]} v \cdot f_i(v)\cdot (1-\str(\Hist[t-1], v)) \;d\mu(v)+ \int_{\{\lambda\} } v \cdot f_i(v)\cdot (\str'(\Hist[t-1], v) \;d\mu(v)\\
         &&\;\;\;\;\;\;\; -\; \int_{[0, \lambda]} v \cdot f_i(v)\cdot \str(\Hist[t-1], v) \;d\mu(v)\\
         &\geq& \lambda\int_{(\lambda, 1]} f_i(v)\cdot (1-\str(\Hist[t-1], v)) \;d\mu(v)+ \lambda\int_{\{\lambda\} } f_i(v)\cdot (\str'(\Hist[t-1], v) \;d\mu(v)\\
         &&\;\;\;\;\;\;\; -\lambda\; \int_{[0, \lambda]}  f_i(v)\cdot \str(\Hist[t-1], v) \;d\mu(v)\\
         &=& \lambda\left(\int_{\lambda}^1 f_i(v)\cdot \str'(\Hist[t-1], \val_i[t]) \;d\mu(v) - \int_{0}^1 f_i(v)\cdot \str(\Hist[t-1], v) \;d\mu(v)\right)\\
         &=& \lambda(\pr(\Req'_i[t] = 1) - \pr(\Req_i[t] = 1)) = 0\\
    \end{eqnarray*}
    
    We now define a coupling of the process when agent $i$ uses $\str$ and $\str'$ such that the request behaviour of each agent is identical in both processes. Since,  $\E\left[\left.\val_i[t]\cdot \ind{\Req'_i[t] = 1}\right|\Hist[t-1]\right] - \E\left[\left.\val_i[t]\cdot \ind{\Req_i[t] = 1}\right|\Hist[t-1]\right] > 0$ and histories in both processes are identical, $$\U_i(\str_1, \cdots, \str_{i-1}, \str',\str_{i+1}, \cdots, \str_n)[t] - \U_i(\str_1, \cdots, \str_{i-1}, \str,\str_{i+1}, \cdots, \str_n)[t]$$ is a sub-martingale. Hence, the result holds by the Azuma-Hoeffding inequality.
\end{proof}

\section{Complete Proofs from Sections \ref{sec:analysis_thrs} and \ref{sec:decomposition}}
\label{appsec:longrun}

We now provide the complete proofs of results from~\cref{sec:analysis_thrs} and ~\cref{sec:decomposition}.

\subsection{Necessary Conditions for Stability}

\begin{lemma} \label{lemma: drift of win difference}
    Let $R, U, V \subset [n]$ be disjoint sets. Let $f_{ R, U}[t] = \frac{\sum_{j \in R} \TotalWin_j[t]}{\sum_{j \in R} \alpha_j} - \frac{\sum_{j \in U} \TotalWin_j[t]}{\sum_{j \in U} \alpha_j}$ be the random variable. Then, on any round where $\frac{\TotalWin_j[t]}{\alpha_j} > \frac{\TotalWin_i[t]}{\alpha_i}$ for all $j \in V$ and $i \in R$,
    $$\E[f_{ R, U}[t+1] - f_{R, U}[t]|\mathcal{F}[t]] \geq \frac{1-\prod_{k \in R}(1-p_k)}{\sum_{j \in R}\alpha_i}\cdot \prod_{k \not\in R \cup V}(1-p_k) -\frac{1-\prod_{k \in U}(1-p_k)}{\sum_{j \in U}\alpha_i}$$ where $\mathcal{F}[t]$ is a filtration for the random process such that $\mChain[t]$ is $\mathcal{F}[t]$ measurable.
\end{lemma}
\begin{proof}
    When agent $j \in R$ requests the resource, they only win if no agent with higher priority request the resource. In the worst case, agents in $R$ has the least possible priority. Since we assume  $\frac{\TotalWin_j[t]}{\alpha_j} > \frac{\TotalWin_i[t]}{\alpha_i}$ for all $j \in V$ and $i \in R$, agents in $R$ has priority over agents in $V$. Hence, in the worst case, all agents in $[n] \setminus (R \cup V)$ have priority over agents in $R$. Hence, an agent in $R$ will only win if no agent in $R \cup V$ requests the resource and an agent in $R$ requests the resource. This happens with probability $(1-\prod_{k \in R}(1-p_k))\cdot \prod_{k \not\in R \cup V}(1-p_k)$ and, on this event, $f_{R, U}[t+1] - f_{R, U}[t] = \frac{1}{\sum_{j \in R}\alpha_i}$. On the other hand, the probability an agent in $U$ win the resource is at most the probability that an agent in $U$ requests the resource, which is given by $1-\prod_{k \in U}(1-p_k)$ and, on this event, $f_{R, U}[t+1] - f_{R, U}[t] = -\frac{1}{\sum_{j \in U}\alpha_i}$
\end{proof}

\ThmNecessaryCondition*
\begin{proof}
    We apply Lemma \ref{lemma: drift of win difference} with $V = \emptyset$ to see that, when the condition does not hold, $\E[f_{R, U}[t+1] - f_{R, U}[t]|\mathcal{F}[t]] \geq \epsilon > 0$ for all $t$. Hence, by the Azuma-Hoeffding inequality, $\frac{\sum_{j \in R} \TotalWin_j[t]}{\sum_{j \in R} \alpha_j} - \frac{\sum_{j \in U} \TotalWin_j[t]}{\sum_{j \in U} \alpha_j} = f_{R, U}[t] \geq (\epsilon + o(1))t$ almost surely.
\end{proof}

\subsection{Sufficient Conditions for Stability}
The following technical result is a minor modification of the result of Pemantle et al. \cite{pemantle1999moment} to get an almost sure bound and will be useful for showing the sufficient condition.
\begin{theorem} Let $a > 0$, $0< J, C \leq \infty$. Let $X_n \geq 0$ be a sequence of random variables with filtration $\mathcal{F}_n$  such that $\E[X_{n+1} - X_{n}|\mathcal{F}_n] \leq 0$ whenever $X_{n} > J$ and $\E[|X_{n+1} - X_{n}||\mathcal{F}_n] \leq C$. Then, $\E[X_n]$ is bounded above by a constant independent of $n$. Hence,  $X_n = O(\sqrt{n\log n})$ almost surely.
\label{Thm: pemantle1999moment almost surely}
\end{theorem}
\begin{proof}
    Fix $n \in N$. Let $r_n = \max\{s \leq n: X_s < J\}$ be the random variable for the last time in which $X_n < M$. Hence, for all $r_n < s \leq n$, $X_s > J$ and $X_{r_n} \leq M$. Observe that, for all $k$, $H^k_l = \sum_{s=k+1}^l (X_{s+1} - X_{s})\ind{X_{s} > J}$ is a super martingale. Let $A_k$ be the event that $r_n = k$ and let $B_k$ be the event that $H^k_n \leq g(n)$. Then conditioned on the events $A_k$ and $B_k$ both occurring,
\begin{equation}
    X_n =  X_k + \sum_{s=k+1}^n (X_{s+1} - X_{s}) \leq  X_k + \sum_{s=k+1}^n (X_{s+1} - X_{s})\ind{X_{s} > J} \leq  J + g(n)
\end{equation}
Hence, on the event $D_n = \cup_{i=k}^n A_k \cap B_k$, $X_n \leq g(n) + J$. Observe that $D^c_n = \cup_{k=1}^n A_k \cap B_k^c$. Hence, $\pr(D^c_n) = \pr\left(\cup_{k=1}^n A_k \cap B_k^c\right) = \sum_{k=1}^n \pr\left(A_k \cap B_k^c\right) \leq \sum_{k=1}^n \pr\left(B_k^c\right) = \sum_{k=1}^n \pr\left(H^k_n \leq g(n)\right)$. Using the Azuma-Hoeffdings inequality, we have that $$\pr\left(H^k_n \leq g(n)\right) \leq \exp\left(-\frac{g(n)^2}{2 (n-k) C^2}\right) \leq \exp\left(-\frac{g(n)^2}{2C^2\cdot n}\right) \leq \exp\left(-3\log n\right) = \frac{1}{n^3}$$ where we take $g(n) = \sqrt{6C^2\cdot n\log n}$. Hence, $\pr(D^c_n) \leq \sum_{k=1}^n \frac{1}{n^3} = \frac{1}{n^2}$. Furthermore, $\sum_{n=1}^\infty \pr(D^c_n) < \infty$. Hence, by the Borel-Cantelli Lemma, only finitely many $D^c_n$ occur almost surely. Thus, almost surely, $X_n \leq J + g(n) = O(\sqrt{n\log n})$ for large $n$.

\end{proof}

To prove the sufficient condition, we will need the following useful lemmas:

\begin{lemma}
    For $w_1, \cdots, w_n \in [0, 1]$ such that $\sum_{i=1}^nw_i = 1$, the following conditions are equivalent:
    \begin{enumerate}
        \item For all $i$, $\sum_{j = 1}^iw_j \geq \sum_{j=1}^i\alpha_i,$ 
        \item For all $i$, $\frac{\sum_{j = 1}^iw_j}{\sum_{j=1}^i\alpha_i} \geq \frac{\sum_{j = i+1}^nw_j}{\sum_{j = i+1}^n\alpha_j}$
    \end{enumerate}
\end{lemma}
\begin{proof}
    To show that $(1)$ implies $(2)$, we see that $(1)$ implies $\sum_{j = i+1}^nw_j \leq 1 - \sum_{j=1}^i\alpha_i = \sum_{j = i+1}^n\alpha_j$ from which $(2)$ is immediate. To show $(2)$ implies $(1)$, we see that 
    \begin{eqnarray*}
    \frac{\sum_{j = 1}^iw_j}{\sum_{j=1}^i\alpha_i} &\geq& \frac{\sum_{j = i+1}^nw_j}{\sum_{j = i+1}^n\alpha_j}\\
     \left(\frac{1}{\sum_{j=1}^i\alpha_i} + \frac{1}{\sum_{j = i+1}^n\alpha_j}\right)\sum_{j = 1}^iw_j &\geq& \frac{\sum_{j = i+1}^nw_j}{\sum_{j = i+1}^n\alpha_j} + \frac{\sum_{j = 1}^iw_j}{\sum_{j = i+1}^n\alpha_j}\\
     \frac{1}{\left(\sum_{j=1}^i\alpha_i\right)\left(\sum_{j = i+1}^n\alpha_j\right)}\sum_{j = 1}^iw_j &\geq& \frac{1}{\sum_{j = i+1}^n\alpha_j}
    \end{eqnarray*} from which $(1)$ is immediate. 
    
\end{proof}

\coedit{
\begin{lemma} \label{sum bound lemma}
    Fix $\gamma_i \in [0, 1]$. Let $\alpha_i \in [0, 1]$ such that $\frac{\gamma_1}{\alpha_1} \leq \cdots \leq \frac{\gamma_n}{\alpha_n}$. Then, for $w_1, \cdots, w_n \in [0, 1]$ such that  $\sum_{j = 1}^nw_j = \sum_{j = 1}^n \alpha_j$. Then,
    $$\sum_{j = 1}^nw_j\cdot \frac{\gamma_j}{\alpha_j} - \sum_{j = 1}^n\gamma_j = \sum_{k=1}^{n-1} \left(\sum_{j = 1}^k \alpha_j - \sum_{j = 1}^kw_j \right)\cdot \left(\frac{\gamma_{k+1}}{\alpha_{k+1}} - \frac{\gamma_{k}}{\alpha_{k}}\right).$$ 
\end{lemma}
\begin{proof}
    Let $E_1 = 0$ and, for $1 < k \leq n$, let $$E_k = \left( \sum_{j = 1}^{k-1}w_j - \sum_{j = 1}^{k-1} \alpha_j \right)\cdot \left(\frac{\gamma_{k-1}}{\alpha_{k-1}} - \frac{\gamma_{k}}{\alpha_{k}}\right).$$ 
    By induction on $p$, we will show that $$\sum_{j = 1}^p w_j\cdot \frac{\gamma_j}{\alpha_j} - \sum_{j = 1}^p\gamma_j = \left(\sum_{j = 1}^p w_j - \sum_{j = 1}^p \alpha_j\right)\cdot \frac{\gamma_{p}}{\alpha_{p}} +  \sum_{k=1}^{p} E_k$$ for all $1 \leq p \leq n$. The result holds trivially for $p = 1$. Assume the result hold for $p = K-1$. We now show that the result holds for $p = K$ as follows:

    \begin{eqnarray*}
        \sum_{j = 1}^Kw_j\cdot \frac{\gamma_j}{\alpha_j} - \sum_{j = 1}^K\gamma_j &=& \left(\sum_{j = 1}^{K-1}w_j\cdot \frac{\gamma_j}{\alpha_j} - \sum_{j = 1}^{K-1}\gamma_j \right) + (w_K - \alpha_K)\frac{\gamma_K}{\alpha_K} \\
        &=& \left(\sum_{j =1}^{K-1}w_j - \sum_{j = 1}^{K-1} \alpha_j\right)\cdot \frac{\gamma_{K-1}}{\alpha_{K-1}}+ \sum_{j=1}^{K-1} E_k + (w_K - \alpha_K)\frac{\gamma_K}{\alpha_K}  \\
        &=& E_K + \left(\sum_{j =1}^{K-1}w_j - \sum_{j = 1}^{K-1} \alpha_j\right)\cdot \frac{\gamma_{K}}{\alpha_{K}} + (w_K - \alpha_K)\frac{\gamma_K}{\alpha_K} + \sum_{j=1}^{K-1} E_k\\
        &=& \left(\sum_{j =1}^{K}w_j - \sum_{j = 1}^{K} \alpha_j\right)\cdot \frac{\gamma_{K}}{\alpha_{K}} + \sum_{j=1}^{K} E_k\\
    \end{eqnarray*} as claimed. The main result then follows by observing that $\sum_{j = 1}^n \alpha_j - \sum_{j = 1}^nw_j = 0$ and re-indexing the terms in $\sum_{k=2}^n E_k$.
\end{proof}
}

\begin{lemma} \label{lemma: top heavy sum}
    Let $a_i, b_i$ be sequences of positive number such that $\sum_{i=1}^n a_i = \sum_{i=1}^n b_i$ and $\frac{a_j}{a_i} \leq \frac{b_j}{b_i}$ for all $i < j$. Then, for all $k \in [n]$, $\sum_{i=1}^k a_i \geq \sum_{i=1}^k b_i$.
\end{lemma}
\begin{proof}
    Fix $k \in [n]$. For all $i < k+1$ and $j \geq k+1$, $\frac{a_j}{a_i} \leq \frac{b_j}{b_i}$ holds. By summing $j$ from $k+1$ to $n$, we have that $\frac{\sum_{l=k+1}^n a_l}{a_i} \leq \frac{\sum_{l=k+1}^n b_l}{b_i} $. Hence, $\frac{a_i}{\sum_{l=k+1}^n a_l} \geq \frac{b_i}{\sum_{l=k+1}^n b_l}$. We now sum $i$ from $1$ to $k$ to achieve $\frac{\sum_{l=1}^{k} a_i}{\sum_{l=k+1}^n a_l} \geq \frac{\sum_{l=1}^{k}b_i}{\sum_{l=k+1}^n b_l}$. Hence,
    $$\frac{\sum_{l=k+1}^n a_l} {\sum_{l=1}^{k} a_l}+  1 \leq \frac{\sum_{l=k+1}^n b_l}{\sum_{l=1}^{k}b_l} + 1 \Rightarrow 
        \frac{\sum_{l=1}^{n} a_l}{\sum_{l=1}^{k}a_l} \leq \frac{\sum_{l=1}^{n}b_l}{\sum_{l=1}^{k}b_l} \Rightarrow 
        \sum_{l=1}^{k}a_l \geq \sum_{l=1}^{k}b_l.$$
\end{proof}

We now prove \cref{lemma: Sufficient condition}. To do this, we will need to describe the state space and state transitions in $\mChain^S[t]$. 

For any $x \inR^n$, let $I_i(\omega) = \{j \in [n]: x_j < x_i \text{ or } x_j = x_i \text{ and } j < i\}$ denote the set of agents with higher priority than $i$ in state $x$ (i.e. the set of agents that will receive the resource over agent $i$ if they request while the system is in state $x$).
Observe that agent $i$ wins the resource on a round if and only if they request and no agent in $I_i(\mChain[t])$ requests, which happens with probability $p_i\prod_{k \in I_i(\mChain[t])}(1-p_k)$. 
Define $\textbf{f}^S_i = \frac{1}{\alpha_i(S)}\textbf{e}_i - \vec{\textbf{1}}_S$, where $\textbf{e}_i$ is the $i$-th standard basis vector in $\R^n$, and $\vec{\textbf{1}}_S$ be the incidence vector of $S$ (i.e., $1$ for each $i\in S$ and $0$ otherwise). Then, with probability $p_i\prod_{k \in I_i(\mChain[t])}(1-p_k)$, $\mChain^S[t+1] = \mChain^S[t] + \textbf{f}^S_i$ for all $i \in S$. Otherwise $\mChain^S[t+1] = \mChain^S[t]$. We may also assume the system is initialized as $\mChain[0] = \vec{\textbf{0}}$. Observe that this implies that the state space of $\mChain^S[t]$ is given by 
$\States(S) = \left\{\displaystyle\sum_{i\in S} \gamma_i \cdot \textbf{f}^S_i: \gamma_i \inZ_{\geq 0}\right\} \subset \R^n$ where $\gamma_i$ can be thought of as the number of rounds each agent has won.

\ThmSufficientCondition*
\begin{proof}
%Let $\beta_i = \frac{\alpha_i}{\sum_{k \in S} \alpha_k}$. 
\coedit{Let $|S| = m$.} Let $G(\vec{x}) = \sum_{i \in S} \alpha_i(S)\cdot x_i^2$. For $i \in [n]$, on round $t+1$, with probability $p_i \prod_{k \in I_i(\mChain[t])}(1-p_k)$, agent $i \in S$ wins the rounds. Furthermore, conditional on agent $i \in S$ winning the resource on round $t + 1$, 

\begin{eqnarray*}
    G(\mChain^S[t+1]) - G(\mChain^S[t]) &=& \alpha_i(S)\left(\mChain^S_i[t]+\frac{1}{\alpha_i(S)}-1\right)^2 + \sum_{j \in S - \{i\}}\alpha_j(S)(\mChain^S_j[t]-1)^2\\
    && \;\;\;\;\;\;\;-\; \sum_{j \in S }\alpha_j(S) \mChain^S_j[t]^2\\
    &=& 2 \alpha_i(S)\left(\frac{1}{\alpha_i(S)}-1\right)\mChain^S_i[t]+\alpha_i(S)\left(\frac{1}{\alpha_i(S)}-1\right)^2\\
    && \;\;\;\;\;\;\;+\; \sum_{j \in S - \{i\}}\alpha_j(S)(1 - 2\mChain^S_j[t])\\
    &=& 2 \mChain^S_i[t] -2\sum_{j \in S} \alpha_j(S)\mChain^S_j[t] +\alpha_i(S)\left(\frac{1}{\alpha_i(S)}-1\right)^2 + \sum_{j \in S - \{i\}}\alpha_j(S)\\ 
    &=& 2 \mChain^S_i[t] +\alpha_i(S)\left(\frac{1}{\alpha_i(S)}-1\right)^2 + \sum_{j \in S - \{i\}}\alpha_j(S)\\
    &=& 2 \mChain^S_i[t]+\nu_i
\end{eqnarray*}
where $\nu_i = \frac{1}{\alpha_i(S)} - 1$ and the fourth equality holds because $\sum_{i \in S} \alpha_i(S)\cdot \mChain^S_i[t] = \sum_{i \in S}  \TotalWin_i[t] - \TotalWin_S = 0$. If no agent in $S$ wins, $G(\mChain^S[t+1]) - G(\mChain^S[t]) = 0$.

Let $w \in [0, 1]^n$ be such that $w_i = p_i\cdot  \prod_{k \in I_i(\mChain[t])}(1-p_k)$. Then,
$$\E[\nabla G(\mChain^S[t])] = \sum_{i \in S} \nu_i \cdot w_i + 2\sum_{i \in S} \mChain^S_i[t] \cdot w_i.$$

Let  $\mChain^S[t] = \sum_{i\in S}\gamma_i\cdot \textbf{f}^S_i$ where $\gamma_i \in \R_{\geq 0}$. (Note that $x \in \States(S)$ if and only if $\gamma_i \in \Z_{\geq 0}$.) We may assume that $\gamma_i = 0$ for some $i$ (since $\sum_{i\in S}\alpha_i(S)\cdot \textbf{f}^S_i = 0$). We now observe that
\begin{eqnarray*}
    \inner{\mChain^S[t]}{w} &=& \sum_{i \in S}\gamma_i\cdot \inner{\textbf{f}^S_i}{w}=  \sum_{i \in S}\gamma_i\cdot \frac{1}{\alpha_i(S)}\cdot w_i - \sum_{i \in S}\gamma_i\cdot\sum_{j \in S}w_j\\
    &=& \sum_{i \in S}\gamma_i\cdot \frac{1}{\alpha_i(S)}\cdot w_i - \sum_{i \in S}w_i\cdot \sum_{j \in S}\gamma_j = \sum_{i \in S}w_i\left(\gamma_i\cdot \frac{1}{\alpha_i(S)} - \sum_{j \in S}\gamma_j\right)
\end{eqnarray*}

\coedit{We now assume, without lose of generality, that $S = [m]$. Let $L = \sum_{i\in S} w_i$. Let $\hat{w} = \frac{w}{L}$. Then, 
$$\inner{\mChain^S[t]}{w} = L\sum_{i=1}^m \frac{w_i}{L} \left(\frac{\gamma_i}{\alpha_i(S)} - \sum_{j=1}^m \gamma_j\right) = L\left(\sum_{i=1}^m \frac{\hat{w_i}\cdot \gamma_i}{\alpha_i(S)}  - \sum_{j=1}^m \gamma_j\right).$$} 
Furthermore, assume that for all $i, j \in S$, for $i < j$ , $i \in I_j(x)$. \coedit{Then, it must be the case that $\frac{\gamma_i}{\alpha_i(S)} \leq \frac{\gamma_j}{\alpha_j(S)}$ and $\gamma_1 = 0$. Thus, by \Cref{sum bound lemma},
$$\inner{\mChain^S[t]}{w} = -L\left(\sum_{k=1}^{m-1} \left( \sum_{j = 1}^k\hat{w}_j - \sum_{j = 1}^k \alpha_j(S)\right)\cdot \left(\frac{\gamma_{k+1}}{\alpha_{k+1}(S)} - \frac{\gamma_{k}}{\alpha_{k}(S)}\right)\right).$$}

For $U \subset [n]$, let $I_j(x, U) =  I_j(x) \cap U$. Then $I_j(x, S) = [j-1]$. Let $\bar{S} = [n] \setminus S$. We must also have that, for $i < j$, $I_i(x, \bar{S}) \subset I_j(x, \bar{S})$ since any agent that has greater priority than $i$ must also have greater priority than $j$. 

Hence, $\hat{w}_i = \frac{1}{L} \cdot p_i\cdot  \prod_{k = 1}^{i-1}(1-p_k)\prod_{k \in I_i(x, \bar{S})}(1-p_k)$. We now claim that $\sum_{k=1}^i\hat{w}_k > \sum_{k = 1}^{i} \alpha_k(S)$ for all $i$. Let $\bar{w}_i = p_i\cdot \frac{\prod_{k = 1}^{i-1}(1-p_k)}{1-\prod_{k \in S}(1-p_k)}$. Then, for $i < j$, observe that 

\begin{eqnarray*}
    \frac{\hat{w}_j}{\hat{w}_i} &=& \frac{p_j\prod_{k = 1}^{j-1}(1-p_k)}{p_i\prod_{k = 1}^{i-1}(1-p_k)}\cdot \frac{\prod_{k \in I_i(x, \bar{R})}(1-p_k)}{\prod_{k \in I_i(x, \bar{R})}(1-p_k)}\\
    &=& \frac{p_j\prod_{k = 1}^{j-1}(1-p_k)}{p_i\prod_{k = 1}^{i-1}(1-p_k)} \cdot \prod_{k \in I_i(x, \bar{R})-I_i(x, \bar{R})}(1-p_k) \leq \frac{p_j\prod_{k = 1}^{j-1}(1-p_k)}{p_i\prod_{k = 1}^{i-1}(1-p_k)} = \frac{\bar{w}_j}{\bar{w}_i}
\end{eqnarray*}

Furthermore, $\sum_{i\in S} \hat{w}_i = \sum_{i\in S} \bar{w}_i = 1$. Hence, by Lemma \ref{lemma: top heavy sum}, for all $i < |S|$, $\sum_{k=1}^i\hat{w}_i \geq \sum_{k=1}^i\bar{w}_i$. We now assume the stability criterion holds with strict inequality for all $R \subsetneq S$. Then
$$\sum_{k=1}^i\hat{w}_k \geq \sum_{k=1}^i\bar{w}_k =  \frac{1-  \prod_{k = 1}^{i}(1-p_k)}{1-  \prod_{k \in S}(1-p_k)} > \sum_{k = 1}^{i} \alpha_k(S)$$ using the assumption. 
\coedit{Let $\epsilon = \min_{i < n}\left\{\sum_{k = 1}^{i} \alpha_k(S) - \sum_{k=1}^i\hat{w}_k\right\} > 0.$ Then,
$$\inner{\mChain^S[t]}{w} = -L\cdot \epsilon \cdot \left(\frac{\gamma_{m}}{\alpha_{m}(S)} - \frac{\gamma_{1}}{\alpha_{1}(S)}\right) = -L\cdot \epsilon \cdot \frac{\gamma_{m}}{\alpha_{m}(S)}.$$

}

%\newpage
%This implies that $\sum_{k = i}^{|S|} \alpha_k(S) > \sum_{k=i}^{|S|}\hat{w}_k$ for all $i > 1$. Let $$\epsilon = \min_{i > 1}\left\{\sum_{k = i}^{|S|} \alpha_k(S) - \sum_{k=i}^{|S|}\hat{w}_k\right\} > 0.$$ 
%Hence, $\hat{w}$ satisfies the condition in Lemma \ref{sum bound lemma}. Hence, $$\sum_{i=1}^{|S|} \frac{\hat{w_i}\cdot \gamma_i}{\alpha_i(S)} = \sum_{i=2}^{|S|} \frac{\hat{w_i}\cdot \gamma_i}{\alpha_i(S)} \leq \sum_{j=2}^{|S|} \gamma_j - \epsilon \cdot \frac{\gamma_{|S|}}{\alpha_{|S|}(S)} = \sum_{j=1}^{|S|} \gamma_j - \epsilon \cdot \frac{\gamma_{|S|}}{\alpha_{|S|}(S)}$$ where use the fact that $\gamma_1 = 0$.

%Thus, $\inner{\mChain^S[t]}{w} \leq -L \cdot \epsilon \cdot \frac{\gamma_{|S|}}{\alpha_{|S|}(S)}$. 
We now consider all $x$ such that $\frac{\gamma_{m}}{\alpha_{m}(S)} \geq \frac{1+\sum_{i=1}^m \nu_i}{2\epsilon\cdot L}$. Hence, $2\inner{\mChain^S[t]}{w} \leq -1-\sum_{i=1}^m \nu_i$. From this, we see that 
$$\E[\nabla G(\mChain^S[t])] < \sum_{i \in S} \nu_i + 2\inner{\mChain^S[t]}{w} < -1.$$ There are only finitely many states $x \in \States(S)$ where $\frac{\gamma_i}{\alpha_i(S)} \leq \frac{1+\sum_{i=1}^m \nu_i}{2\epsilon\cdot L}$ for all $i \in S$. Hence, whenever, $G(\mChain^S[t])$ is sufficiently large, $\E[\nabla G(\mChain^S[t])] < -1$. Hence, by Theorem \ref{Thm: pemantle1999moment almost surely}, $\norm{\mChain^S[t]}^2 \leq \frac{G(\mChain^S[t])}{\min_{i \in S}\alpha_i(S)}  = o(t)$ almost surely. Hence, we conclude that 
\begin{eqnarray*}
    \left|\frac{\TotalWin_i[t]}{\alpha_i} -  \frac{\TotalWin_j[t]}{\alpha_j}\right| &=& \frac{1}{\sum_{k \in S} \alpha_i} \left|\frac{\TotalWin_i[t]}{\alpha_i(S)} - \frac{\TotalWin_j[t]}{\alpha_j(S)}\right|\\
    &\leq& \frac{1}{\sum_{k \in S} \alpha_i}\left( \left|\frac{\TotalWin_i[t]}{\alpha_i(S)} - \TotalWin_R[t]\right| + \left|\frac{\TotalWin_j[t]}{\alpha_j(S)} - \TotalWin_R[t]\right|\right)\\
    &\leq& O(1) \cdot \norm{\mChain^S[t]}^2 = o(t)
\end{eqnarray*}

We now remove the assumption that the stability criterion holds with strict inequality for all $R \subsetneq S$. Notice that the set of request probabilities $(p_1, \cdots, p_n)$ where the stability criterion holds with strict inequality for all $R \subsetneq S$ is a dense subset of the set of request probabilities $(p_1, \cdots, p_n)$ where the stability criterion holds generally. 

Fix $\vec{p} = (p_1, \cdots, p_n)$ where the stability criterion holds. Let $\vec{p}_l$ be a request probability vector such that the stability criterion holds with strict inequality for each $\vec{p}_l$ and $\norm{\vec{p}_l - \vec{p}} \leq \frac{1}{l}$. Then 
$$\left|\frac{\TotalWin_i[t](\vec{p})}{\alpha_i} -  \frac{\TotalWin_j[t](\vec{p})}{\alpha_j}\right| \leq \left|\frac{\TotalWin_i[t](\vec{p}_l)}{\alpha_i} -  \frac{\TotalWin_j[t](\vec{p})}{\alpha_j}\right| + \left|\frac{\TotalWin_i[t](\vec{p})}{\alpha_i} -  \frac{\TotalWin_i[t](\vec{p}_l)}{\alpha_i}\right|+ \left|\frac{\TotalWin_j[t](\vec{p})}{\alpha_j} -  \frac{\TotalWin_j[t](\vec{p}_l)}{\alpha_j}\right|$$

We now apply Lemma \ref{lemma: W is cont} to achieve that
$$\lim\sup_{t \rightarrow \infty}\left|\frac{1}{t}\left(\frac{\TotalWin_i[t](\vec{p})}{\alpha_i} -  \frac{\TotalWin_j[t](\vec{p})}{\alpha_j}\right)\right| \leq n\left(\frac{1}{\alpha_i} + \frac{1}{\alpha_j}\right)\frac{1}{l} $$ Since this holds, for all $l$, it must hold that 
$$\left|\frac{\TotalWin_i[t](\vec{p})}{\alpha_i} -  \frac{\TotalWin_j[t](\vec{p})}{\alpha_j}\right| \leq \lim\sup_{t \rightarrow \infty}\left|\frac{\TotalWin_i[t](\vec{p})}{\alpha_i} -  \frac{\TotalWin_j[t](\vec{p})}{\alpha_j}\right| \leq o(t).$$

\end{proof}

\subsection{Decomposition of \texorpdfstring{$\mChain[t]$}{the Markov process}}
\SplitChainLemma*

\begin{proof}
For simplicity of presentation, we will index the rows of $\mChainY^{S}(\omega)[t]$ using indices in $S$ and the rows of $\mChainY^{[n] \setminus S}(\omega')[t]$ using indices in $[n] \setminus S$. By the splitting property, almost surely, there exist $T$ such that $\frac{\TotalWin_j[t]}{\alpha_j} - \frac{\TotalWin_j[t]}{\alpha_i} \geq \Omega(t)$ for $t \geq T$. Consider any such sample path. Let $\omega = \mChain^S[t]$ and $\omega' = \mChain^{[n] \setminus S}[t]$.

We now couple $\mChain[t]$ and $\mChainY^{S}(\omega)[t]$ such that, for all agent $i \in S$, the value of agent $i$ in round $t$ in the instances of the DMMF mechanism corresponding to $\mChain$ is equal to the value of agent $i$ in round $t - T$ in the instances of the DMMF mechanism corresponding to $\mChainY^{S}(\omega)$. This means the behaviour of each agent in round $t$ in $\mChain$ is identical to the behaviour of each agent in round $t - T$ in $\mChainY^{S}(\omega)$ (though who wins in each instance may be different). 

We couple $\mChain[t]$ and $\mChainY^{[n] \setminus S}(\omega')[t]$ such that, for all agent $i \in [n] \setminus S$, the value of agent $i$ in round $t$ in the instances of the DMMF mechanism corresponding to $\mChain$ is equal to the value of agent $i$ in round $t - \TotalAnyReq_S[t] - (T-\TotalAnyReq_S[T])$ in the instances of the DMMF mechanism corresponding to $\mChainY^{[n] \setminus S}(\omega')$.

We now show inductively that $\mChain^{S}[t] = \mChainY^{S}(\omega)[t-T]$ for $t \geq T$. First, $\mChainY^{S}(\omega)[0] = \omega = \mChain^{S}[T]$, showing the base case. Assume the result holds for $t - 1 \geq T$. Then, if agent $i \in S$ wins the resource in the round $t-T$ in $\mChainY^{S}(\omega)$ then agent $i$ wins the resource in round $t$ in $\mChain$. This is because the request behaviour of agents is the same and, by the  induction hypothesis, the allocation represented by $\mChain[t-1]$ and $\mChainY^{S}(\omega)[t-1 -T]$ are identical; hence, the agent priorities are the same. Thus, the allocations on round $t$ are the same ie. $\mChain[t] = \mChainY^{S}(\omega)[t -T]$, proving the claim.

We now show inductively that $\mChain^{[n] \setminus S}[t] = \mChainY^{[n] \setminus S}(\omega')\left[t - \TotalAnyReq_S[t] - (T - \TotalAnyReq_S[T])\right]$.  First, $\mChainY^{[n] \setminus S}(\omega')[0] = \omega' = \mChain^{[n] \setminus S}[T]$, showing the base case. Assume this holds for $t - 1 \geq T$. If $\TotalAnyReq_S[t] =  \TotalAnyReq_S[t-1] +1$ then an agent in $S$ requested the resource. Hence, no agent in $[n] \setminus S$ can win the resource. Thus, 
\begin{eqnarray*}
    \mChain^{[n] \setminus S}[t] &=& \mChain^{[n] \setminus S}[t-1]\\
    &=& \mChainY^{[n] \setminus S}(\omega')\left[t-1 - \TotalAnyReq_S[t-1] -(T-\TotalAnyReq_S[T])\right]\\
    &=& \mChainY^{[n] \setminus S}(\omega')\left[t - \TotalAnyReq_S[t] - (T-\TotalAnyReq_S[T])\right]
\end{eqnarray*}

On the other hand, when no agent in $S$ requests the resource then $\TotalAnyReq_S[t] =  \TotalAnyReq_S[t-1]$. In this case, the request behaviour of agents is the same in $\mChainY^{[n] \setminus S}$ and $\mChain^{[n] \setminus S}$. Furthermore, by the  induction hypothesis, the allocation represented by $\mChain[t-1]$ and $\mChainY^{[n] \setminus S}(\omega')[t-1 - \TotalAnyReq_S[t-1] - (T-\TotalAnyReq_S[T])]$ are identical; hence, the agent priorities are the same. Thus, the allocations are the same ie. $\mChain[t] = \mChainY^{[n] \setminus S}(\omega')[t - \TotalAnyReq_S[t] - (T-\TotalAnyReq_S[T])]$, proving the claim.
\end{proof}

\UtilityAfterSplit*
\begin{proof}
    Consider any sample path where Lemma \ref{lemma: process splittng lemma} holds. Then $\mChainY^{S}(\omega)[t-T] = \mChain^{S}[t]$ for $t > T$ represents an instance of the DMMF mechanism that satisfies the stability criterion. In this instance of the DMMF, the fair share of each agent $i \in S$ is $\alpha_i(S)$. Hence, by Lemma \ref{thm: condition for stablity}, the process $\mChainY^{S}(\omega)$ is stable and $$\TotalWin_i[t] = \frac{\alpha_i}{\sum_{k \in S}\alpha_k}\left(1-\prod_{i \in S}(1-p_i)\right)\cdot (t-T) + o(t) = \frac{\alpha_i}{\sum_{k \in S}\alpha_k}\left(1-\prod_{i \in S}(1-p_i)\right)\cdot t + o(t)$$ for all $i \in S$ almost surely. We note that we incur a shift of $T$ in time scale as we go from $\mChainY^{S}(\omega)[t]$ to $\mChain^{S}[t]$. This error is however $O(1) = o(t)$. W

     We make the same argument for $\mChainY^{[n] \setminus S}\left(\omega'\right)\left[t - \TotalAnyReq_S[t] - (T-\TotalAnyReq_S[T])\right] = \mChain^{[n] \setminus S}[t] $. Thus, we have that 
     \begin{equation} \label{eq: Wins of higher set 1}
         \TotalWin_i[t] = \frac{\alpha_i}{\sum_{k \in S}\alpha_k}\left(1-\prod_{i \in S}(1-p_i)\right)\cdot (t - \TotalAnyReq_S[t] - (T-\TotalAnyReq_S[T])) + o(t)
     \end{equation}
     almost surely. 
     
     On any sample path where Lemma \ref{lemma: process splittng lemma} holds, for $t > T$, $\AnyReq_S[t]$, the indicator for the event that an agent in $S$ wins round $t$, is equivalent to the indicator for the event that an agent in $S$ requests the resource. This is Bernoulli variable with probability $1-\prod_{i \in S}(1-p_i)$. Hence, by the law of large numbers, tt must be that $\TotalAnyReq_S[t] = \TotalAnyReq_S[T] + \sum_{\tau=T+1}^t \AnyReq_S[t] = (1-\prod_{i \in S}(1-p_i))\cdot t + o(t).$ We then achieve the result by substituting the almost sure expression into \cref{eq: Wins of higher set 1}

\end{proof}

\subsection{Existence of a Splitting Set and Splitting Partition}
In this section, we shall show that there exist $S \subseteq [n]$ that has the splitting property.

\begin{lemma} \label{lemma: positive drift random process}
    Fix $a > 0$ and $J, C < \infty$. Let $X_n \inR$ be a sequence of random variables with filtration $\mathcal{F}_n$  such that $\E[X_{n+1} - X_{n}|\mathcal{F}_n] \geq \epsilon >0$ whenever $X_{n} < J \cdot n$ and $\E[|X_{n+1} - X_{n}||\mathcal{F}_n] \leq C$. Then, $X_n > \Omega(n)$ for all $n$ almost surely. 
\end{lemma}
\begin{proof}
    Fix $n \in N$. Let $r_n = \max\{s \leq n: X_s \geq J \cdot s\}$ be the random variable for the last time in which $X_s \geq J \cdot s$ up to round $s$. Hence, for all $r_n < s \leq n$, $X_s < J\cdot n$. Thus, $\E[X_{n+1} - X_{n}|\mathcal{F}_n] \geq \epsilon >0$ for all $r_n < s \leq n$ and $X_{r_n} \geq J \cdot r_n$. Observe that, for all $k$, $H^k_l = \sum_{s=k+1}^l (X_{s+1} - X_{s} -\epsilon)\ind{X_{s} < J \cdot s}$ is a sub-martingale. Let $A_k$ be the event that $r_n = k$ and let $B_k$ be the event that $H^k_n \geq -\sqrt{5C^2n\log n}$. Then conditioned on the events $A_k$ and $B_k$ both occurring,
\begin{eqnarray*}
    X_n &=&  X_k + \sum_{s=k+1}^n (X_{s+1} - X_{s})\\
    &\geq&  X_k + \sum_{s=k+1}^n (X_{s+1} - X_{s} - \epsilon)\ind{X_{s} > J} + (n-k)\epsilon\\
    &\geq&  (J-\epsilon) \cdot k + n\cdot \epsilon -\sqrt{5C^2n\log n} \geq \min\{J, \epsilon\} \cdot n - \sqrt{5C^2n\log n} = \Omega(n)
\end{eqnarray*}
Hence, on the event $D_n = \cup_{i=k}^n A_k \cap B_k$, $X_n \geq  \Omega(n)$. Observe that $D^c_n = \cup_{k=1}^n A_k \cap B_k^c$. Hence, $\pr(D^c_n) = \pr\left(\cup_{k=1}^n A_k \cap B_k^c\right) = \sum_{k=1}^n \pr\left(A_k \cap B_k^c\right) \leq \sum_{k=1}^n \pr\left(B_k^c\right) = \sum_{k=1}^n \pr\left(H^k_n \leq -\sqrt{5nC^2\log n}\right)$. Using the Azuma-Hoeffdings inequality, we have that $\pr\left(H^k_n \leq -\sqrt{5C^2n\log n}\right) \leq \frac{1}{n^3}$. Hence, $\pr(D^c_n) \leq \sum_{k=1}^n \frac{1}{n^3} = \frac{1}{n^2}$. Furthermore, $\sum_{n=1}^\infty \pr(D^c_n) < \infty$. Hence, by the Borel-Cantelli Lemma, only finitely many $D^c_n$ occur almost surely. Thus, almost surely, $X_n \geq  \Omega(n)$ for large $n$. 

\end{proof}

\begin{lemma} \label{lemma: minimum value is lower bound}
    For $S \subset [n]$,  $\frac{1-\prod_{k \in S}(1-p_k)}{\sum_{k\in S} \alpha_k} \geq \min_{i \in S}\frac{p_i}{\alpha_i}$.
\end{lemma}
\begin{proof}
    For all $U \subset [n]$, let $N(U) = \prod_{k \in U}(1-p_k)$ and $\alpha_U = \sum_{k\in U} \alpha_k$. We now show that, for all $U \subset [n]$ and $s \in [n] \setminus U$ where $\frac{1-N(U)}{\alpha_U} \geq \frac{p_s}{\alpha_s}$, $\frac{1-N(U)}{\alpha_U} \geq \frac{1-N(U \cup \{s\})}{\alpha_{U \cup \{s\}}}$. Indeed, observe that 
    \begin{eqnarray*}
        \frac{1-N(U \cup \{s\})}{\sum_{k\in U \cup \{s\}} \alpha_k} &=& \frac{p_s + (1-N(U))(1-p_s)}{\alpha_{U}}\\
        &=& \frac{1-N(U)}{\alpha_{U}}\cdot \frac{\frac{\alpha_{U}p_s}{1-N(U)} +\alpha_{U}(1-p_s)}{\alpha_{U} + \alpha_s}\\
        &\leq& \frac{1-N(U)}{\alpha_{U}}\cdot \frac{\alpha_s +\alpha_{U}(1-p_s)}{\alpha_{U} + \alpha_s} \leq \frac{1-N(U)}{\alpha_{U}}
    \end{eqnarray*}

    Assume the elements in $S$ are given by $k_1, \cdots, k_{|S|}$ where $\frac{p_{k_i}}{\alpha_{k_i}} \geq \frac{p_{k_{i+1}}}{\alpha_{k_{i+1}}}$. For all $i \leq |S|$, $U_i = \{k_1, \cdots, k_i\}$. Then, by induction, we can show that, for all $i \leq |S|$,  $\frac{1-N(U_i)}{\alpha_{U_i}} \geq \frac{p_{k_i}}{\alpha_{k_i}}$. Hence, 
    $$\frac{1-\prod_{k \in S}(1-p_k)}{\sum_{k\in S} \alpha_k} = \frac{1-N(S)}{\alpha_{S}} = \frac{1-N(U_{|S|})}{\alpha_{U_{|S|}}} \geq \frac{p_{k_{|S|}}}{\alpha_{k_{|S|}}} = \min_{i \in S}\frac{p_i}{\alpha_i}$$ proving the result.
\end{proof}

\begin{lemma} \label{lemma: Maximizer is stable}
    Fix $S \subset [n]$. For $U \subset S$, let $F_S(U) = \frac{1-\prod_{k \in U}(1-p_k)}{\sum_{j \in U}\alpha_i}\cdot \prod_{k \in S-U}(1-p_k)$. Let $S_{\max}$ be a maximizer of $F_S(U)$. Then, for all $U \subset S_{\max}$, $$\frac{1- \prod_{k \in U}(1-p_k)}{1-\prod_{k \in S_{\max}}(1-p_k)} \geq \frac{\sum_{j \in U}\alpha_{j}}{\sum_{j \in  S_{\max}}\alpha_{j}}.$$ Hence, $S_{\max}$ satisfies the stability criterion.
    
\end{lemma}
\begin{proof}
    Let $N(U) = \prod_{k \in U}(1-p_k)$. Let $U \subset S_{\max}$ and $\alpha_U = \sum_{k\in U} \alpha_k$. Let $V = S_{\max} - U$. By maximality, $F(S) > F(V)$. Hence, 
    \begin{eqnarray*}
        \frac{1-N(S_{\max})}{\alpha_{S_{\max}}}\cdot N(S- S_{\max}) &\geq& \frac{1-N(U)}{\alpha_{V}}\cdot N(S-V)\\
        \frac{1-N(S_{\max})}{\alpha_{S_{\max}}} &\geq& \frac{N(U)-N(S_{\max})}{\alpha_{V}}\\
         \alpha_{S_{\max}}N(S_{\max}) - \alpha_{S_{\max}}N(U) &\geq& \alpha_{V}N(S_{\max}) - \alpha_{V} \\
         \alpha_{S_{\max}}- \alpha_{S_{\max}}N(U) &\geq& -(\alpha_{S_{\max}}-\alpha_{V})N(S_{\max})+  \alpha_{S_{\max}}- \alpha_{V} \\
         \frac{1- N(U)}{1-N(S_{\max})} &\geq& \frac{\alpha_{U}}{\alpha_{S_{\max}}}
    \end{eqnarray*} which proves the claim.
\end{proof}

\SplittingSetsExist*
\begin{proof}
Let $i^* = \arg\min_{i \in [n]} \frac{p_i}{\alpha_i}$. Note that $S^*$ must exist since $\{i^*\}$ satisfies the stability criterion and $i^* \in \{i^*\}$. We now show by induction that there exists a sequences of subsets $U_1, \cdots, U_K$ such that $U_i \subsetneq U_{i+1}$ and $\frac{\TotalWin_j[t]}{\alpha_j} - \frac{\TotalWin_{i^*}[t]}{\alpha_{i^*}} \geq (C_j+o(1))t$ for all $j \in U_i$ for all $i \in [K]$ where $C_j > 0$. Furthermore, we will show that $[n] \setminus U_K$ satisfies the stability criterion and $i^* \in [n] \setminus U_i$ for all $i$. 

Let $U_1 = \emptyset$. Then $U_1$ indeed satisfies our condition vacuously. We now assume we have constructed $U_1, \cdots, U_{k-1}$ that satisfies our condition. If  $[n] \setminus U_{k-1}$ is stable then $U_1, \cdots, U_{k-1}$ is a sequence of the desired form. Hence, we assume $V = [n] \setminus U_{k-1}$ is not stable. There must exists $S \subset V$ such that $\frac{1-\prod_{k \in S}(1-p_k)}{\sum_{j \in S}\alpha_j} < \frac{1-\prod_{k \in V}(1-p_k)}{\sum_{j \in V}\alpha_j}$. 
Thus, it must hold that 
$$\frac{p_{i^*}}{\alpha_{i^*}}\leq \min_{i \in S} \frac{p_i}{\alpha_i} \leq \frac{1-\prod_{k \in S}(1-p_k)}{\sum_{j \in S}\alpha_j} < \frac{1-\prod_{k \in V}(1-p_k)}{\sum_{j \in V}\alpha_j}.$$ 
Hence, 
\begin{eqnarray*}
    \frac{p_{i^*}}{\alpha_{i^*}} &<& (1-p_{i^*})\cdot \frac{1-\prod_{k \in V - \{i^*\}}(1-p_k)}{\sum_{j\in V - \{i^*\}}\alpha_j} = (1-p_{i^*})\cdot F_{V - \{i^*\}}(V - \{i^*\})\\
    &\leq& (1-p_{i^*})\cdot F_{V - \{i^*\}}(U^*) =  \frac{1-\prod_{k \in U^*}(1-p_k)}{\sum_{j\in U^*}\alpha_j} \cdot \prod_{k \in V- U^*}(1-p_{k})
\end{eqnarray*}
where $U^*$ is the maximizer of $F_{V - \{i^*\}}(U) = \frac{1-\prod_{k \in U}(1-p_k)}{\sum_{j \in U}\alpha_i}\cdot \prod_{k \in V - \{i^*\}-U}(1-p_k)$. By Lemma \ref{lemma: Maximizer is stable}, $U^*$ is stable.

Let $f_{i^*, U^*}(X[t]) = \frac{\sum_{j \in U^*}\TotalWin_{j}[t]}{\sum_{j \in U^*}\alpha_{j}} - \frac{\TotalWin_{i^*}[t]}{\alpha_{i^*}}$. Then, on any round where $\frac{\TotalWin_{j^*}[t]}{\alpha_{j^*}} \leq \frac{\TotalWin_{j}[t]}{\alpha_{j}}$ for all $j \in U_{k-1}$, 
$$\E[\Delta f_{i^*, U^*}(X[t])] \geq \frac{1-\prod_{k \in U^*}(1-p_k)}{\sum_{j\in U^*}\alpha_j} \cdot \prod_{k \in V- U^*}(1-p_{k}) - \frac{p_{i^*}}{\alpha_{i^*}} = \epsilon > 0$$ by Lemma \ref{lemma: drift of win difference}. 
On the other hand, when $\frac{\TotalWin_{j^*}[t]}{\alpha_{j^*}} \geq \frac{\TotalWin_{j}[t]}{\alpha_{j}}$ for some $j^* \in U^*$ and $j \in U_{k-1}$, we have that 
\begin{eqnarray*}
    f_{i^*, U^*}(X[t]) &=& \frac{\sum_{j \in U^*}\TotalWin_{j}[t]}{\sum_{j \in U^*}\alpha_{j}} - \frac{\TotalWin_{i^*}[t]}{\alpha_{i^*}}\\
    &=& \frac{\TotalWin_{j^*}[t]}{\alpha_{j^*}} - \frac{\TotalWin_{i^*}[t]}{\alpha_{i^*}} + o(t) \\
    &=& \left(\frac{\TotalWin_{j^*}[t]}{\alpha_{j^*}} - \frac{\TotalWin_{j}[t]}{\alpha_{j}}\right) + \left(\frac{\TotalWin_{j}[t]}{\alpha_{j}} - \frac{\TotalWin_{i^*}[t]}{\alpha_{i^*}}\right) + o(t) \geq (C + o(1))t
\end{eqnarray*} where use the fact that $U^*$ is stable and apply Theorem \ref{lemma: Sufficient condition}. Hence, we conclude that the random variable $f_{i^*, U^*}(\mChain[t])$ satisfies the condition of Lemma \ref{lemma: positive drift random process}. Hence, $\frac{\TotalWin_{j}[t]}{\alpha_{j}} - \frac{\TotalWin_{i^*}[t]}{\alpha_{i^*}} = \frac{\sum_{j \in U^*}\TotalWin_{j}[t]}{\sum_{j \in U^*}\alpha_{j}} - \frac{\TotalWin_{i^*}[t]}{\alpha_{i^*}} + o(t) = f_{i^*, U^*}(\mChain[t]) + o(t) \geq  (C + o(1))t$ for some $C > 0$ almost surely for all $j \in U^*$. Hence, for $U_k = U_{k-1} \cup U^*$ has the desired property, completing the induction argument.

Thus we have that $[n] \setminus U_k$ has the splitting property. Since $S^*$ is stable, it cannot be that there exist $j \in S^*$ such that $\frac{\TotalWin_{j}[t]}{\alpha_{j}} - \frac{\TotalWin_{i^*}[t]}{\alpha_{i^*}} \geq  (C_{j} + o(1))t$. Hence, $S^* \subset [n] \setminus U_k$. However, $[n] \setminus U_k$ is stable and contains $i^*$. Hence, $[n] \setminus U_k \subset S^*$. Thus, we conclude that $S^* = [n] \setminus U_k$ has the splitting property.

We now show that $S^*$ is the \textit{unique} maximal subset of $[n]$ containing $i^*$ that satisfies the stability criterion. Assume there exists another such $S' \subset [n]$. Then, by stability, for all $j \in S' \setminus S^*$, $\left|\frac{\TotalWin_{j}[t]}{\alpha_{j}} - \frac{\TotalWin_{i^*}[t]}{\alpha_{i^*}}\right| = o(t)$. However, by the splitting property,  $\left|\frac{\TotalWin_{j}[t]}{\alpha_{j}} - \frac{\TotalWin_{i^*}[t]}{\alpha_{i^*}}\right| = \Omega(t)$. Hence, $S' \setminus S^* = \emptyset$. The same argument shows that $S^* \setminus S'  = \emptyset$. Thus, $S' = S^*$.

\end{proof}

\SplittingPartitionsExist*
\begin{proof}
    We show inductively the we can partition $[n]$ into $C_1, \cdots, C_k$ and $U_k$ such that $C_i$ is stable and for all $i$ and $C_i$ has the splitting property with respect to $\bigcup_{k=i}^nC_k \cup U$. Hence, $\bigcup_{k=1}^i C_i$ has the splitting property for all $i$. Clearly, $U_0 = [n]$ has the property.
    
    We now assume we have $C_1, \cdots, C_{k-1}$ and $U_{k-1}$ with the desired property. If $U_{k-1}$ is stable then $(C_1, \cdots, C_{k-1}, C_k)$ where $C_k = U_{k-1}$ is a splitting partition. Assume $U_{k-1}$ is not stable. Since $\bigcup_{i=1}^{k-1} C_i$ has the splitting property. By Lemma \ref{lemma: process splittng lemma}, $\mChain^{U_{k-1}}[t]$ eventually behaves as a instance the DMMF mechanism with only agents in $U_{k-1}$. Hence, we can apply Lemma \ref{lemma: splitting sets exist} on this instance. Let $i_k = \arg\min_{i \in U_{k-1}} \frac{p_i}{\alpha_i}$. Let $C_k$ be a maximal subset of $U_{k-1}$ that is stable and contains $i_k$. Then, by Lemma \ref{lemma: splitting sets exist}, $C_k$ is stable and has the splitting property wrt. $U_{k-1}$. Hence, we conclude that  $(C_1, \cdots, C_{k-1}, C_k)$ where $C_k = U_{k-1}$  and $U_k = U_{k-1} \setminus C_{k}$ has the desired property. 

    Eventually, we must achieve $U_{k}$ that is stable. Hence, we have constructed a splitting partition.
\end{proof}

\FinalWinRateCharaterization*
\begin{proof}
Recall \cref{def: Y subprocess}. Let $B_i = \bigcup_{k=1}^i C_{k}$.  By the definition of a splitting subset, $B_{i}$ has the splitting property. Hence, we can apply \cref{lemma: process splittng lemma} to see that $\mChain^{B_i}[t] = \mChainY^{B_i}\left(\mChain^{B_i}[T]\right)\left[t - T\right]$ for some $T$. Since, $\mChainY^{B_i}\left(\mChain^{B_i}[T]\right)$ is an instance of the DMMF mechanism and $B_{i-1}$ has the splitting property wrt. $B_i$, we can apply \cref{lemma: process splittng lemma} again, we see that  $$\mChain^{C_i}[t] = (\mChain^{B_i})^{C_i}[t] = \mChainY^{C_i}\left(\mChain^{C_i}[T']\right)\left[t - \TotalAnyReq_{B_{i-1}}[t] - (T'-\TotalAnyReq_S[T'])\right]$$ for some $T'$ Since $C_i$ is stable, we can apply \cref{cor: utility after split} to achieve the result.
\end{proof}

\subsection{Continuity of \texorpdfstring{$\TotalWin_i[t]$}{the total number of wins} and \texorpdfstring{$\U_i(p_1, \cdots, p_n)$}{utility in the Threshold game}}

\begin{lemma} \label{Bounded Difference Lemma}
    Fix $R_i \subset [T]$ for all $i$. Let $W_i(R_1, \cdots, R_n)$  be the number of rounds won by agent $i$ when $\Req_j[t] = \mathds{1}\{t \in R_j\}$ for all $j \in [n]$ and $t \in [T]$. Let $R'_i\subset [T]$. Then, $$\left|W_i(R_1, \cdots, R_n) - W_i(R'_1, \cdots, R'_n)\right| \leq \sum_{j=1}^n\left|R_j \triangle R'_j\right|.$$
\end{lemma}
\begin{proof}
We can think of $R_i$ as a request sequence for agent $i$. Let $W_i(R_1, \cdots, R_n)[t]$  be the number of rounds won by agent $i$ among the first $t$ rounds when request sequence $R_j$ is used by agent $j$. Fix $i \in [n]$. Let $R^1_i \subset [T]$ such that $|R_i \triangle R^1_i| = 1$. Let $U_{j}[t] = W_j(R_1, \cdots, R_i,\cdots R_n)[t] - W_j(R_1, \cdots, R^1_i,\cdots R_n)[t]$. 

Let $\hat{t} \in R_i \triangle R^1_i$ be the unique round where the request behaviour of agent $i$ differs between the request sequences. Observe that $U_{j}[t] = 0$ for all $i$ for all $t < \hat{t}$ since the agent behaviour has not changed. We now show, by induction, that $U_{j}[t] = 1$ for at most 1 $j$, $U_{j}[t] = -1$ for at most 1 $j$ and $U_{j}[t] = 0$ for all other $j$. 

In round $\hat{t}$, if the winner of the round when the request sequence profile is  is $(R_1, \cdots, R_n)$ is the same as the winner when the request sequence profile is $(R_1, \cdots, R^1_i,\cdots R_n)$ then $U_{j}[t] = 0$ for all $t \in [T]$ since the behaviour of agents after round $\hat{t}$ and, hence, the allocation is the same in both profiles.

On the other hand, if in round $\hat{t}$, the winner differs between request profiles. It must be that either agent $i$ won with request profile $(R_1, \cdots, R_n)$ but did not request in request profile $(R_1, \cdots, R^1_i,\cdots R_n)$ or agent $i$ did not request in request profile $(R_1, \cdots, R_n)$ but won in request profile $(R_1, \cdots, R^1_i,\cdots R_n)$. In the former case, $U_{i}[\hat{t}] = 1$. If another agent $l$ requested on round $\hat{t}$, then the new winner has $U_{l}[\hat{t}] = -1$. Similarly, in the latter case, $U_{i}[\hat{t}] = -1$ and $U_{l}[\hat{t}] = 1$ where agent $l$ is the new winner. Furthermore, $U_{l}[\hat{t}] = 0$ for all other agents. This establishes the base case.

Assume the induction hypothesis holds for some $t-1 \geq \hat{t}$. If the winner in round $t$ does not change between request profiles, the claim would hold at round $t$. If instead, agent $l$ won in request profile $(R_1, \cdots, R_n)$ but agent $m$ won in request profile $(R_1, \cdots, R^1_i,\cdots R_n)$, it must be that either agent $U_{l}[t-1] = -1$ or $U_{m}[t-1] = 1$. In the former case, we then have that $U_{l}[t] = 0$ and $U_{m}[t]= U_{m}[t-1] - 1 \in \{0, -1\}$ and, in the latter case, we then have that $U_{m}[t] = 0$ and $U_{l}[t]= U_{l}[t-1] + 1 \in \{0, 1\}$. Hence, the induction hypothesis holds for $t$. Hence, by induction, we see that 
$$\left|W_i(R_1, \cdots, R_n) - W_i(R_1, \cdots,R^1_i, \cdots, R_n)\right| \leq 1$$ for all $i$. 

Let $R^0_i, R^1_i, R^2_i, \cdots, R^{\left|R_i \triangle R'_i\right|}_i$ such that $R^0_i = R_i$, $R^{\left|R_i \triangle R'_i\right|}_i = R'_i$ and $\left|R^k_i \triangle R^{k+1}_i\right| = 1$ for $k \geq 0$. Then, 
\begin{eqnarray*}
    \left|W_i(\vec{R}) - W_i(R'_i,\vec{R}_{-i})\right| &\leq& \sum_{k=0}^{\left|S_i \triangle R'_i\right| - 1}\left|
    W_i(R^{k}_i,\vec{R}_{-i}) - W_i(R^{k+1}_i,\vec{R}_{-i})\right|\\
    &\leq& \left|S_i \triangle S'_i\right|
\end{eqnarray*}
We now conclude by observing that
\begin{eqnarray*}
    \left|W_i(R_1, \cdots, R_n) - W_i(R'_1, \cdots, R'_n)\right| &\leq& \sum_{j=1}^{n}\left|W_i(R'_1, \cdots,R'_{j-1}, R_j, \cdots, R_n) - W_i(R'_1, \cdots, R'_j,R_{j+1} \cdots, R_n)\right|\\
    &\leq& \sum_{j=1}^{n}\left|R_j \triangle R'_j\right|
\end{eqnarray*}
\end{proof}

\begin{lemma}\label{lemma: W is cont}
    Let $\TotalWin_i[t](\strThresh_{p_1}, \cdots, \strThresh_{p_n})$ be the number of rounds won by agent $i$ in the first $t$ rounds when the threshold strategy profile $\vec{\str} = (\strThresh_{p_1}, \cdots, \strThresh_{p_n})$ is used. Then, on almost all sample paths, $$\lim\sup_{t \rightarrow \infty}\left|\frac{\TotalWin_i[t](\strThresh_{p_1}, \cdots, \strThresh_{p_n}) - \TotalWin_i[t](\strThresh_{p'_1}, \cdots, \strThresh_{p'_n})}{t}\right| \leq n\epsilon$$ whenever $\norm{(p'_1 - p_1, \cdots, p'_n-p_n)} \leq \epsilon$ on almost all sample paths of values.
\end{lemma}
\begin{proof}
    Let $\vec{\str} = (\strThresh_{p_1}, \cdots, \strThresh_{p_n})$. We show that $\left|\TotalWin_i[t](\vec{\str}) - \TotalWin_i[t](\strThresh_{p_j+\epsilon},\vec{\str}_{-j})\right| \leq \epsilon \cdot t + o(t)$ almost surely.

    Fix a sample path of values. Under the different between the request behavior of agents under the strategy profiles $(\vec{\str})$ and $(\strThresh_{p_j+\epsilon},\vec{\str}_{-j})$ is that agent $j$ requests, almost surely, $\epsilon\cdot t + o(t)$ more rounds in the latter case by the law of large numbers. When agent $i$ requests in the former case, they will also request in the latter case for all $i$. Hence, the difference in the request sequences as size $\epsilon\cdot t + o(t)$ almost surely, Hence, by Lemma \ref{Bounded Difference Lemma}, on almost all sample paths, 
    $\left|\TotalWin_i[t](\vec{\str}) - \TotalWin_i[t](\strThresh_{p_j+\epsilon},\vec{\str}_{-j})\right| \leq \epsilon \cdot t + o(t)$. Using similar argument, we can show that $\left|\TotalWin_i[t](\vec{\str}) - \TotalWin_i[t](\strThresh_{p_j-\epsilon},\vec{\str}_{-j})\right| \leq \epsilon \cdot t + o(t)$. Hence, if $|p_j - p'_j| < \epsilon$, $$\lim\sup_{t \rightarrow 
 \infty} \left|\frac{\TotalWin_i[t](\vec{\str}) -  \TotalWin_i[t](\strThresh_{p'_j},\vec{\str}_{-j})}{t}\right| \leq \epsilon.$$ We then repeat this argument for all $j$ to achieve that $$\lim\sup_{t \rightarrow 
 \infty} \left|\frac{\TotalWin_i[t](\vec{\str}) -  \TotalWin_i[t](\strThresh_{p'_1}, \cdots, \strThresh_{p'_n})}{t}\right| \leq n\epsilon$$ when $\norm{(p'_1 - p_1, \cdots, p'_n - p_n)} < \epsilon$. The result immediately follows. 
 \end{proof}

    %Assume $\TotalWin_i(p_1, \cdots, p_n)$ is defined on a dense set. Let $\vec{P}$ be a profile such that $\TotalWin_i(\vec{P})$ is not defined. Fix $\epsilon > 0$. Let $\vec{P'}(\epsilon)$ such that $\TotalWin_i(\vec{P'}(\epsilon))$ is defined and $\left|\TotalWin_i(\vec{P})[t] - \TotalWin_i(\vec{P'}(\epsilon))[t]\right| \leq \epsilon \cdot t + o(t)$ on all almost all sample path. Then on almost all sample path, 

\begin{lemma}\label{lemma: U is cont}
    Let $\U_i(p_1, \cdots, p_n) = \lim_{t \rightarrow \infty}\frac{\U_i[t](\strThresh_{p_1}, \cdots, \strThresh_{p_n})}{t}$. Then $\U_i(p_1, \cdots, p_n)$ is continuous on almost all sample paths of values and, almost surely,   $$\U_i(p_1, \cdots, p_n) =\frac{\alpha_i}{\sum_{k \in C_{u}} \alpha_k} \left(1 - \prod_{k \in C_{u}}(1-p_k)\right)\prod_{v < u}\prod_{k \in C_{v}}(1-p_k) \cdot \ValueFunc_{i}(p_i)$$ where $(C_1, \cdots, C_m)$ be a splitting partition with respect to the threshold strategy profile $(\strThresh_{p_1}, \cdots, \strThresh_{p_n})$.
\end{lemma}

\begin{proof}
    
    In \cref{thm: characterize number of wins}, we have shown that $\lim_{t \rightarrow \infty}\frac{\TotalWin_i[t](\strThresh_{p_1}, \cdots, \strThresh_{p_n})}{t}$ exists and is almost surely a fixed value $\TotalWin_i(p_1, \cdots, p_n)$. Lemma \ref{lemma: W is cont} tells us that $\TotalWin_i(p_1, \cdots, p_n)$ is almost surely continuous. By the continuity of $\ValueFunc_{i}(p_i)$, we conclude that
    $$\U_i(\vec{p}) = \ValueFunc_{i}(p_i) \cdot \lim_{t \rightarrow 
 \infty} \frac{\TotalWin_i[t](\vec{\str}) + o(t)}{t} = \ValueFunc_{i}(p_i) \cdot \TotalWin_i[t](p_1, \cdots, p_n),$$ almost surely, is continuous in $p_j$. 
\end{proof}

\section{Appendix C}\label{appsec: proof of no nash}

\ThmNoPureNash*
\begin{proof}
    Assume there exist a pure Nash Equilibrium. Let $(p_1, p_2)$ be the request probability of each agent in the equilibrium. 

    We claim that both agents win the same number of rounds at equilibrium; in particular, the process must be stable. Assume $(p_1, p_2)$ did not satisfy the stability criterion. Assume WLOG that $p_1 < p_2$. Then $p_2(1-p_1) > p_1$. Thus, by \cref{thm:necessary condition 1}, $\TotalWin_2[t] - \TotalWin_1[t] \geq \Omega(t)$ for large $t$. Hence, agent 1 can increase their utility by requesting more while staying in the regime where the the stability criterion does not hold. By the continuity of $\U(p_1, p_2)$ (from \cref{lemma: U is cont}), this argument tells us that agent $1$ should increase $p_1$ until $p_2(1-p_1) \leq p_1$. Hence, indeed at equilibrium, $(p_1, p_2)$ satisfy the stability criterion.
    
    Consider the distribution $\valDist_i(q, \epsilon)$ where, for $v \sim \valDist_i(q, \epsilon)$, $\pr(v = 1) = q$ and $\pr(v = \epsilon) = 1-q$. Given that this distribution is discrete, the $p$-threshold strategy is defined as follows: for $p \leq q$, on any round where $v = 1$, the agent requests the resource with probability $\frac{p}{q}$ and never requests when $v = \epsilon$; for $p \geq q$, on any round where $v = 1$, the agent requests the resource and, on any round where $v = \epsilon$, the agent requests the resource with probability $\frac{p-q}{1-q}$. 

    We first observe that for any agent, using the $q$-threshold strategy dominates using the $p$-threshold strategy for any $p < q$. This is because the value when the agent requests is 1 for all $p \leq q$ and an agent can only win more rounds by asking more. Hence, they can win more rounds while maintaining the utility the achieve when they win. Thus, when considering equilibrium solutions, it is sufficient to have assume $p_1, p_2 \in  [q, 1]$.
    
    Observe that when an agent is using the threshold strategy with probability $p > q$, $\ValueFunc(p) = \frac{q + \epsilon(p-q)}{p}.$ By \cref{thm: condition for stablity},
    \begin{eqnarray*}
        \U_i(p_1, p_2) =  \frac{1-(1-p_1)(1-p_2)}{2}\cdot \ValueFunc(p_i) = \frac{1-(1-p_1)(1-p_2)}{2}\cdot \frac{q + \epsilon(p_i-q)}{p_i}
    \end{eqnarray*}

    We first show that $\U_1(p_1, p_2)$ is convex in $p_1$. Indeed,
    \begin{eqnarray*}
        \frac{\partial \U_1(p_1, p_2)}{\partial p_1} &=&  \frac{\epsilon (1-p_2)}{2} - \frac{q(1-\epsilon)p_2}{2p_1^2}\\
        \frac{\partial^2 \U_1(p_1, p_2)}{\partial p_1^2} &=& \frac{q(1-\epsilon)p_2}{p_1^3} \geq 0
    \end{eqnarray*}

    Furthermore, for $p_2 \geq \frac{1}{2}$, $\U_1(p_1, p_2)$ is decreasing in $p_1$ for $\epsilon = \frac{q}{2+q}$. Indeed,
    \begin{eqnarray*}
        \frac{\partial \U_1(p_1, p_2)}{\partial p_1} &<& \frac{\epsilon (1-p_2)}{2} - \frac{q(1-\epsilon)p_2}{2}\\
         &<& \frac{\epsilon}{2} - \frac{q(1-\epsilon)}{4} = \frac{q}{2(2+q)} -  \frac{q}{4}\left(1-\frac{q}{2+q}\right) = 0
    \end{eqnarray*}

We now set $q = \frac{1}{4}$ and $\epsilon = \frac{q}{2+q}$. By the the convexity of $\U_1(p_1, p_2)$, for any $p_2$, the best response $p_1$ in the region where the stability criterion holds is on one of the boundaries of the region. Hence, either $p_2(1-p_1) = p_1$ or $p_1(1-p_2) = p_2$. We now show that at any Nash Equilibrium, one agent is requests every round. Assume WLOG that $p_1 < p_2$. Then $p_2(1-p_1) = p_1$. Observe that when $p_2(1-p_1) > p_1$, the stability criterion fails and, hence, $\TotalWin_2[t] - \TotalWin_1[t] \geq \Omega(t)$. Thus, agent 2 eventually only wins when agent 1 does not request and they request. Since agent 2 has no control over when agent 1 requests, they increase their utility by asking in every round. By the continuity of $\U_2(p_1, p_2)$, this must also hold when $p_2(1-p_1) = p_1$. Hence, for $(p_1, p_2)$ to be a Nash Equilibrium, it must holds that $p_2 = 1$.

Since $\U_1(p_1, 1)$ is decreasing, the best response of agent 1 is to set $p_1$ such that  $p_2(1-p_1) = p_1$ for $p_2 = 1$. Hence, the best response of agent 1 to agent 2 requesting in every round is $p_1 = \frac{1}{2}$. Thus, the only possible request probability profile that could be a Nash equilibrium is $(\frac{1}{2}, 1)$. However, by symmetry, $\U_2(\frac{1}{2}, p_2)$ is decreasing in $p_2$. Hence, requesting in every round is not a best response for agent 2 to agent 1 requesting with probability $\frac{1}{2}$. Thus, $(\frac{1}{2}, 1)$ is not a Nash equilibrium.
\end{proof}

\section{Proofs of Subsection \ref{ssec:followthedeviator}} \label{sec:app:followthedeviator}

In this section, we shall first prove our claims about the convergence of the random process $M_{\eta,\zeta}[t]$ for appropriate chosen $\eta$ and $\zeta$. Afterwards, we prove the utility bounds when the agents follow \algorithmName.

In this section, $J[t] = \frac{M_{\eta,\zeta}[t] - p^*}{1-\eta(t)}$. Furthermore, let $\zeta(t) = 1 - t^{-\alpha}$ and $\eta(t) = \log(t)^{-\frac{1}{2}+\epsilon}$ for $0 < \epsilon <  \frac{1}{4}$. 

\begin{lemma}\label{lemma: bound on scale ratio}
        Let $f(\xi) = \frac{1-\xi}{1-M_{\eta,\zeta}[t] - \eta(t)\cdot J[t]}$. For some $C_1 > 0$ and $C_2 < \infty$ and sufficiently large $t$, $C_1 < f(\xi) < C_2$  for all $\xi$ between $M_{\eta,\zeta}[t]$ and $M_{\eta,\zeta}[t] + \eta(t)\cdot J[t]$. 
    \end{lemma}
    \begin{proof}
        Clearly since $f(\xi)$ is linear, it is maximized (and minimized) at an extremal value. Furthermore, $f(M_{\eta,\zeta}[t] + \eta(t)\cdot J[t]) = 1$. Hence, it is sufficient to show a bound when on $f(M_{\eta,\zeta}[t])$. We first show the result for $p^* = 1$. Indeed,
        $$\frac{1-M_{\eta,\zeta}[t]}{1-M_{\eta,\zeta}[t] - \eta(t)\cdot J[t]} = \frac{1-M_{\eta,\zeta}[t]}{1-M_{\eta,\zeta}[t] + \eta(t)\cdot \frac{1-M_{\eta,\zeta}[t]}{1-\eta(t)}}  = \frac{1}{1 + \frac{\eta(t)}{1-\eta(t)}} \rightarrow 1 $$ where we use the fact that since $\zeta(t) < 1$ for all $t\in\N$, $M_{\eta,\zeta}[t] \neq 1$ for any $t \in\N$

        We now assume $p^* < 1$. 
        We first bound $f(M_{\eta,\zeta}[t])$ from below. Notice that when $J[t] \geq 0$, $\frac{1-M_{\eta,\zeta}[t]}{1-M_{\eta,\zeta}[t] - \eta(t)\cdot J[t]} \geq 1$. Hence, we assume $J[t] < 0$. Then

    \begin{eqnarray*}
        f(M_{\eta,\zeta}[t]) &=& \frac{1}{1 + \frac{\eta(t)\cdot |J[t]|}{1-M_{\eta,\zeta}[t]}} \geq \frac{1}{1 + \frac{\eta(t)}{1-(1 - \eta(t)) \left(1-t^{-\alpha}\right) - p^* \cdot \eta(t)}} = \frac{1}{1 + \frac{1}{\frac{t^{-\alpha}}{\eta(t)} + 1 -t^{-\alpha}  - p^*}} \rightarrow \frac{1}{1 + \frac{1}{1 - p^*}}
    \end{eqnarray*} 

    We now bound $f(M_{\eta,\zeta}[t])$ from above above. Notice that when $J[t] \leq 0$, $\frac{1-M_{\eta,\zeta}[t]}{1-M_{\eta,\zeta} - \eta(t)\cdot J[t]} \leq 1$. Hence, we assume $J[t] > 0$.
    \begin{eqnarray*}
        f(M_{\eta,\zeta}[t]) &=& \frac{1}{1 - \frac{\eta(t)\cdot J[t]}{1-M_{\eta,\zeta}[t]}} = \frac{1}{1 - \frac{\eta(t)}{1-\eta(t)}\cdot \frac{M_{\eta,\zeta}[t] - p}{1-M_{\eta,\zeta}[t]}} = \frac{1}{1 - \frac{\eta(t)}{1-\eta(t)}\cdot \left(-1 + \frac{1 - p^*}{1-M_{\eta,\zeta}[t]}\right)}\\
        &=& \frac{1}{1 - \frac{\eta(t)}{1-\eta(t)}\cdot \left(-1 + \frac{1 - p^*}{\frac{t^{-\alpha}}{\eta(t)} -t^{-\alpha}  + 1- p^*}\right)} \rightarrow 1
    \end{eqnarray*} 
    proving the result when $p^* < 1$.
    
    \end{proof}

Recall that $\AnyReq[t]$ is the indicator random variable for the event that an agent wins round $t$ (ie. at least one agent requests in round $t$). 
\begin{lemma}\label{lem: change of time index conversion}
   Fix $\alpha < \frac{1}{2}$. For all $t \inN$,
    \begin{eqnarray*}
        \Phi^{-1}\left(\frac{\zeta(t+1)\TotalAnyReq[t]}{t}\right) &=& \Phi^{-1}\left(\frac{\zeta(t)\TotalAnyReq[t-1]}{t-1}\right) + \frac{1 + o(1)}{n(1-M_{\eta,\zeta}[t] - \eta(t)\cdot J[t])^{n-1}} \cdot \frac{\zeta(t)\AnyReq[t] -  \Phi\left(M_{\eta,\zeta}[t]\right)}{t}\\
        &&\;\;\;\;\;\; - \;\theta\left(\frac{ \eta(t)\cdot J[t]}{t}\right)  + O(t^{-\frac{\alpha}{n} - 1})
    \end{eqnarray*} where $\AnyReq[t] = \ind{\Req_i[t] = 1 \text{ for some $i \in [n]$}}$.
\end{lemma}

\begin{proof}
We first observe that 
 \begin{eqnarray*}
     \zeta(t+1)\cdot \frac{\TotalAnyReq[t]}{t} &=& \zeta(t+1)\cdot \frac{\TotalAnyReq[t-1]}{t-1} + \zeta(t+1)\cdot \frac{\AnyReq[t] - \frac{\TotalAnyReq[t-1]}{t-1}}{t}\\
     &=& \zeta(t) \frac{\TotalAnyReq[t-1]}{t-1}+ \zeta(t+1)\cdot \frac{\AnyReq[t] - \frac{\TotalAnyReq[t-1]}{t-1}}{t} +  (\zeta(t+1) - \zeta(t)) \frac{\TotalAnyReq[t-1]}{t-1} \\
     &=& \zeta(t) \frac{\TotalAnyReq[t-1]}{t-1} + \zeta(t+1)\cdot \frac{\AnyReq[t] - \frac{\TotalAnyReq[t-1]}{t-1}}{t} +  O(t^{-1-\alpha})
 \end{eqnarray*} where we use the bound $\zeta(t+1) - \zeta(t) =(t+1)^{-\alpha} - t^{-\alpha} = O(t^{-1-\alpha})$.
    We take a second order Taylor expansion of $\Phi^{-1}(x)$ about $\zeta(t) \frac{\TotalAnyReq[t-1]}{t-1}$ to observe that
    \begin{eqnarray*}
    \Phi^{-1} \left(\zeta(t+1) \frac{\TotalAnyReq[t]}{t}\right) &=& \Phi^{-1} \left(\zeta(t) \frac{\TotalAnyReq[t-1]}{t-1}\right) + O(t^{-2}) \cdot (\Phi^{-1})''(\xi_t) \\
    &&\;\;\;\;\; +\; (\Phi^{-1})' \left(\zeta(t) \frac{\TotalAnyReq[t-1]}{t-1}\right)\left(\zeta(t+1)\cdot \frac{\AnyReq[t] - \frac{\TotalAnyReq[t-1]}{t-1}}{t} +  O(t^{-1-\alpha})\right)
\end{eqnarray*} where $\xi_t$ is between $\zeta(t+1) \frac{\TotalAnyReq[t]}{t}$ and $\zeta(t) \frac{\TotalAnyReq[t-1]}{t-1}$.
Observe that $\Phi^{-1}\left(\zeta(t) \frac{\TotalAnyReq[t-1]}{t-1}\right) = M_{\eta,\zeta}[t] + \eta(t)\cdot J[t]$. 
We then see that $(\Phi^{-1})'(x) = \frac{1}{n}\cdot (1-x)^{\frac{1-n}{n}} = \frac{1}{n(1-\Phi^{-1}(x))^{n-1}}$. 
Hence, 
$$(\Phi^{-1})' \left(\zeta(t)\frac{\TotalAnyReq[t-1]}{t-1}\right) = \frac{1}{n\left(1-\Phi^{-1}\left(\zeta(t)\frac{\TotalAnyReq[t-1]}{t-1}\right)\right)^{n-1}} = \frac{1}{n\left(1-M_{\eta,\zeta}[t] - \eta(t)\cdot J[t]\right)^{n-1}}.$$ 
Furthermore, 
$$(\Phi^{-1})' \left(\zeta(t)\frac{\TotalAnyReq[t-1]}{t-1}\right) \leq (\Phi^{-1})' \left(\zeta(t)\right) = \frac{1}{n}\cdot t^{\frac{\alpha(n-1)}{n}} = O(t^{\frac{\alpha(n-1)}{n}})$$
Similarly, we achieve $(\Phi^{-1})''(x)= \frac{n-1}{n^2} (1-x)^{\frac{1-2n}{n}}$. Hence, 
$$(\Phi^{-1})''\left(\zeta(t)\frac{\TotalAnyReq[t-1]}{t-1}\right) \leq  (\Phi^{-1})''(1-t^{-\alpha}) = \frac{n-1}{n^2} \cdot t^{\frac{\alpha(2n-1)}{n}} = O(t^{\frac{\alpha(2n-1)}{n}}).$$ 
Hence, 
\begin{eqnarray*}
    \Phi^{-1} \left(\zeta(t+1)\frac{\TotalAnyReq[t]}{t}\right) &=& \Phi^{-1} \left(\zeta(t)\frac{\TotalAnyReq[t-1]}{t-1}\right) + \frac{\zeta(t+1)}{n(1-M_{\eta,\zeta}[t] - \eta(t)\cdot J[t])^{n-1}} \cdot \left(\frac{\AnyReq[t] - \frac{\TotalAnyReq[t-1]}{t-1}}{t}\right)\\
    &&\;\;\; +\; O(t^{-1-\alpha}) \cdot O(t^{\alpha\cdot \frac{n-1}{n}}) + O(t^{-2(1-\alpha) -\frac{\alpha}{n}})
\end{eqnarray*}
For $\alpha < \frac{1}{2}$, we see that $O(t^{-1-\frac{\alpha}{n}}) \cdot O(t^{\frac{\alpha(n-1)}{n}}) + O(t^{-2(1-\alpha) -\frac{\alpha}{n}}) = O(t^{-1-\frac{\alpha}{n}})$. 

Observe that $\frac{\TotalAnyReq[t-1]}{t-1} = \frac{\Phi\left(M_{\eta,\zeta}[t] + \eta(t)\cdot J[t]\right)}{\zeta(t)}$ (by the definition of $M_{\eta,\zeta}[t]$). By the mean value theorem, $$\Phi\left(M_{\eta,\zeta}[t] + \eta(t)\cdot J[t]\right) = \Phi\left(M_{\eta,\zeta}[t]\right) + \eta(t)\cdot J[t] \cdot \Phi'(\xi)$$ where $\xi$ is between $M_{\eta,\zeta}[t]$ and $M_{\eta,\zeta}[t] + \eta(t)\cdot J[t]$. We now see that
\begin{eqnarray*}
    \frac{(1 + o(1))\Phi'(\xi)}{n(1-M_{\eta,\zeta}[t] - \eta(t)\cdot J[t])^{n-1}} = (1+o(1))\cdot \left(\frac{1-\xi}{1-M_{\eta,\zeta}[t] - \eta(t)\cdot J[t]}\right)^{n-1} = \theta(1)
\end{eqnarray*} by Lemma \ref{lemma: bound on scale ratio}, from which the result follows.

\end{proof}

The following technical lemma will be the key tool we shall use to demonstrate the almost sure converge of $M_{\eta, \zeta}[t]$ to the desired values.

\begin{lemma} \label{lemma: random process convergence}
    Assume $Z[t]$ is a random process such the $\mathcal{F}_t$ be a filtration such that $Z[t]$ is $\mathcal{F}_{t-1}$ measurable. Furthermore, assume 
    \begin{eqnarray*}\label{eqn: next time step relation}
        Z[t+1]=Z[t]\cdot \left(1 - \Omega\left(\frac{\eta(t)}{t}\right)\right)  + C[t] \cdot (\AnyReq[t] - \E[\AnyReq[t]|\mathcal{F}_{t-1}]) + O\left(\frac{g_1(t)}{t}\right)
    \end{eqnarray*} where $C[t]$ is a $\mathcal{F}_{t-1}$-measurable random variable such that $|C[t]| \leq C \cdot \frac{g_2(t)}{t}$ for some constant $C$ where $g_1(t) = o(1)$ and  $g_2(t) = \omega(1)$. 
    Assume there exist $f(t) = o(1)$ such that $\frac{g_2(t)}{tf(t)} \leq C\cdot t^{-\frac{1}{2}-\epsilon}$ for some $\epsilon > 0$ and $O\left(\frac{g_1(t)}{tf(t)}\right) = o\left(\frac{\eta(t)}{t}\right)$. 
    Let 
    $$h(t) =  \max_{k \in [1, t]}\left\{O(f(k))\cdot \exp\left( - \sum_{s=k+1}^{t-1}\Omega\left(\frac{\eta(s)}{s}\right)  + \sqrt{C^*\log(t-k+5)}\right)\right\}.$$ 
    Then, almost surely, $|Z[t]| \leq  h(t)$ for sufficiently large $t$. %and $\E[|Z[t]|] \leq h(t) + O(t^{-2})$.
\end{lemma}
\begin{proof}

    Let $V_r[q] = \sum_{s=r}^{r+q}\frac{C[s]}{Z[s]}\cdot (\AnyReq[s] - \E[\AnyReq[s]|\mathcal{F}_{s-1}]) \cdot \ind{|Z[s]| > f(s)}$. Observe that $V_r[q]$ is a martingale with $|V_r[q] - V_r[q-1]| \leq \frac{|C[q+t]|}{f(q+t)} = C\cdot \frac{g_2(q+t)}{(q+t)f(q+t)} \leq  C\cdot (q+t)^{-\frac{1}{2}-\epsilon} $.

    Let $S = \left\{s \inN: |Z[s]| \leq f(s) \right\}$ be a random set of indices. Fix $t \inN$. Let $r_t = \max\{r \in S \cup \{0\}: r \leq t\}$. Let $L_k = \{r_t = k\}$ and $B_{k} = \left\{\left|V_{k+1}[t-k-2]\right| \leq \sqrt{C^*\log(t-k+5)}\right\}$ be random events for some fixed $C^*$. Observe that conditioned on the event $L_k$, for $s > k$
    \begin{eqnarray*}
        \frac{Z[s+1]}{Z[s]} &=& 1 - \Omega\left(\frac{\eta(s)}{s}\right)  + \frac{C[s]}{Z[s]} \cdot (\AnyReq[s] - \E[\AnyReq[s]|\mathcal{F}_{s-1}]) + \frac{O(\frac{g_1(t)}{t})}{Z[s]}\\
        &=& 1 - \Omega\left(\frac{\eta(s)}{s}\right)  + \frac{C[s]}{Z[s]} \cdot (\AnyReq[s] - \E[\AnyReq[s]|\mathcal{F}_{s-1}])\cdot \ind{|Z[s]| > f(s)}\\
        &&\;\;\;\;\;\;+\; \frac{O(\frac{g_1(t)}{t})}{Z[s]}\cdot \ind{|Z[s]| > f(s)}\\
        &=& 1 - \Omega\left(\frac{\eta(s)}{s}\right)  + \frac{C[s]}{Z[s]} \cdot (\AnyReq[s] - \E[\AnyReq[s]|\mathcal{F}_{s-1}])\cdot \ind{|Z[s]| > f(s)}
    \end{eqnarray*} where we use the assumption that $O\left(\frac{g_1(t)}{tf(t)}\right) = o\left(\frac{\eta(t)}{t}\right)$, hence, $\Omega\left(\frac{\eta(s)}{s}\right) - o\left(\frac{\eta(t)}{t}\right) = \Omega\left(\frac{\eta(s)}{s}\right)$

    Observe $|1-x| \leq e^{-x}$ for $|x| < 1$. Since $\frac{Z[s+1]}{Z[s]} = 1 - o(1)$, for $s > T^*$ for some fixed $T^*$,
\begin{eqnarray*}
        \frac{|Z[s+1]|}{|Z[s]|} & \leq & \exp\left( - \Omega\left(\frac{\eta(s)}{s}\right)  + \frac{C[s]}{Z[s]} \cdot (\AnyReq[s] - \E[\AnyReq[s]|\mathcal{F}_{s-1}])\cdot \ind{|Z[s]| > f(s)}\right)
    \end{eqnarray*}

    We now apply this bound repeatedly to achieve
    \begin{eqnarray*}
        |Z[t]| &=& |Z[k+1]|\cdot \prod_{s=k+1}^{t-1}\frac{|Z[s+1]|}{|Z[s]|}\\
        &\leq& |Z[k+1]|\cdot \exp\left( - \sum_{s=k+1}^{t-1}\Omega\left(\frac{\eta(s)}{s}\right)  + \sum_{s=k+1}^{t-1}\frac{C[s]}{Z[s]} \cdot (\AnyReq[s] - \E[\AnyReq[s]|\mathcal{F}_{s-1}])\cdot \ind{|Z[s]| > f(s)}\right)\\
        &=& |Z[k+1]|\cdot \exp\left( - \sum_{s=k+1}^{t-1}\Omega\left(\frac{\eta(s)}{s}\right)  + V_{k+1}[t-k-2]\right)
    \end{eqnarray*}

    Observe that from the definition of $k$, $|Z[k+1]| = O(|Z[k]|) = O(f(k))$. Thus,
    \begin{eqnarray*}
        |Z[t]| &\leq& O(f(k))\cdot \exp\left( - \sum_{s=k+1}^{t-1}\Omega\left(\frac{\eta(s)}{s}\right)  + V_{k+1}[t-k-2]\right)
    \end{eqnarray*} 
    When we further condition on the event that $B_k$ occurs, 
    \begin{eqnarray*}
        |Z[t]| &\leq& O(f(k))\cdot \exp\left( - \sum_{s=k+1}^{t-1}\Omega\left(\frac{\eta(s)}{s}\right)  + \sqrt{C^*\log(t-k+5)}\right) \leq h(t)
    \end{eqnarray*} 

Hence, when the event $A_t = \bigcup_{k=1}^{t} L_k\cap B_k$ occurs, $|Z[t]| \leq h(t)$. Observe that exactly one of the $L_k$ must occur. Hence, $A_t^c = \bigcup_{k=1}^t L_k\cap B_k^c$. Hence, $\pr(A_t^c) = \sum_{k=1}^t \pr(L_k\cap B_k^c) \leq \sum_{k=1}^t \pr( B_k^c)$. For $k \leq t-2$, the Azuma–Hoeffding inequality tells us 
\begin{eqnarray*}
    \pr(B^c_k) = \pr\left(\left|V_{k+1}[t-k-2]\right| > \sqrt{C^*\log(t-k+5)}\right) &\leq& 2\exp\left(-\frac{C^*\log(t-k + 5)}{2\cdot C^2\sum_{s=k+1}^{t-1}s^{-1-2\epsilon}}\right)\\
    &\leq& 2\exp\left(-\frac{C^*\log(t-k + 5)}{2\cdot C^2\sum_{s=k+1}^{t-1}s^{-\frac{3}{2}}}\right)
\end{eqnarray*}
Observe that 
$$\sum_{s=k}^{t}s^{-1-2\epsilon} \leq \int_{s=k}^t s^{-1-2\epsilon}\; ds = 2\epsilon\left(k^{-2\epsilon} - t^{-2\epsilon}\right) = 2t^{-2\epsilon}\left(\left(\frac{t}{k}\right)^{2\epsilon} - 1\right).$$ 
Hence, 
$\pr(B^c_k) \leq 2\exp\left(-\frac{C^*\log(t-k + 5)\sqrt{t}}{4\cdot C^2\cdot t^{-2\epsilon}\left(\left(\frac{t}{k}\right)^{2\epsilon} - 1\right)}\right).$
One can show that this function is maximized at $k=1$. Thus, $$\pr(B^c_k) \leq 2\exp\left(-(1+o(1))\log(t-2) \cdot \frac{C^*}{4\cdot C^2}\right) \leq O(1) \cdot (t-2)^{-(1+o(1))\frac{C^*}{4\cdot C^2}} = O(t^{-3})$$ where we take $C^* = 12\cdot C^2$. Hence, $\pr(A_t^c) \leq O(t^{-2})$. We now observe that $\sum_{t=1}^\infty \pr(A_t^c) < \infty$. Hence, by the Borel-Cantelli Lemma, for sufficiently large $t$, $A_t$ occurs almost surely. Hence, almost surely, $|Z[t]| \leq h(t)$ for sufficiently large $t$ proving the claim. 

%The second claim holds as follows:
%$$\E[|Z[t]|] = \E[|Z[t]||A_t]\pr(A_t) + \E[|Z[t]||A^c_t]\pr(A^c_t)  \leq h(t) + O(t^{-2})$$

\end{proof}

\begin{proposition} \label{prop: simplify convergence bound}
    Let $$h(t) =  \max_{k \in [1, t]}\left\{O(f(k))\cdot \exp\left( - \sum_{s=k+1}^{t-1}\Omega\left(\frac{\eta(s)}{s}\right)  + \sqrt{C^*\log(t-k+5)}\right)\right\}.$$ Then, for $f(t) = t^{-\beta}$ for some $\beta > 0$ and $\eta(t) = o(1)$, $$h(t) \leq O\left(\exp\left(- \sum_{s=1}^{t-1}\Omega\left(\frac{\eta(s)}{s}\right)  + O\left(\sqrt{\log(t)}\right)\right)\right).$$ Furthermore, for $f(t) = \frac{1}{\log(k)^{\beta}}$ for some $\beta > 0$ and $\eta(t) = O(1)$,
    $h(t) \leq o(1).$
\end{proposition}
\begin{proof}
Consider the case when $f(t) = t^{-\beta}$ for some $\beta > 0$ and $\eta(t) = o(1)$. Then, $\sum_{s=1}^{k}\Omega\left(\frac{\eta(s)}{s}\right) = o(\log(k))$. Hence, observe that
\begin{eqnarray*}
    h(t) &=& \max_{k \in [1, t]}\left\{k^{-\beta}\cdot \exp\left( - \sum_{s=k+1}^{t-1}\Omega\left(\frac{\eta(s)}{s}\right)  + \sqrt{C^*\log(t-k+5)}\right)\right\}\\
    &\leq&  \max_{k \in [1, t]}\left\{\exp\left(-\beta\log k - \sum_{s=k+1}^{t-1}\Omega\left(\frac{\eta(s)}{s}\right)  + \sqrt{C^*\log(t-k+5)}\right)\right\}\\
    &=&  \max_{k \in [1, t]}\left\{\exp\left(-\beta\log k - \sum_{s=1}^{t-1}\Omega\left(\frac{\eta(s)}{s}\right) + \sum_{s=1}^{k}\Omega\left(\frac{\eta(s)}{s}\right)  + \sqrt{C^*\log(t-k+5)}\right)\right\}\\
    &=&  \max_{k \in [1, t]}\left\{\exp\left(-\beta\log k(1+o(1)) - \sum_{s=1}^{t-1}\Omega\left(\frac{\eta(s)}{s}\right)  + \sqrt{C^*\log(t-k+5)}\right)\right\}\\
    &\leq&  \exp\left(- \sum_{s=1}^{t-1}\Omega\left(\frac{\eta(s)}{s}\right)  + O\left(\sqrt{\log(t)}\right)\right).
\end{eqnarray*}

We now consider the case when $f(t) = \frac{1}{\log(k)^{\beta}}$ for some $\beta > 0$ and $\eta(t) = O(1)$.
\begin{eqnarray*}
    h(t) &=& \max_{k \in [1, t]}\left\{\frac{1}{\log(k)^{\beta}}\cdot \exp\left( - \sum_{s=k+1}^{t-1}\Omega\left(\frac{1}{s}\right)  + \sqrt{C^*\log(t-k+5)}\right)\right\}\\
    &=& \max_{k \in [1, t]}\left\{\exp\left( - \sum_{s=k+1}^{t-1}\Omega\left(\frac{1}{s}\right)  + \sqrt{C^*\log(t-k+5)} - \beta\log\log(k)\right)\right\}\\
    &\leq& \exp\left( O(1) - \Omega(\log\log(t))\right) = o(1)
\end{eqnarray*}
\end{proof}

\ThmThresholdConvergesWhenAllRequest*
\begin{proof}
    Let $Z[t] = \frac{M_{\eta,\zeta}[t] - p^*}{1-\eta(t-1)}$. We now apply Lemma \ref{lem: change of time index conversion}. Observe that $J[t] = Z[t]$. Hence, we achieve
    \begin{eqnarray*}
        Z[t+1] &=& \Phi^{-1}\left(\frac{\zeta(t)\TotalAnyReq[t-1]}{t-1}\right) -p^* + \frac{1 + o(1)}{n(1-M_{\eta,\zeta}[t] - \eta(t)\cdot Z[t])^{n-1}} \cdot \frac{\zeta(t)\AnyReq[t] -  \Phi\left(M_{\eta,\zeta}[t]\right)}{t}\\
        &&\;\;\;\;\;\; - \;\theta\left(\frac{ \eta(t)\cdot Z[t]}{t}\right)  + O(t^{-\frac{\alpha}{n} - 1})\\
        &=& Z[t]  + \frac{1 + o(1)}{n(1-M_{\eta,\zeta}[t] - \eta(t)\cdot Z[t])^{n-1}} \cdot \frac{\zeta(t)\AnyReq[t] -  \Phi\left(M_{\eta,\zeta}[t]\right)}{t}\\
        &&\;\;\;\;\;\; - \; \theta\left(\frac{ \eta(t)\cdot Z[t]}{t}\right)  + O(t^{-\frac{\alpha}{n} - 1})\\
    \end{eqnarray*}

    Let $\mathcal{F}_t$ be the appropriate filtration of the random process. Observe that $\E[\AnyReq[t]|\mathcal{F}_{t-1}] = 1 - (1-M_{\eta,\zeta}[t])^n = \Phi\left(M_{\eta,\zeta}[t]\right)$. Hence,
     \begin{eqnarray*}
        Z[t+1] &=& Z[t]  + \frac{1+o(1)}{n(1-M_{\eta,\zeta}[t] - \eta(t)\cdot Z[t])^{n-1}} \cdot \frac{\AnyReq[t] - \E[\AnyReq[t]|\mathcal{F}_{t-1}]}{t} -  \theta\left(\frac{ \eta(t)\cdot Z[t]}{t}\right)\\
        &&\;\;\;\;\;+\; \frac{1 + o(1)}{n(1-M_{\eta,\zeta}[t] - \eta(t)\cdot Z[t])^{n-1}} \cdot\frac{(\zeta(t)-1)\AnyReq[t]}{t} + O(t^{-\frac{\alpha}{n} - 1})
    \end{eqnarray*}

    Let $C[t]  = \frac{1+o(1)}{n(1-M_{\eta,\zeta}[t] - \eta(t)\cdot Z[t])^{n-1}}\cdot \frac{1}{t}= \frac{1+o(1)}{n\left(1-\Phi^{-1}\left(\zeta(t) \frac{\TotalAnyReq[t-1]}{t-1}\right)\right)^{n-1}} \cdot \frac{1}{t}$. Then, 
    $$C[t] \leq \frac{1+o(1)}{n\left(1-\Phi^{-1}\left(\zeta(t)\right)\right)^{n-1}} \cdot \frac{1}{t} = \frac{1+o(1)}{n\left(1-\zeta(t)\right)^\frac{n-1}{n}} \cdot \frac{1}{t} = \frac{1 + o(1)}{n}\cdot t^{\frac{\alpha(n-1)}{n}-1} = O(t^{-1 + \frac{\alpha(n-1)}{n}} )$$

    Furthermore, $C[t]\cdot (1-\zeta(t)) \leq O(t^{-1 - \frac{\alpha}{n}} )$. Hence, 
    \begin{eqnarray*}
        Z[t+1] &=& Z[t]  + C[t] \cdot (\AnyReq[t] - \E[\AnyReq[t]|\mathcal{F}_{t-1}]) -  \theta\left(\frac{ \eta(t)}{t}\right)\cdot Z[t] + O(t^{-\frac{\alpha}{n} - 1})
    \end{eqnarray*}
    We now apply Lemma \ref{lemma: random process convergence} with $f(t) = t^{-\frac{\alpha}{2n}}$ and Proposition \ref{prop: simplify convergence bound}, to achieve that 
    \begin{eqnarray*}
        |Z[t]| &\leq& \max_{k \in [1, t]}\left\{O(k^{-\frac{\alpha}{2n}})\cdot \exp\left( - \sum_{s=k+1}^{t-1}\Omega\left(\frac{\eta(s)}{s}\right)  + \sqrt{C^*\log(t-k+5)}\right)\right\}\\
        &\leq& O\left(\exp\left(- \sum_{s=1}^{t-1}\Omega\left(\frac{\eta(s)}{s}\right)  + O\left(\sqrt{\log(t)}\right)\right)\right)
    \end{eqnarray*} almost surely. Observe that taking $\eta(t) = \frac{1}{log(t)^{\frac{1}{2}-\epsilon}}$ for any $\epsilon > 0$ is sufficient to have $|Z[t]| = o(1)$ almost surely. Hence, $|M_{\eta,\zeta}[t] - p^*| = (1+o(1))|Z[t]| \rightarrow 0$ almost surely.
    
\end{proof}

\ThmThresholdConvergesWhenDeviator*
\begin{proof}
    Let $Z[t] = \frac{M_{\eta,\zeta}[t] - \hat{p}}{1-\eta(t)}$. We now apply Lemma \ref{lem: change of time index conversion}. Hence, we achieve

        \begin{eqnarray*}
        Z[t+1] &=& \frac{M_{\eta,\zeta}[t+1] - p^*}{1-\eta(t+1)} + \frac{p^*-\hat{p}}{1-\eta(t+1)}\\         
        &=& \Phi^{-1}\left(\frac{\zeta(t)\TotalAnyReq[t-1]}{t-1}\right)-p^* + \frac{1 + o(1)}{n(1-M_{\eta,\zeta}[t] - \eta(t)\cdot Z[t])^{n-1}} \cdot \frac{\zeta(t)\AnyReq[t] -  \Phi\left(M_{\eta,\zeta}[t]\right)}{t}\\
        &&\;\;\;\;\;\; - \;\theta\left(\frac{ \eta(t)\cdot J[t]}{t}\right) + \frac{p^*-\hat{p}}{1-\eta(t+1)}   + O(t^{-\frac{\alpha}{n} - 1})\\
        &=& Z[t]  + \frac{1 + o(1)}{n(1-M_{\eta,\zeta}[t] - \eta(t)\cdot J[t])^{n-1}} \cdot \frac{\zeta(t)\AnyReq[t] -  \Phi\left(M_{\eta,\zeta}[t]\right)}{t}\\
        &&\;\;\;\;\;\; - \;\theta\left(\frac{ \eta(t)\cdot J[t]}{t}\right)  + \frac{(p^*-\hat{p})\cdot (\eta(t) - \eta(t+1))}{1-\eta(t+1)} + O(t^{-\frac{\alpha}{n} - 1})\\
        &=& Z[t]  + \frac{1 + o(1)}{n(1-M_{\eta,\zeta}[t] - \eta(t)\cdot J[t])^{n-1}} \cdot \frac{\zeta(t)\AnyReq[t] -  \Phi\left(M_{\eta,\zeta}[t]\right)}{t}\\
        &&\;\;\;\;\;-\; \theta\left(\frac{ \eta(t)\cdot Z[t]}{t}\right)  + O(\eta'(t))
    \end{eqnarray*} where we use the fact that $\eta(t) - \eta(t+1) = O(\eta'(t))$  by the mean value theorem and $O(t^{-\frac{\alpha}{n} - 1}) = O(\eta'(t))$.
    
Let $\mathcal{F}_t$ be the appropriate filtration of the random process. Observe that $\E[\AnyReq[t]|\mathcal{F}_{t-1}] = 1 - (1-M_{\eta,\zeta}[t])^{n-1}(1-\hat{p})$. Hence, 
 \begin{eqnarray*}
        Z[t+1] &=& Z[t]  + \frac{1 + o(1)}{n(1-M_{\eta,\zeta}[t] - \eta(t)\cdot J[t])^{n-1}} \cdot \frac{1 - (1-M_{\eta,\zeta}[t])^{n-1}(1-\hat{p}) - 1+(1-M_{\eta,\zeta}[t])^n}{t}\\
        &&\;\;\;\;\; - \; \theta\left(\frac{ \eta(t)\cdot J[t]}{t}\right) +  C[t] \cdot \frac{\AnyReq[t] - \E[\AnyReq[t]|\mathcal{F}_{t-1}]}{t} + O(\eta'(t))\\
        &=& Z[t]  + \left(\frac{1-M_{\eta,\zeta}[t]}{1-M_{\eta,\zeta}[t] - \eta(t)\cdot J[t]}\right)^{n-1} \cdot \frac{\hat{p}-M_{\eta,\zeta}[t]}{n\cdot t} - \theta\left(\frac{ \eta(t)\cdot J[t]}{t}\right)\\
        &&\;\;\;\;\;+\;  C[t] \cdot \frac{\AnyReq[t] - \E[\AnyReq[t]|\mathcal{F}_{t-1}]}{t} + O(\eta'(t))\\
        &=& Z[t]  - \theta\left( \frac{1}{t} \right)\cdot Z[t] +  C[t] \cdot \frac{\AnyReq[t] - \E[\AnyReq[t]|\mathcal{F}_{t-1}]}{t} + O(\eta'(t)) + O\left(\frac{ \eta(t)}{t}\right)
    \end{eqnarray*} 
    where we apply Lemma \ref{lemma: bound on scale ratio}. Observe that $\eta'(t) = O\left(\frac{1}{t\log(t)^{\frac{3}{2}-\epsilon}}\right) \leq O\left(\frac{ \eta(t)}{t}\right)$. Hence, 

\begin{eqnarray*}
        Z[t+1] &=& Z[t]  - \theta\left( \frac{1}{t} \right)\cdot Z[t]   +  C[t] \cdot \frac{\AnyReq[t] - \E[\AnyReq[t]|\mathcal{F}_{t-1}]}{t} + O\left(\frac{ \eta(t)}{t}\right)
    \end{eqnarray*} 
    We now apply Lemma \ref{lemma: random process convergence} with $f(t) = \frac{1}{\log(t)^{\frac{1}{4}-\epsilon}}$ and Proposition \ref{prop: simplify convergence bound} to achieve that 
    $$|Z[t]| \leq \max_{k \in [1, t]}\left\{O\left(\frac{1}{\log(k)^{\frac{1}{4}-\epsilon}}\right)\cdot \exp\left( - \sum_{s=k+1}^{t-1}\Omega\left(\frac{1}{s}\right)  + \sqrt{C^*\log(t-k+5)}\right)\right\} = o(1)$$ almost surely.
Hence, $|M_{\eta,\zeta}[t] - p^*| = (1+o(1))|Z[t]| \rightarrow 0$ almost surely.
    
\end{proof}

\ThmFTDIsEquilibrium*
\begin{proof}
    Consider any sample path where $M_{\eta,\zeta}[t] \rightarrow p$. Fix $\epsilon > 0$. 
    Let $T_\epsilon$ such that $\left|M_{\eta,\zeta}[t] - p\right| < \epsilon$ for $t > T_\epsilon$. We now consider the difference in the behaviour of agent $i$ when using $\strThresh_{p}$ instead of $\str^{\ftd}_{\eta, \zeta}$. Observe that the request behaviour differs with probability at most $2\epsilon$ in each round. Almost surely, they differ in at most $2 \epsilon t + o(t)$ rounds. Hence, by Lemma \ref{Bounded Difference Lemma}, for all $j \in [n]$
    $$\lim_{t \rightarrow \infty}\sup\left|\frac{\TotalWin_j[t](\str^{\ftd}_{\eta, \zeta}, \vec{\str}_{-i})}{t} - \frac{\TotalWin_j[t](\strThresh_{p},\vec{\str}_{-i})}{t}\right|\leq 2\epsilon$$ where $\vec{\str}_{-i}$ is an arbitrary profile of strategies for agents except $i$. Since this holds for all $\epsilon$,
    $$\lim_{t \rightarrow \infty}\sup\left|\frac{\TotalWin_j[t](\str^{\ftd}_{\eta, \zeta}, \vec{\str}_{-i})}{t} - \frac{\TotalWin_j[t](\strThresh_{p},\vec{\str}_{-i})}{t}\right|= 0.$$

    Hence, $\TotalWin_j[t](\str^{\ftd}_{\eta, \zeta}, \vec{\str}_{-i}) =\TotalWin_j[t](\strThresh_{p},\vec{\str}_{-i}) + o(t)$ almost surely.

    Let $\Win_j[t](\vec{\str})$ be the indicator for whether at $j$ wins round $t$ when agents use the strategy profile $\vec{\str}$.
    Observe that 
    \begin{eqnarray*}
        \U_i[t](\str^{\ftd}_{\eta, \zeta}, \vec{\str}_{-i}) &=& \sum_{s=1}^t\Win_j[s](\str^{\ftd}_{\eta, \zeta}, \vec{\str}_{-i}) \cdot \ValueFunc_{i}(M_{\eta,\zeta}[s]) +o(t)\\
        &=& \sum_{s=T_{\epsilon}}^t\Win_j[s](\str^{\ftd}_{\eta, \zeta}, \vec{\str}_{-i}) \cdot \ValueFunc_{i}(p) - \sum_{s=T_{\epsilon}}^t\Win_j[s](\str^{\ftd}_{\eta, \zeta}, \vec{\str}_{-i}) \cdot \left(\ValueFunc_{i}(M_{\eta,\zeta}[s]) - \ValueFunc_{i}(p)\right) + o(t)\\
        &=& \TotalWin_j[t](\strThresh_{p},\vec{\str}_{-i}) \cdot \ValueFunc_{i}(p) - \sum_{s=T_{\epsilon}}^t\Win_j[s](\str^{\ftd}_{\eta, \zeta}, \vec{\str}_{-i}) \cdot \left(\ValueFunc_{i}(M_{\eta,\zeta}[s]) - \ValueFunc_{i}(p)\right) + o(t)
    \end{eqnarray*} where we use $\sum_{s=T_{\epsilon}}^t\Win_j[s](\str^{\ftd}_{\eta, \zeta}, \vec{\str}_{-i}) \cdot \ValueFunc_{i}(p) = \left(\TotalWin_j[T](\str^{\ftd}_{\eta, \zeta}, \vec{\str}_{-i}) - \TotalWin_j[T_{\epsilon}](\str^{\ftd}_{\eta, \zeta}, \vec{\str}_{-i})\right)\cdot \ValueFunc_{i}(p) = \TotalWin_j[T](\str^{\ftd}_{\eta, \zeta}, \vec{\str}_{-i})\cdot \ValueFunc_{i}(p) - O(1)$. We now see that 
    \begin{eqnarray*}
    \left|\U_i[t](\str^{\ftd}_{\eta, \zeta}, \vec{\str}_{-i}) - \U_i[t](\strThresh_{p}, \vec{\str}_{-i})\right| &\leq& \left|\sum_{s=T_{\epsilon}}^t\Win_j[s](\str^{\ftd}_{\eta, \zeta}, \vec{\str}_{-i}) \cdot \left(\ValueFunc_{i}(M_{\eta,\zeta}[s]) - \ValueFunc_{i}(p)\right)\right| + o(t)\\
    &\leq& \sum_{s=T_{\epsilon}}^t  \left|\ValueFunc_{i}(M_{\eta,\zeta}[s]) - \ValueFunc_{i}(p)\right| + o(t) \leq \epsilon' \cdot t + o(t) 
    \end{eqnarray*} where we use the continuity of $\ValueFunc_{i}(p)$ to have that $\epsilon'$ is a constant such that $\epsilon' \rightarrow 0$ as $\epsilon \rightarrow 0$. Hence, $$\lim\sup_{t \rightarrow \infty} \left|\frac{\U_i[t](\str^{\ftd}_{\eta, \zeta}, \vec{\str}_{-i})}{t} - \frac{\U_i[t](\strThresh_{p}, \vec{\str}_{-i})}{t}\right| \leq \epsilon'$$ for all $\epsilon' > 0$. Thus, $\U_i[t](\str^{\ftd}_{\eta, \zeta}, \vec{\str}_{-i}) = \U_i[t](\strThresh_{p}, \vec{\str}_{-i}) + o(t)$ almost surely.

    By Lemma \ref{Thm:ThresholdConvergesWhenAllRequest} and \ref{Thm:ThresholdConvergesWhenDeviator} and repeated application of the logic above, we see that
$$\U_i[t](\str^{\ftd}_{\eta, \zeta}, \cdots, \str^{\ftd}_{\eta, \zeta}) = \U_i[t](\strThresh_{p^*}, \cdots, \strThresh_{p^*}) + o(t)$$ 
and 
$$\U_i[t](\str^{\ftd}_{\eta, \zeta}, \cdots,\strThresh_{p}, \cdots, \str^{\ftd}_{\eta, \zeta}) = \U_i[t](\strThresh_{p}, \cdots, \strThresh_{p}) + o(t).$$ 
The result follows because, by the definition of $p^*$, $$\U_i[t](\strThresh_{p^*}, \cdots, \strThresh_{p^*}) \geq \U_i[t](\strThresh_{p}, \cdots, \strThresh_{p}) - o(t).$$

\end{proof}

Finally, we prove \cref{thm:utility_comparison} which we restate here.

\ThmUtilityComparison*

\begin{proof}
    From \cref{Thm:ThresholdConvergesWhenAllRequest} we know that $M_{\eta,\zeta}[t] \rightarrow p^*$ almost surely.
    Using \cref{thm: characterize number of wins,eq: expected utility expression} we get that for the utility of every agent 
    \begin{equation*}
        \lim_{t \to \infty} \frac{U_i [t]}{t}
        =
        \ValueFunc_i(p^*)
        \frac{1 - \qty(1 - p^*)^n}{n}
    \end{equation*}
    almost surely.
    First, for arbitrary distributions we use \cref{prop:symmetricUtilMax} and consider the request probability $p = \frac{1}{n}$ which leads to less utility than $p^*$.
    Using the fact that $\ValueFunc_i(1/n) = n v_i^\star$, where $v_i^\star$ is agent $i$'s ideal utility, and get
    \begin{equation*}
        \lim_{t \to \infty} \frac{U_i [t]}{t}
        \ge
        n v_i^\star
        \frac{1 - \qty( 1 - \frac{1}{n} )^n}{n}
        \ge
        v_i^\star \qty( 1 - \frac{1}{e} )
    \end{equation*}
    where we used the fact that $\qty( 1 - \frac{1}{n} )^n \le \frac{1}{e}$.
    This proves the theorem for arbitrary distributions.
    
    For uniform distributions we use that $\ValueFunc_i(p) = 1 - \frac{p}{2}$ and considering the request probability $\frac{\log n}{n}$, we get
    \begin{equation*}
        \lim_{t \to \infty} \frac{U_i [t]}{t}
        \geq
        \qty( 1 - \frac{\log n}{2n} )\cdot 
        \frac{1 - \qty(1 - \frac{\log n}{n})^n}{n}
        =
        v_i^\star \cdot \frac{\qty(2n - \log n)\qty(1 - \qty( 1 - \frac{\log n}{n} )^n )}{2n - 1}
    \end{equation*}
    where we used the fact that $v_i^\star = \frac{1}{n} - \frac{1}{2n^2}$.
    We now have that
    \begin{equation*}
        \frac{\qty(2n - \log n)\qty(1 - \qty( 1 - \frac{\log n}{n} )^n )}{2n - 1}
        \ge
        \qty( 1 - \frac{\log n}{n}) \qty(1 - \qty( 1 - \frac{\log n}{n} )^n )
        =
        1 - O\qty( \frac{\log n}{n} )
    \end{equation*}
    which follows from the fact that $\qty( 1 - \frac{\log n}{n} )^n = O(\frac{1}{n})$.
    This proves the theorem for uniform distributions.
\end{proof}

%\newpage
%\input{Notes/Proof Appendix 2 backup}

%\newpage
%
%\newpage

%\input{Paper Draft/Appendix/Proof Appendix E}

\end{document}